\documentclass[prd,,aps,showpacs,nofootinbib,amsmath,amssymb]{revtex4}
\usepackage{amssymb,epsfig}
\usepackage{mathrsfs}
\usepackage[usenames,dvipsnames]{color}
\usepackage[bookmarks]{hyperref}
\usepackage[usenames,dvipsnames]{color}

\newcommand{\Real}{\mathbb{R}}

\newcommand{\dvol}{\mbox{dvol}}

\newcommand{\proof}{\noindent {\bf Proof. }}
\newcommand{\proofof}[1]{\noindent {\bf Proof of #1. }}
\newcommand{\qed}{\hfill \fbox{} \vspace{.3cm}}

\newtheorem{lemma}{Lemma}
\newtheorem{proposition}{Proposition}

\newtheorem{theorem}{Theorem}

\begin{document}

\title{Accretion of a relativistic, collisionless kinetic gas into a Schwarzschild black hole}

\author{Paola Rioseco and Olivier Sarbach}
\affiliation{$^1$Instituto de F\'isica y Matem\'aticas,
Universidad Michoacana de San Nicol\'as de Hidalgo,
Edificio C-3, Ciudad Universitaria, 58040 Morelia, Michoac\'an, M\'exico.}

\begin{abstract}
We provide a systematic study for the accretion of a collisionless, relativistic kinetic gas into a nonrotating black hole. To this end, we first solve the relativistic Liouville equation on a Schwarzschild background spacetime. The most general solution for the distribution function is given in terms of appropriate symplectic coordinates on the cotangent bundle, and the associated observables, including the particle current density and stress energy-momentum tensor, are determined. Next, we explore the case where the flow is steady-state and spherically symmetric. Assuming that in the asymptotic region the gas is described by an equilibrium distribution function, we determine the relevant parameters of the accretion flow as a function of the particle density and the temperature of the gas at infinity. In particular, we find that in the low temperature limit the tangential pressure at the horizon is about an order of magnitude larger than the radial one, showing explicitly that a collisionless gas, despite exerting kinetic pressure, behaves very differently than an isotropic perfect fluid, and providing a partial explanation for the known fact that the accretion rate is much lower than in  the hydrodynamic case of Bondi-Michel accretion. Finally, we establish the asymptotic stability of the steady-state spherical flows by proving pointwise convergence results which show that a large class of (possibly nonstationary and nonspherical) initial conditions for the distribution function lead to solutions of the Liouville equation which relax in time to a steady-state, spherically symmetric configuration.
\end{abstract}

\date{\today}


\maketitle

\section{Introduction}

The relativistic kinetic theory of gases started over 100 years ago with papers by J\"uttner ~\cite{fJ11a,fJ11b} who generalized the well-known Maxwell-Boltzmann distribution function to the special relativistic case. Over the years, the theory was further developed and cast into the context of general relativity, see for example~\cite{jS34,gTjW61,wI63,CercignaniKremer-Book}. By now, the formal structure of the theory is well understood, and manifestly covariant formulations based on the tangent bundle associated with the spacetime manifold have been given which exploit their natural metric and symplectic structures, see for instance~\cite{rL66,jE71,jE73,oStZ13,oStZ14a,oStZ14b} and references therein.

Recently, there has been a growing interest in applications of the theory, both from a mathematical and an astrophysical point of view. From the point of view of mathematical relativity, there has been a lot of activity in studying the qualitative properties of the solutions of the Einstein-Liouville system\footnote{In this article, we denote by the ``Liouville equation'' the collisionless Boltzmann equation, which is also known as the ``Vlasov equation" in the literature.} of equations, describing a self-gravitating collisionless gas, see~\cite{hA11} for a recent review on this topic. These studies include the analysis of the well posedness of the Cauchy problem and the existence of global in time solutions and their asymptotic properties with applications to the nonlinear stability of Minkowski spacetime~\cite{gRaR92,mD06,dFjJjS15,mT16} and the future stability of the Universe~\cite{Ringstrom-Book,hAhR16,dF16}. Other works analyze the complete gravitational collapse of a collisionless kinetic gas (see for instance~\cite{aRjV10,hA12,hA14} and \cite{aAmC14} and references therein for numerical work on critical collapse), which constitutes a more general model than the simple dust collapse one which, as is well-known, leads to the formation of naked shell-focusing singularities~\cite{dElS79,dC84,nOoS11}. Further mathematical results establish the existence of static, spherically symmetric solutions~\cite{gR93,hAgR06,hAdFmT15} (see~\cite{hAmEgR09} for related numerical work) or axisymmetric solutions which are either static or stationary (see~\cite{hAmKgR11,hAmKgR14} for recent work and~\cite{eAhAaL16} for a related numerical study). A further interesting, more difficult problem is the stability analysis of these solutions. For a numerical investigation in spherical symmetry, see~\cite{hAgR06b}. For work on the relativistic Boltzmann equation including the collision term, see for instance~\cite{dByC73,pNnNaR04,hLaR13}.

Regarding applications towards astrophysics, the relativistic kinetic theory of gases  is playing an increasingly important role in view of recent and near-future observations of supermassive black holes in the center of galaxies, like the ones in the Milky Way and M87, at scales smaller than their gravitational radius~\cite{EHT}. These observations often require specific models for describing the matter and plasma surrounding such black holes, and clearly a fully consistent description should be based on a general relativistic formulation. Other astrophysical applications of the relativistic kinetic theory include the description of a distribution of stars around a supermassive black hole (see~\cite{pP72,jBrW76} for standard references based on the Newtonian approximation and~\cite{sSsT86} for a relativistic treatment based on $N$-body simulations) and the modeling of dark matter (see, for instance~\cite{sCtSsPgPyS15}). The relativistic Boltzmann equation also plays an important role in the modeling of the early Universe~\cite{cMeB95}; see also~\cite{dBgDuHmMjN16,dBgDuHmMjN16b} for related recent analytic work.

Although none of these astrophysical applications will be directly addressed in the present work (only a simpler astrophysical example will be discussed in section~\ref{Sec:Accretion}), they serve as a motivation for a thorough mathematical investigation for the propagation of a relativistic kinetic gas on a curved spacetime geometry. In this work, we initiate such a systematic study for the case in which the spacetime is generated by an isolated black hole. More specifically, we work under the assumption that the gas is sufficiently diluted for its self-gravity to be neglected. Assuming that no other matter sources are present and that the accretion rate is small enough such that the black hole can be considered to be stationary, the no-hair theorems~\cite{lrr-2012-7} imply that the spacetime geometry belongs to the two-parameter family of Kerr solutions, which are characterized by their mass and angular momentum. For simplicity, we further assume that the rotation of the black hole can be neglected, in which case the background spacetime reduces to the Schwarzschild geometry. Finally, we work on the hypothesis that collisions between the gas particles can be neglected. Therefore, the gas in our model is described by a one-particle distribution function satisfying the relativistic Liouville equation on a Schwarzschild background. A study towards more realistic models relaxing some of these assumptions will be provided in future work.

The main results presented in this article are the following: (i) we derive the most general collisionless distribution function on a Schwarzschild background. This derivation is based on the theory of integrable Hamiltonian systems and parallels the earlier presentation in~\cite{oStZ14b} for the Kerr background. However, in contrast to the result derived in~\cite{oStZ14b}, here we also compute the spacetime observables (namely, the current density and the stress energy-momentum tensor). Furthermore, in this work the distribution function and the observables are expressed in terms of horizon-penetrating coordinates which facilitates their interpretation on the horizon. We emphasize that although the spacetime metric is static and spherically symmetric, our solutions for the distribution function and the corresponding observables do not necessarily posses any symmetries (in particular they could be time-dependent and nonspherical). (ii) As a specific example of astrophysical interest, we specialize our result to the case where the distribution function is static and spherically symmetric and we provide explicit expressions for the observables assuming that in the asymptotic region the distribution function is described by an equilibrium distribution function. Next, we apply these results to the spherical accretion problem and compute the accretion and compression rates of the gas. Finally, in the low-temperature limit, we compare our results to previous calculations based on the Newtonian approximation~\cite{ZelNovik-Book,Shapiro-Book} and find agreement. By computing the radial and tangential pressures at the horizon, we also shed light on the reason for which the accretion rate is much lower than in the corresponding Bondi-Michel accretion models~\cite{hB52,fM72,eCoS15a} in the hydrodynamic case. (iii) Finally, we study the nonlinear stability of the static, spherically symmetric solutions described in (ii). To this end, we consider initial data for the distribution function on a hypersurface in phase space of constant time $t$ which is not necessarily spherically symmetric, but converges (in an appropriate sense) to a function $f_\infty(E)$ depending only on the energy $E$ of the gas particles in the asymptotic region $r\to \infty$. Further assuming that this initial distribution function is uniformly bounded by an equilibrium distribution function and only populates unbounded trajectories, we analyze the  behavior of the observables along the world lines of future-directed timelike observers reaching timelike infinity. Assuming that these observers have a constant asymptotic, purely radial velocity, we prove that the components of the particle current density and stress energy-momentum tensor with respect to a Fermi-propagated frame converge pointwise to those computed from $f_\infty(E)$, thus establishing the stability of the static, spherically symmetric solution belonging to $f_\infty(E)$. In particular, the observables decay pointwise to zero if $f_\infty = 0$, meaning that in this case the gas disperses completely.

The remainder of this article is organized as follows. In section~\ref{Sec:Cotangent} we provide a brief review of the geometrical structures that are relevant for the formulation of relativistic kinetic theory in this work. Contrary to many previous works in the literature which are based on the tangent bundle $T M$, in this article we find it more convenient to formulate the theory on the cotangent bundle $T^*M$ associated with the spacetime manifold $(M,g)$ since it is more naturally adapted to the Hamiltonian formalism on which this work is based. Our review includes a discussion of the symplectic form and the Liouville vector field on $T^* M$ and the properties of the complete lift $\hat{\xi}$ of a Killing vector field $\xi$ on $(M,g)$. These notions from classical Hamiltonian mechanics are key for the construction of the new symplectic coordinates $(Q^\mu,P_\mu)$, $\mu = 0,1,2,3$, which are required for the explicit representation formula of the distribution function.

Next, in section~\ref{Sec:Distribution}, we analyze in detail the structure of the invariant submanifolds of $T^* M$ which are determined by the constants of motion. An important part of our analysis is the restriction of the phase space to unbounded trajectories emanating from the asymptotic region which are relevant for the accretion process. Exploiting the fact that the geodesic motion in the Schwarzschild spacetime constitutes an integrable Hamiltonian system, we derive the most general solution of the Liouville equation on a Schwarzschild background in horizon-penetrating coordinates. In general, this solution depends on the seven phase space variables $(Q^1,Q^2,Q^3,P_\mu)$; hence, it does not need to be stationary nor spherically symmetric. However, a time-independent and spherical distribution function is obtained after requiring the distribution function to be invariant with respect to the complete lifts of the Killing vector fields of the Schwarzschild spacetime. In this case, the distribution function depends only on the mass, energy and total angular momentum of the particle. 

Next, in section~\ref{Sec:Observables} we specify the observables on the spacetime manifold $M$ which, at each $x\in M$, are obtained from a fibre integral of the distribution function over the canonical momentum $p\in T_x^* M$. Besides giving results which are applicable to the general solution of the Liouville equation, we also specialize these expressions to the case where the distribution function depends only on the energy $E$ of the gas particles, in which case the fibre integrals simplify considerably. Furthermore, we compare the structure of the observables to those belonging to an isotropic, perfect fluid.

In section~\ref{Sec:Accretion} we apply our results to the accretion problem. First, we focus on the spherical, steady-state case in which all the gas particles have the same rest mass. Further, we assume that the gas is in thermodynamic equilibrium at temperature $T$ at infinity, implying that the distribution function is proportional to $e^{-\frac{E}{k_B T}}$, with $k_B$ Boltzmann's constant. We compute the particle and energy fluxes $j_{n}$ and $j_{\varepsilon}$, respectively, and also the particle density $n_{\infty}$, energy density $\varepsilon _{\infty}$ and pressure $p_{\infty}$ at infinity which, as a consequence of our assumptions, obey the ideal gas equation. Next, we calculate the radial and tangential kinetic pressure of the gas, $p_{rad}$ and $p_{tan}$, and the particle and energy densities $n_{H}$ and $\varepsilon_{H}$ on the horizon.  In particular, we note that in the low temperature limit the tangential pressure is almost ten times larger than the radial one, implying that the gas does not behave as an isotropic perfect fluid at the horizon, although it does so in the asymptotic region. The fact that the tangential pressure becomes more important as the horizon is approached fits well with the known fact that the accretion rate for a collisionless gas is much lower than in the hydrodynamic case.

Second, in section~\ref{Sec:Accretion} we present a study for the stability of the steady-state, spherical accretion flow. In order to achieve this, it is necessary to relax the symmetry assumptions on the distribution function, and our only essential hypothesis is that the distribution function is bounded by an equilibrium distribution function and that it converges to some functions $f_{\pm \infty}(E)$ of the energy $E$ when $r\to \infty$ along the initial time slice $t = 0$ in phase space, where the $\pm$ sign refers to the sign of the radial velocity of the particles at infinity. Using Lebesgue's dominated convergence theorem, we compute the asymptotic behavior of the components of the observables with respect to a Fermi-propagated frame along the world lines of timelike observers. For static observers, we prove that the observables converge pointwise to the corresponding static and spherical observables associated with the asymptotic function $f_{-\infty}(E)$. For outgoing timelike observers with asymptotically constant, purely radial velocity, we show that the observables also converge to those describing a stationary and spherically symmetric gas; however, in this case the result depends on both $f_{-\infty}(E)$ and $f_{+\infty}(E)$. For the physically most relevant case $f_{-\infty} = f_{+\infty}$ in which the asymptotic distribution function does not discriminate between in- and outgoing particle, our result implies the asymptotic stability of the steady-state, spherical accretion flow.

Finally, conclusions and an outlook on future work are given in section~\ref{Sec:Conclusions}. Technical details, calculations and proofs of certain auxiliary results are given in appendices.

With the exception of some results given in section~\ref{SubSec:SSS} regarding the steady-state, spherical accretion configurations, throughout this work we use geometrized units where the gravitational constant and speed of light are set equal to one. $(M,g)$ denotes a smooth, four-dimensional Lorentzian manifold where we adopt the signature convention $(-,+,+,+)$ for the metric. We denote by ${\cal X}(M)$ the class of smooth vector fields on the differentiable manifold $M$. For $X\in {\cal X}(M)$ the symbols $i_X$ and $\pounds_X$ denote, respectively, the inner product and Lie derivative with respect to $X$.

\section{Relativistic kinetic theory on the cotangent bundle}
\label{Sec:Cotangent}

The purpose of this section is to provide a brief review for the formal structure of relativistic kinetic theory and to fix the notation used in this article. For more detailed reviews we refer the reader to the early work by Ehlers~\cite{jE71} and to the more recent work in~\cite{oStZ13,oStZ14b} and references therein.

\subsection{Relativistic phase space}

Kinetic theory is based on the assumption that the ensemble-averaged properties of the gas can be described by a one-particle distribution function $f$. Accordingly, $f: \Gamma\to \Real$ is a nonnegative function on the one-particle phase space $\Gamma$ (the ``$(q,p)$-space'' at each fixed time). In the relativistic case, $\Gamma$ is a subset of the cotangent bundle
\begin{equation}
T^* M := \{ (x,p) : x\in M, p\in T_x^* M \}
\end{equation}
associated with the spacetime manifold $M$ which consists of points of the form $(x,p)$, where $x\in M$ is spacetime event, and $p\in T_x^*M$ is a co-vector at this event, representing the canonical momentum of the gas particle at $x$. Instead of working on the cotangent bundle, one usually formulates kinetic theory on the tangent bundle $TM$, where $p\in T_x M$ is a tangent vector at $x$, representing the physical momentum of the particle. Since the present article is strongly based on the Hamiltonian formulation, the choice of the cotangent bundle is more natural. However, we stress that both formulations are equivalent because of the  existence of the spacetime metric $g$ which provides a natural isomorphism between $TM$ and $T^*M$.

We recall:

\begin{lemma}
Let $M$ be a $n$-dimensional differentiable manifold (orientable or not). Then, $T^* M$ is a $2n$-dimensional differentiable and orientable manifold.
\end{lemma}

\proof The proof is standard (see, for example~\cite{DoCarmo-Book1,oStZ14b} for the tangent bundle case). Given local coordinates $(x^\mu)$ on $M$ one defines corresponding {\em adapted local coordinates} $(x^\mu,p_\mu)$ on $T^* M$ by expanding each $p\in T_x^* M$ in the form $p = p_\mu dx^\mu_x$. Taking an atlas of local coordinate charts on $M$ one obtains a corresponding atlas on $T^* M$, and it is simple to verify that all the transition maps are differentiable and have unit determinant, implying that the transition maps are volume and orientation-preserving. As we will see shortly, this property stems from the fact that the adapted local coordinates $(x^\mu,p_\mu)$ are symplectic coordinates.
\qed

For the following, we assume that spacetime $(M,g)$ is time-oriented and that all gas particles have positive mass. In this case, the one-particle phase space is the following submanifold of $T^* M$:\footnote{We define $p\in T_x^* M$ to be future-directed timelike if the corresponding tangent vector $g_x^{-1}(p,\cdot)\in T_x M$ is future-directed timelike.}
$$
\Gamma := \{ (x,p) \in T^* M : \hbox{$p$ future-directed timelike} \}.
$$
If one further assumes that all the gas particles have the same positive rest mass $m > 0$, one obtains instead the {\em mass shell}
$$
\Gamma_m := \{ (x,p) \in T^* M : g^{-1}_x(p,p) = -m^2, \hbox{ $p$ future-directed} \}.
$$
For generality, in this article we mostly work on $\Gamma$ although in some applications towards the end we consider a simple gas, that is, a kinetic gas composed of identical particles of the same positive rest mass in which case we restrict ourselves to $\Gamma_m$.

\subsection{Symplectic structure and Liouville vector field}

The cotangent space $T^* M$ admits a natural symplectic structure which can be described as follows. First, the {\em Poincar\'e (or canonical) one-form} on $T^* M$ is introduced, which is defined by
\begin{equation}
\Theta_{(x,p)}(X) := p\left( d\pi_{(x,p)}(X) \right),\qquad X\in {\cal X}(T^* M),
\label{Eq:PoincareForm}
\end{equation}
where here $\pi: T^*M \to M$, $(x,p)\mapsto x$ is the natural projection map and $d\pi_{(x,p)}: T_{(x,p)}(T^* M)\to T_x M$ its differential at $(x,p)$. In adapted local coordinates,
$$
\Theta_{(x,p)} = p_\mu dx^\mu_{(x,p)}.
$$
Its exterior differential
$$
\Omega_s := d\Theta = dp_\mu\wedge dx^\mu
$$
is seen to define a symplectic structure on $T^* M$, that is, a closed, non-degenerated two-form on $T^* M$. This symplectic structure allows one to assign to any function $H$ on $T^* M$ a unique vector field $X_H\in {\cal X}(T^* M)$ called the {\em associated Hamiltonian vector field} which is defined by
\begin{equation}
dH = -i_{X_H}\Omega_s = \Omega_s(\cdot,X_H).
\end{equation}
In adapted local coordinates $(x^\mu,p_\mu)$ one has
$$
X_H = \frac{\partial H}{\partial p_\mu}\frac{\partial}{\partial x^\mu} 
 - \frac{\partial H}{\partial x^\mu}\frac{\partial}{\partial p_\mu},
$$
and the integral curves $\gamma(\lambda)$ of $X_H$ are described by {\em Hamilton's equations}
\begin{equation}
\frac{dx^\mu}{d\lambda}(\lambda) = \frac{\partial H}{\partial p_\mu}(\gamma(\lambda)),\qquad
\frac{dp_\mu}{d\lambda}(\lambda) = -\frac{\partial H}{\partial x^\mu}(\gamma(\lambda)).
\end{equation}

The {\em Poisson bracket} between two smooth functions $F,G: T^*M\to \Real$ on the cotangent bundle is defined as
\begin{equation}
\{ F, G \} := \Omega_s(X_F,X_G) = -dF(X_G) = dG(X_F).
\end{equation}
In particular when $G = H$ is the Hamiltonian of the system, it follows that $\{ F, H \} = 0$ if and only if $F$ is constant along the Hamiltonian flow, which is the case if and only if $H$ is invariant under the canonical flow generated by $F$. For later use we also note the identity\footnote{See for instance page 217 in Ref.~\cite{Arnold-Book}.}
\begin{equation}
[X_F,X_G] = X_{\{ F, G \}}.
\label{Eq:CommPoisson}
\end{equation}

Of particular interest to this article is the free-particle Hamiltonian
$$
H(x,p) := \frac{1}{2} g_x^{-1}(p,p) = \frac{1}{2} g^{\mu\nu}(x) p_\mu p_\nu;
$$
the associated Hamiltonian vector field $L := X_H$ is called the {\em Liouville vector field}, and in adapted local coordinates it reads
$$
L = g^{\mu\nu}(x) p_\nu\frac{\partial}{\partial x^\mu} - \frac{1}{2}p_\alpha p_\beta
 \frac{\partial g^{\alpha\beta}(x)}{\partial x^\mu}\frac{\partial}{\partial p_\mu}.
$$
The corresponding integral curves, when projected onto $(M,g)$, describe spacetime geodesics. For a given positive mass $m > 0$, the Liouville vector field $L$ at each point $(x,p)\in \Gamma_m$ of the mass shell $\Gamma_m$ is nonvanishing and tangent to $\Gamma_m$. Therefore, when restricted to the relativistic phase space $\Gamma$ the Liouville vector field does not have any critical points, and the projections of the integral curves of $L$ describe future-directed timelike geodesics of $(M,g)$.

The {\em Liouville equation}, which describes the evolution of a collisionless distribution function $f: \Gamma\to \Real$ is
\begin{equation}
L[f] = 0.
\label{Eq:Liouville}
\end{equation}
In the next section, we find the most general solution of this equation for the case where $(M,g)$ describes the geometry of a Schwarzschild black hole. This will be achieved by introducing new symplectic coordinates on $\Gamma$ which trivialize the Liouville vector field $L$.

\subsection{Complete lifts of vector fields}
\label{SubSec:Symmetries}

Suppose $\xi\in {\cal X}(M)$ is a vector field on spacetime generating a one-parameter group of diffeomorphism $\varphi^\lambda: M\to M$. We may lift $\varphi^\lambda$ to a corresponding group $\hat{\varphi}^\lambda : T^*M\to T^*M$ on the cotangent bundle by defining
$$
\hat{\varphi}^\lambda(x,p) := \left( \varphi^\lambda(x), 
\left[ (d\varphi^\lambda_x)^* \right]^{-1}(p) \right),\qquad (x,p)\in T^* M,
$$
where $d\varphi^\lambda_x : T_x M\to T_{\varphi^\lambda(x)} M$ denotes the differential of the map $\varphi^\lambda$ at $x$ and $(d\varphi^\lambda_x)^* : T^*_{\varphi^\lambda(x)} M\to T^*_x M$ its adjoint. It is simple to verify that $\hat{\varphi}^0$ is the identity map on $T^*M$ and that $\hat{\varphi}^\lambda\circ\hat{\varphi}^\mu = \hat{\varphi}^{\lambda+\mu}$ for all $\lambda,\mu\in\Real$, and thus $\hat{\varphi}^\lambda$ defines a one-parameter group of diffeomorphism, as claimed. The {\em complete lift of $\xi$} is defined as
$$
\hat{\xi}_{(x,p)} := \left. \frac{d}{d\lambda} \right|_{\lambda=0} \hat{\varphi}^\lambda(x,p),
\qquad
(x,p)\in T^* M.
$$
Of course, the same definition applies to a vector field $\xi\in {\cal X}(M)$ which only generates a local one-parameter group of diffeomorhism, and hence the complete lift can be introduced in the same way for any $\xi\in {\cal X}(M)$.

In adapted local coordinates one finds the following explicit expression:
\begin{equation}
\hat{\xi}_{(x,p)}  = \xi^\mu(x)\left. \frac{\partial}{\partial x^\mu} \right|_{(x,p)}
 - p_\alpha\frac{\partial\xi^\alpha}{\partial x^\mu}(x) 
 \left. \frac{\partial}{\partial p_\mu} \right|_{(x,p)},\qquad 
\xi_x = \xi^\mu(x) \left. \frac{\partial}{\partial x^\mu} \right|_x.
\label{Eq:hatxiCoords}
\end{equation}

The most important properties of the complete lift are summarized in the following proposition.

\begin{proposition}[cf. Proposition~5 and Lemmas 9 and 10 in~\cite{oStZ14b} and Lemma C.9 in~\cite{dFjJjS15}]
\label{Prop:CompleteLift}
\noindent
\begin{enumerate}
\item[(i)] The complete lift preserves the Lie-brackets, that is
$[\hat{\xi},\hat{\eta}] = \hat{\zeta}$ with $\zeta = [\xi,\eta]$ for all $\xi,\eta\in {\cal X}(M)$.
\item[(ii)] If $\xi\in {\cal X}(M)$, then $\hat{\xi}$ is the infinitesimal generator of a symplectic flow on $T^* M$, that is, $\pounds_{\hat{\xi}}\Omega_s = 0$. Moreover,
$$
\hat{\xi} = X_F,\qquad
F := \Theta(\hat{\xi}) = p(\xi),
$$
which shows that this flow is generated by the function $F = p(\xi)$.
\item[(iii)] Let $\xi\in {\cal X}(M)$ and $m > 0$. Then, $\hat{\xi}$ is tangent to the mass shell $\Gamma_m$ at each point $(x,p)\in \Gamma_m$ if and only if $\xi$ is a Killing vector field of $(M,g)$.
\end{enumerate}
\end{proposition}

{\bf Remarks}:
\begin{enumerate}
\item Unlike the tangent bundle formulation (see Proposition~5 in~\cite{oStZ14b}), in the cotangent bundle formulation the vector field $\xi$ is not required to be a Killing vector field in order for $\hat{\xi}$ to generate a symplectic flow.
\item If $\xi\in {\cal X}(M)$ is a Killing vector field of $(M,g)$, the properties (ii) and (iii) of the proposition imply that
$$
\{ F, H \} = dH(X_F) = dH(\hat{\xi}) = \pounds_{\hat{\xi}} H = 0,
$$
and thus $F$ is conserved along the Liouville flow. In this case it follows that the Hamiltonian vector fields $X_F = \hat{\xi}$ and $X_H = L$ associated with $F$ and $H$ commute, see equation~(\ref{Eq:CommPoisson}):
$$
[\hat{\xi},L] = 0.
$$
\end{enumerate}

\proofof{Proposition~\ref{Prop:CompleteLift}} Property (i) can be verified by a straightforward calculation using equation~(\ref{Eq:hatxiCoords}). As for (ii) and (iii) we first calculate
$$
i_{\hat{\xi}}\Omega_s = i_{\hat{\xi}}(dp_\mu\wedge dx^\mu)
 = dp_\mu(\hat{\xi}) dx^\mu - dx^\mu(\hat{\xi}) dp_\mu.
$$
Using the coordinate expression~(\ref{Eq:hatxiCoords}) for $\hat{\xi}$ we obtain
$$
i_{\hat{\xi}}\Omega_s = -p_\alpha\frac{\partial\xi^\alpha}{\partial x^\mu} dx^\mu
 - \xi^\mu dp_\mu = -d(p_\alpha \xi^\alpha) = -dF,
$$
which shows (ii). In particular, the fact that $d\Omega_s = 0$ implies
$$
\pounds_{\hat{\xi}}\Omega_s = (di_{\hat{\xi}} + i_{\hat{\xi}} d)\Omega_s = -d(dF) = 0.
$$

Finally, we note that since $\Gamma_m$ is a level surface of the free-particle Hamiltonian $H$, $\hat{\xi}$ is tangent to $\Gamma_m$ if and only if $\pounds_{\hat{\xi}} H = dH(\hat{\xi}) = 0$ at each $(x,p)\in \Gamma_m$. On the other hand, the definition of the Liouville vector field and (ii) imply that
$$
dH(\hat{\xi}) = i_{\hat{\xi}}\Omega_s(L)
  = -p_\alpha\frac{\partial\xi^\alpha}{\partial x^\mu} g^{\mu\beta} p_\beta
 + \xi^\mu\frac{1}{2} p_\alpha p_\beta\frac{\partial g^{\alpha\beta}}{\partial x^\mu}
 = \frac{1}{2} p_\alpha p_\beta (\pounds_\xi g^{\alpha\beta}).
$$
Hence, if $\xi$ is a Killing vector field of $(M,g)$ then $dH(\hat{\xi}) = 0$ and $\hat{\xi}$ is tangent to $\Gamma_m$. Conversely, if $dH(\hat{\xi}) = 0$ at each point $(x,p)\in \Gamma_m$ then, for any $x\in M$ it follows that $p_\alpha p_\beta (\pounds_\xi g^{\alpha\beta}) = 0$ for all future-directed timelike covectors $p\in T_x^* M$ normalized such that $g^{-1}_x(p,p) = -m^2$, which implies that $\pounds_\xi g^{\alpha\beta}(x) = 0$, and thus that $\xi$ is a Killing vector field on $(M,g)$.
\qed

As an example which will turn out to be relevant for the next section, consider a spherically symmetric, four-dimensional spacetime $(M,g)$. In terms of standard spherical coordinates, the infinitesimal generators $\xi_1$, $\xi_2$, $\xi_3$ of the action of the rotation group $SO(3)$ on $M$ can be represented in the following way:
\begin{eqnarray}
\xi_1 + i\xi_2 &=& e^{i\varphi}\left(
 i\frac{\partial}{\partial\vartheta} - \cot\vartheta\frac{\partial}{\partial\varphi} \right),
\label{Eq:xi12}\\
\xi_3 &=& \frac{\partial}{\partial\varphi}.
\label{Eq:xi3}
\end{eqnarray}
They satisfy the commutation relations $[\xi_1,\xi_2] = -\xi_3$ (and cyclic permutations of $123$) and are tangent to the invariant two-spheres (the orbits of the rotation group). The associated conserved quantities are
\begin{eqnarray}
\ell_1 + i\ell_2 &:=& p(\xi_1 + i\xi_2) = e^{i\varphi}\left( i p_\vartheta - \cot\vartheta p_\varphi \right),\\
\ell_3 &:=& p(\xi_3) = p_\varphi,
\end{eqnarray}
and consequently, the total angular momentum
$$
\ell := \sqrt{\ell_1^2 + \ell_2^2 + \ell_3^2} 
 = \sqrt{p_\vartheta^2 + \frac{p_\varphi^2}{\sin^2\vartheta}},
$$
is also conserved along the Liouville flow. The complete lifts of the vector fields $\xi_a$ on $T^* M$ are given by
\begin{eqnarray}
\hat{\xi}_1 + i\hat{\xi}_2 &=& e^{i\varphi}\left[ 
 i\frac{\partial}{\partial\vartheta} - \cot\vartheta\frac{\partial}{\partial\varphi} 
- \frac{p_\varphi}{\sin^2\vartheta}\frac{\partial}{\partial p_\vartheta} 
+ \left( p_\vartheta + i\cot\vartheta p_\varphi \right)\frac{\partial}{\partial p_\varphi} 
 \right],
\label{Eq:hatxi12}\\
\hat{\xi}_3 &=& \frac{\partial}{\partial\varphi}.
\label{Eq:hatxi3}
\end{eqnarray}
They satisfy the same commutation relations as the $\xi_a$'s; however, now the three vector fields $\hat{\xi}_1$, $\hat{\xi}_2$, $\hat{\xi}_3$ are pointwise linearly independent, which means that the orbits of $SO(3)$ in $T^* M$ are three-dimensional submanifolds (as opposed to the orbits in $M$ which are two-dimensional). Therefore, a function on $T^* M$ which is invariant with respect to $SO(3)$ is subject to three independent constraints, whereas a $SO(3)$-invariant function on $M$ is only subject to two.

Finally, note that since $d(\ell^2) = 2(\ell_1 d\ell_1 + \ell_2 d\ell_2 + \ell_3 d\ell_3)$ the Hamiltonian vector field associated with the square of the total angular momentum is
$$
X_{\ell^2} = 2\left( \ell_1\hat{\xi}_1 + \ell_2\hat{\xi}_2 + \ell_3\hat{\xi}_3 \right).
$$

\section{General solution of the Liouville equation on a Schwarzschild background}
\label{Sec:Distribution}

In this section, we discuss the most general solution of the Liouville equation~(\ref{Eq:Liouville}) describing the accretion into a Schwarzschild black hole. This is achieved using standard techniques from the classical theory of integrable systems, see for example~\cite{Arnold-Book}.

\subsection{Horizon-penetrating coordinates}

In standard coordinates $(\bar{t},r,\vartheta,\varphi)$ the Schwarzschild metric describing a black hole of mass $M_H > 0$ is given by
$$
g = -N(r) d\bar{t}^2 + \frac{dr^2}{N(r)} 
 + r^2\left( d\vartheta^2 + \sin^2\vartheta\, d\varphi^2 \right),
\qquad N(r) := 1 - \frac{2M_H}{r}.
$$
Since in this work we shall be interested in computing the observables associated with the kinetic gas in the exterior region $r > 2M_H$ as well as on the future horizon, for what follows we work in regular coordinates $(t,r,\vartheta,\varphi)$. Here, the new time coordinate $t$ is defined by
$$
t = \bar{t} + \int\limits^r \left[ \frac{1}{N(r)} - \eta(r) \right] dr,
$$
with a smooth function $\eta: (0,\infty)\to \Real$. In terms of the new coordinates, the metric takes the form
$$
g = -N dt^2 + 2(1 - N\eta) dt dr + \eta(2 - N\eta) dr^2
 + r^2\left( d\vartheta^2 + \sin^2\vartheta\, d\varphi^2 \right),
$$
and there are no coordinate singularities for $r > 0$ except at the poles $\vartheta = 0,\pi$.\footnote{These coordinate singularities can be removed by replacing $(r,\vartheta,\varphi)$ with Cartesian-like coordinates $(x,y,z) = r(\cos\varphi\sin\vartheta,\sin\varphi\sin\vartheta,\cos\vartheta)$.} Here, the function $\eta$ reflects the freedom in choosing the foliation of the Schwarzschild spacetime which is regular at the future horizon. When $\eta = 0$ the surfaces of constant time $t$ are incoming null surfaces and the corresponding coordinates $(t,r,\vartheta,\varphi)$ are called (ingoing) Eddington-Finkelstein coordinates (see for instance Ref.~\cite{MTW-Book}). When $\eta(2 - N\eta) > 0$ the constant time surfaces are spacelike. Some interesting choices which have been considered in the literature are:
\begin{itemize}
\item Painlev\'e-Gullstrand coordinates (see for example Ref.~\cite{Faraoni-Book}): $\eta = [1 + \sqrt{1 - N}]^{-1}$.
\item Analytic trumpet slices that were recently proposed in~\cite{kAdTb14}:
$$
\eta(r) = \frac{r^2}{r- R_0}\frac{1}{r - R_0 + \sqrt{2(M_H - R_0)r + R_0^2}},\quad
r > R_0,
$$
with $R_0$ a free parameter satisfying $0 < R_0 \leq M_H$. In this case the function $\eta$ is only defined on the interval $(R_0,\infty)$, but otherwise it satisfies all the required properties mentioned above. Replacing $r$ with the new radial coordinate $\rho = r - R_0$ (denoted by $r$ in~\cite{kAdTb14}) leads to spatially homogenous coordinates.
\end{itemize}
For the calculations below, we shall focus on the simple case $\eta = 1$, in which the inverse metric reads
\begin{equation}
g^{-1} = -\left(1 + \frac{2M_H}{r} \right)\frac{\partial}{\partial t}\otimes\frac{\partial}{\partial t}
 + \frac{4M_H}{r}\frac{\partial}{\partial t}\otimes_s\frac{\partial}{\partial r}
 + \left(1 - \frac{2M_H}{r} \right)\frac{\partial}{\partial r}\otimes\frac{\partial}{\partial r}
 + \frac{1}{r^2}\left( \frac{\partial}{\partial\vartheta}\otimes\frac{\partial}{\partial\vartheta}
 + \frac{1}{\sin^2\vartheta}\frac{\partial}{\partial\varphi}\otimes\frac{\partial}{\partial\varphi} \right),
\label{Eq:SchwarzschildInv}
\end{equation}
where $\otimes_s$ denotes the symmetrized tensor product.

From now on, we consider the region $M$ of the (extended) Schwarzschild manifold which is covered by the horizon-penetrating coordinates $(t,r)$ with $t\in\Real$ and $r > 0$.

\subsection{Hamiltonian flow, conserved quantities and invariant submanifolds}

The free-particle Hamiltonian computed from the inverse metric~(\ref{Eq:SchwarzschildInv}) is
\begin{equation}
H(x,p) = \frac{1}{2}\left[ -\left(1 + \frac{2M_H}{r} \right) p_t^2
 + \frac{4M_H}{r} p_t p_r + \left(1 - \frac{2M_H}{r} \right) p_r^2 + \frac{1}{r^2}
 \left( p_\vartheta^2 + \frac{p_\varphi^2}{\sin^2\vartheta} \right) \right].
\label{Eq:HSchwarzschild}
\end{equation}
Since the underlying spacetime is stationary and spherically symmetric, the following quantities are conserved along the particle trajectories:
\begin{eqnarray}
m = \sqrt{-2H} && \hbox{(rest mass)},\\
E = -p_t && \hbox{(energy)},\\
\ell_z = p_\varphi && \hbox{(azimutal angular momentum)},\\
\ell = \sqrt{p_\vartheta^2 + \frac{p_\varphi^2}{\sin^2\vartheta} } 
 && \hbox{(total angular momentum)}.
\end{eqnarray}
Accordingly, we introduce the smooth functions $F_0,F_1,F_2,F_3: T^* M\to \Real$ on the cotangent bundle defined by\footnote{With respect to Cartesian-like coordinates, one can rewrite $p_\varphi = x p_y - y p_x$ and $\ell^2 = |(x,y,z)\wedge (p_x,p_y,p_z)|^2$ with $\wedge$ denoting the vector product, which shows that these quantities are indeed smooth functions on $T^* M$.}
\begin{equation}
F_0(x,p) := -H(x,p),\qquad
F_1(x,p) := -p_t,\qquad
F_2(x,p) := p_\varphi,\qquad
F_3(x,p) := p_\vartheta^2 + \frac{p_\varphi^2}{\sin^2\vartheta}.
\label{Eq:FalphaDef}
\end{equation}
It is not difficult to verify that these quantities Poisson-commute with each other:
$$
\{ F_\alpha, F_\beta \} = 0, \qquad \alpha,\beta=0,1,2,3.
$$

For the following, we consider for each given value of $(m,E,\ell_z,\ell)$ the (possibly empty) subset
$$
\Gamma_{m,E,\ell_z,\ell} 
 := \left\{ (x,p)\in \Gamma : F_0(x,p) = \frac{1}{2} m^2, F_1(x,p) = E, F_2(x,p) = \ell_z,
  F_3(x,p) = \ell^2 \right\}
$$ 
of the one-particle phase space $\Gamma$. By construction, these sets are invariant under the Hamiltonian flow. The main result of this subsection is the following:

\begin{proposition}
\label{Prop:GammaProp}
Suppose that $E > m > 0$ and that $|\ell_z| < \ell$. In addition, suppose that the condition $\ell\neq \ell_c(E)$ holds (see below for the definition of $\ell_c(E)$). Then, $\Gamma_{m,E,\ell_z,\ell}$ is a smooth, four-dimensional submanifold of $\Gamma$ which is invariant with respect to the Hamiltonian flows associated with $F_0$, $F_1$, $F_2$ and $F_3$. It consists of three connected components, each of which has the topology $\Real^2\times S^1\times S^1$.

Furthermore, the restriction of the Poincar\'e one-form $\Theta$ defined in equation~(\ref{Eq:PoincareForm}) on $\Gamma_{m,E,\ell_z,\ell}$ is closed.
\end{proposition}

\begin{figure}[ht]
\centerline{\resizebox{8.5cm}{!}{\includegraphics{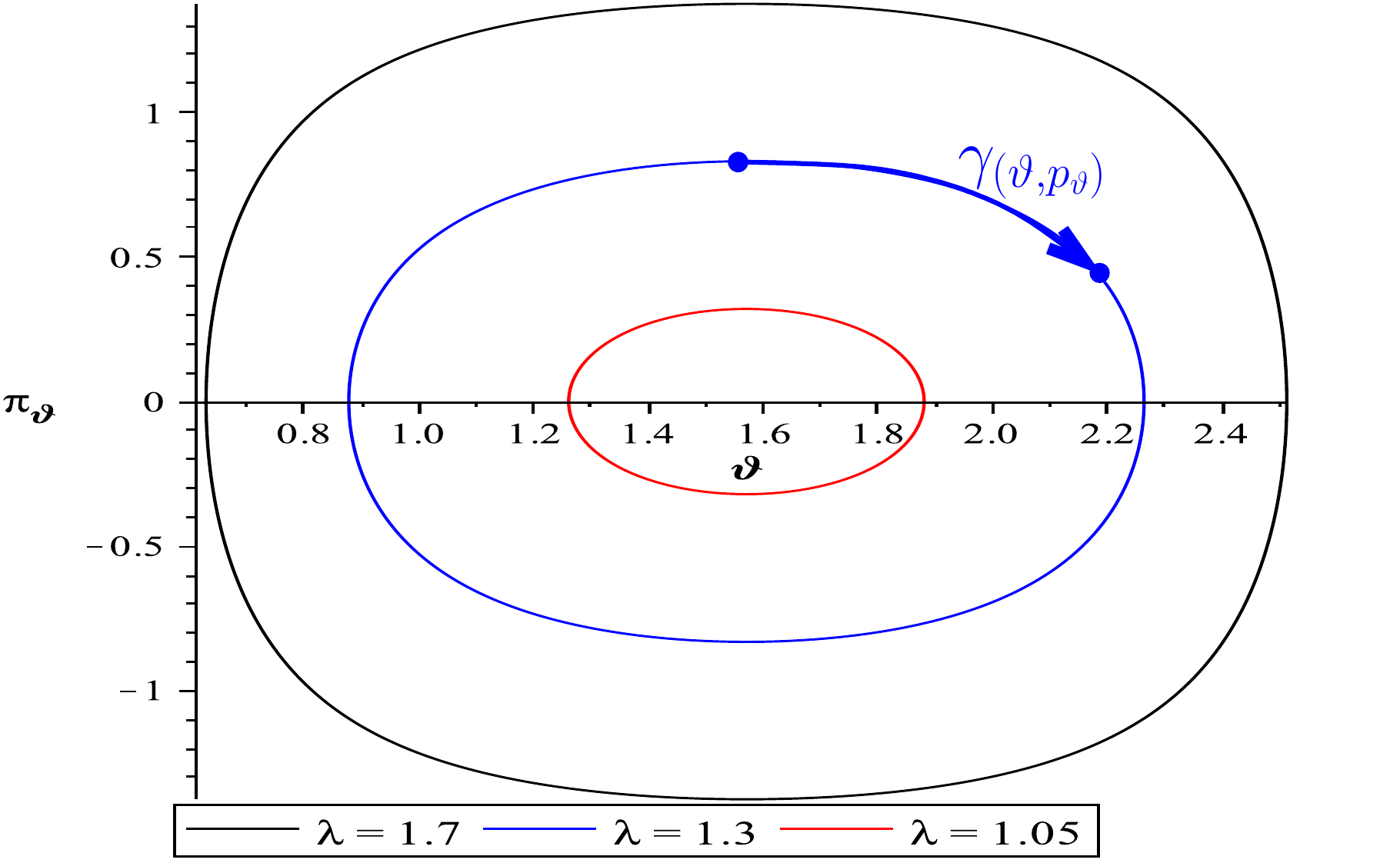}}}
\caption{Phase diagram illustrating the projection of the invariant sets onto the $(\vartheta,p_\vartheta)$-plane. Here, we use the dimensionless variables $\pi_\vartheta = p_\vartheta/(M_H m)$ and $\lambda = \ell/(M_H m)$ and rescaled the variables such that $\ell_z = M_H m$. When $\lambda\to 1$ the curves shrink to a point, corresponding to motion confined to the equatorial plane $\vartheta = \pi/2$.}
\label{Fig:plot_thpth}
\end{figure}

In order to prove the proposition, we separate the adapted local coordinates $(x^\mu,p_\mu)$ into the conjugate pairs $(t,p_t)$, $(\varphi,p_\varphi)$, $(\vartheta,p_\vartheta)$, $(r,p_r)$ and notice that $\Gamma_{m,E,l_z,\ell}$ consists of those points of $\Gamma$ for which these pairs fulfill the following restrictions:
\begin{eqnarray} \label{Eq:Ept}
(t,p_t) &:& p_t = -E,\\
(\varphi,p_\varphi) &:& p_\varphi = \ell_z,\\
(\vartheta,p_\vartheta) &:& p_\vartheta^2 + \frac{\ell_z^2}{\sin^2\vartheta} = \ell^2,
\label{Eq:thetaPlaneRestriction}\\
(r,p_r) &:& \left(1 - \frac{2M_H}{r} \right)  p_r^2 - \frac{4M_H}{r} E p_r 
 - \left(1 + \frac{2M_H}{r} \right) E^2 + m^2 + \frac{\ell^2}{r^2} = 0.
\label{Eq:rPlaneRestriction}
\end{eqnarray}
From these conditions it is clear that the time coordinate $t$ and the azimutal angle $\varphi$ are free, whereas for $0 < |\ell_z| < \ell$ the pair $(\vartheta,p_\vartheta)$ is confined to a closed curve in the $(\vartheta,p_\vartheta)$-plane which winds around the equilibrium point $(\vartheta,p_\vartheta) = (\pi/2,0)$, see figure~\ref{Fig:plot_thpth}.\footnote{The exceptional cases $\ell_z = 0$ and $\ell_z = \pm\ell$ correspond, respectively, to motion confined in a plane which contains de $z$ axis and to motion in the equatorial plane, see Lemma~\ref{Lem:Differentials} below for more information on these cases. Since they give rise to zero measure sets in the phase space, we will not consider these particular cases further.}

Next, let us analyze the set in the $(r,p_r)$-plane described by equation~(\ref{Eq:rPlaneRestriction}) in more detail. Multiplying both sides of equation~(\ref{Eq:rPlaneRestriction}) by $N(r) = 1 - 2M_H/r$ and assuming that $N(r)\neq 0$ we can rewrite this equation as
\begin{equation}
\left[ \left(1-\frac{2M_H}{r} \right) p_r - \frac{2M_H}{r}E \right]^2 + V_{m,\ell}(r) = E^2,
\label{Eq:rPlaneRestrictionBis}
\end{equation}
with the effective potential
\begin{equation}
V_{m,\ell}(r) = \left( 1- \frac{2M_H}{r}\right)\left(m^2 + \frac{\ell^2}{r^2} \right).
\label{Eq:Vmell}
\end{equation}
By Hamilton's equations the particle's radial velocity is
$$
\dot{r} := \frac{dr}{d\lambda} = \frac{\partial H}{\partial p_r} 
 = \left(1-\frac{2M_H}{r} \right) p_r + \frac{2M_H}{r} p_t
 = \left(1-\frac{2M_H}{r} \right) p_r - \frac{2M_H}{r} E,
$$
so as long as $N(r)\neq 0$ the set defined by equation~(\ref{Eq:rPlaneRestriction}) is equivalent to $\dot{r}^2 + V_{m,\ell}(r) = E^2$, and the sign of the expression inside the square parenthesis in equation~(\ref{Eq:rPlaneRestrictionBis}) determines whether the particle is incoming or outgoing. In the $(r,p_r)$-plane, equation~(\ref{Eq:rPlaneRestriction}) can be written as
\begin{equation}
p_r = p_{r\pm}(r) := \frac{\frac{2M_H}{r} E \pm \sqrt{E^2 - V_{m,\ell}(r)}}{N(r)},
\label{Eq:prpm}
\end{equation}
where the $\pm$ sign corresponds to the one of $\dot{r}$.

\begin{figure}[ht]
\centerline{\resizebox{7.0cm}{!}{\includegraphics{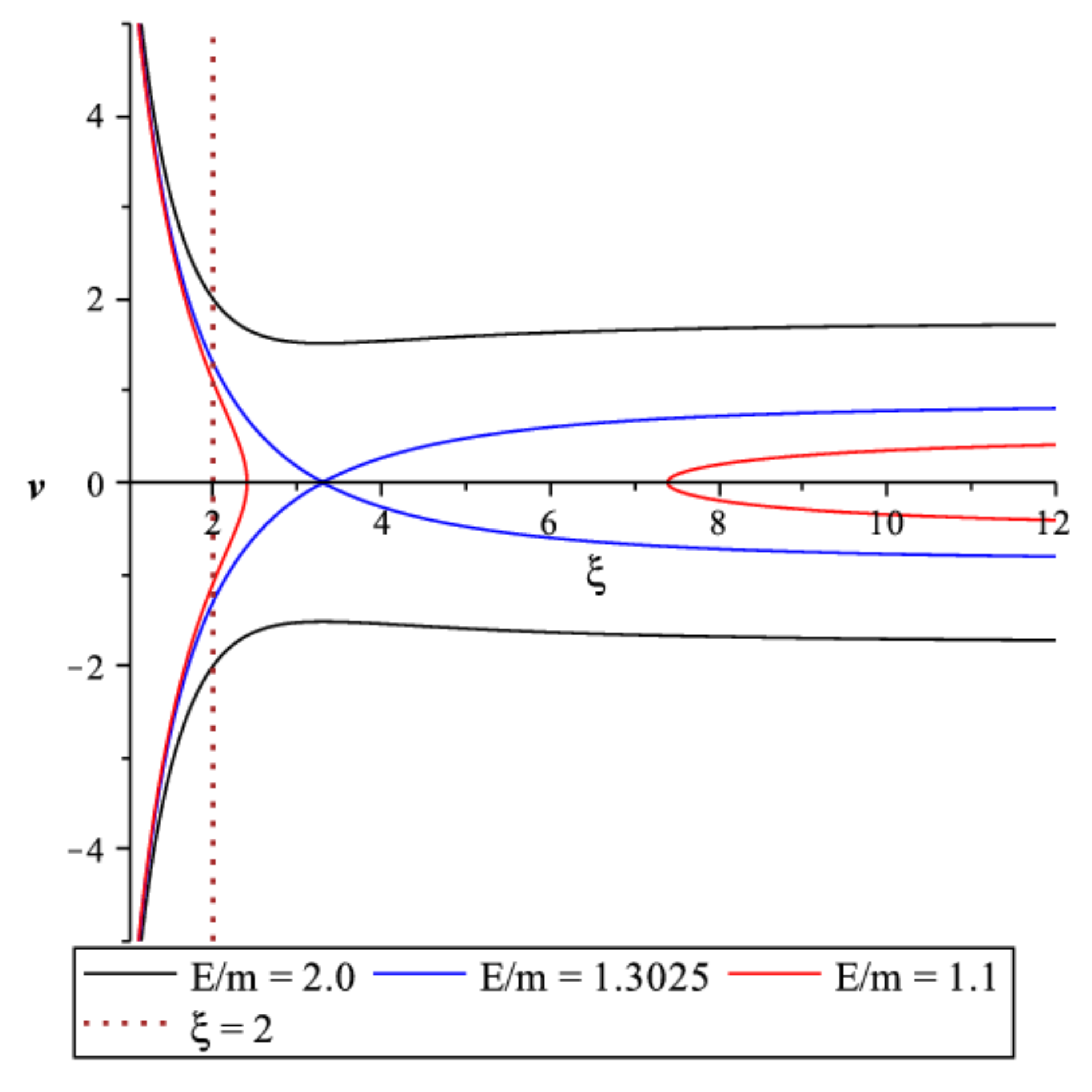}}\resizebox{11.0cm}{!}{\includegraphics{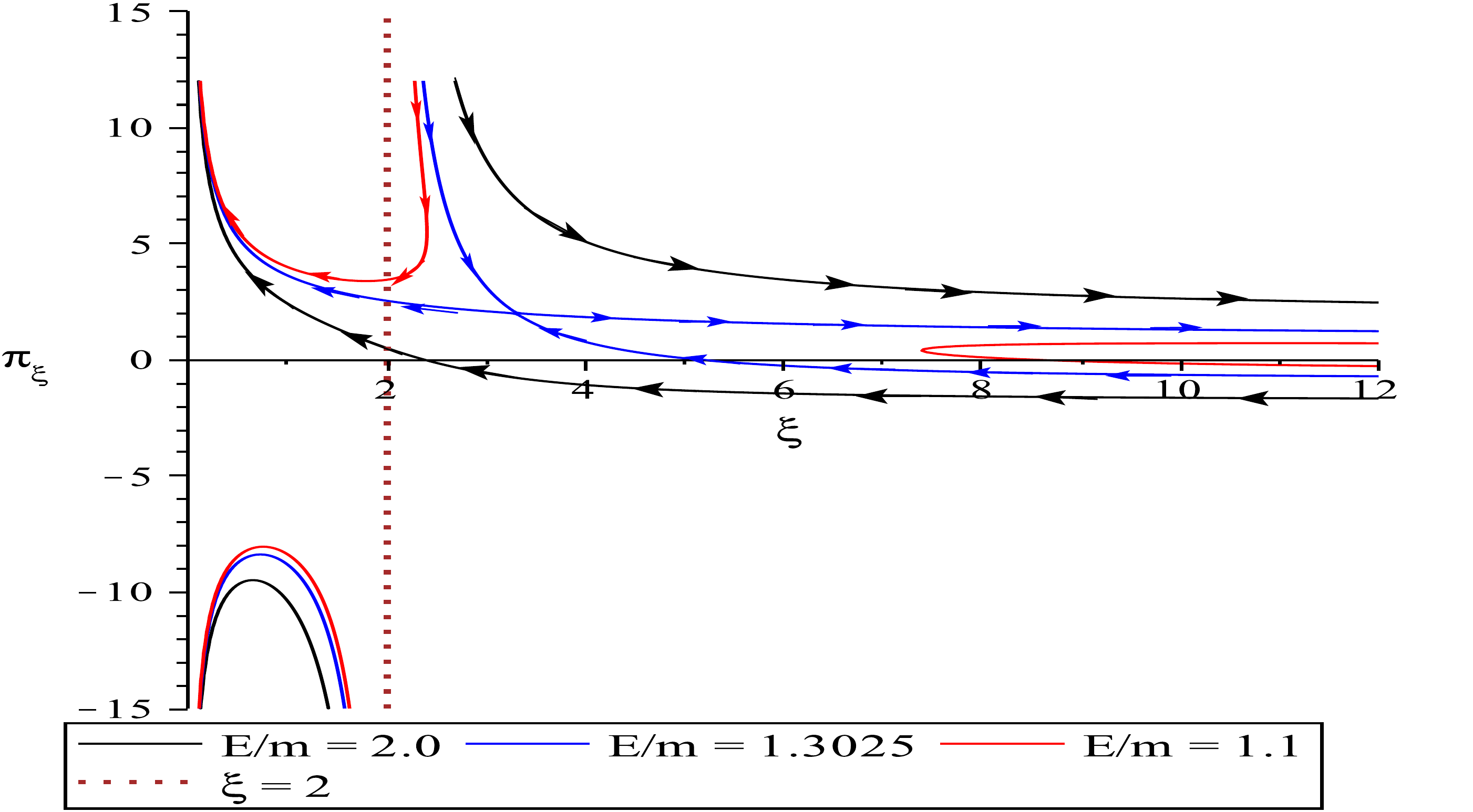}}}
\caption{Phase diagrams illustrating the projection of the invariant sets onto the $(r,p_r)$-plane. Left panel: $v = \dot{r}/m$ vs. $\xi = r/M_H$ for different energy levels $E$ and $\ell = 6M_H m$. The black curve in the region $v < 0$ corresponds to an incoming particle from infinity which is absorbed by the black hole (case (I)), while the black curve in the region $v > 0$ describes an outgoing particle that is emitted from the white hole and escapes to infinity. The red curve in the region $\xi > 6$ describes a particle that is incoming from infinity but has large enough angular momentum $\ell > \ell_c(E)$ to be reflected at the potential barrier  (case (II)), and the red curve in the region $\xi < 3$ describes a particle that is emitted by the white whole, is reflected at the potential barrier and absorbed by the black hole. The blue curve describes the separatrix and corresponds to the energy level $E$ such that $\ell_c(E) = \ell$. Right panel: The same situation showing $\pi_\xi = p_r/m$ vs. $\xi = r/M_H$. Since $\pi_\xi = (1- 2\xi^{-1})^{-1} (v + 2E/(m\xi))$, the curves in the previous diagrams in the region $v > 0$ split into two separate parts, one of them lying entirely in the region $\xi < 2$ and $\pi_\xi < 0$ inside the black hole region.}
\label{Fig:plot_rpr}
\end{figure}

The properties of the set in the $(r,p_r)$ plane described by equation~(\ref{Eq:rPlaneRestriction}) thus depend on the shape of the effective potential and the value of $E$. Since we are only interested in particle trajectories that emanate from the asymptotic region $r\to \infty$, and since the linear momentum $p$ of the particle is restricted to be future-directed, we assume $E > m > 0$ in what follows. For a given $E > m$, denote by $\ell_c(E)$ the value of $\ell$ for which $V_{m,\ell}$ has a centrifugal barrier whose maximum is equal to $E^2$, see App.~\ref{App:EffPot} for details and explicit expressions. We distinguish between the following two cases:
\begin{enumerate}
\item[(I)] \emph{Absorbed particles $0 < \ell < \ell_c(E)$.} In this case, the trajectories describe incoming particles from infinity which are absorbed by the black hole (negative values of $\dot{r}$) or particles which emanate from the white hole and escape to infinity (positive values of $\dot{r}$). There are also other trajectories which take place entirely inside the black hole (see figure~\ref{Fig:plot_rpr}) and will not be considered further. For the purpose of the present article, we focus on the first kind of trajectories which consist of open lines in the $(r,p_r)$-plane.
\item[(II)] \emph{Scattered particles $\ell > \ell_c(E)$.} This corresponds to trajectories which are incoming from infinity, but in contrast to the previous case they carry enough angular momentum to be reflected at the potential barrier and so they escape again to infinity. There are also two further open lines corresponding to particles that are emitted from the white hole and reabsorbed by the black hole, or particle trajectories taking place entirely inside the black hole (see figure~\ref{Fig:plot_rpr}) that will not be considered further for our purposes.
\end{enumerate}

To conclude the proof of Proposition~\ref{Prop:GammaProp}, we show in the next lemma that the differentials $dF_0$, $dF_1$, $dF_2$, and $dF_3$ are linearly independent at each point of $\Gamma_{m,E,\ell_z,\ell}$, implying that $\Gamma_{m,E,\ell_z,\ell}$ is a four-dimensional submanifold of $\Gamma$ with linearly independent tangent vector fields $X_{F_0}$, $X_{F_1}$, $X_{F_2}$, and $X_{F_3}$. Since $\Omega_s(X_{F_\alpha},X_{F_\beta}) = \{ F_\alpha,F_\beta \} = 0$ it then follows that the restriction of $\Omega_s = d\Theta$ to $\Gamma_{m,E,\ell_z,\ell}$ vanishes.

\begin{lemma}
\label{Lem:Differentials}
Let $E, m > 0$. The differentials $dF_0$, $dF_1$, $dF_2$ and $dF_3$ are linearly independent from each other unless one of the following two cases occur:
\begin{enumerate}
\item[(a)] Motion confined to the equatorial plane: $|\ell_z| = \ell$,
\item[(b)] Circular trajectories:
$$
V_{m,\ell}(r) = E^2,\qquad
\frac{d}{dr} V_{m,\ell}(r) = 0.
$$
\end{enumerate}
\end{lemma}

\proof We first note that $dF_1 = -dp_t\neq 0$, $dF_2 = dp_\varphi\neq 0$. Next, we compute\footnote{Note that $\sin\vartheta\neq 0$ as long as $\ell_z\neq 0$. When $\ell_z = 0$ we can use Cartesian coordinates at the poles to show that $dF_3$ is linearly independent of $dF_2$.}
$$
dF_3 = 2\left( p_\vartheta dp_\vartheta -
  \frac{\ell_z^2}{\sin^2\vartheta}\cot\vartheta d\vartheta
 + \ell_z\frac{dF_2}{\sin^2\vartheta} \right),
$$
from which we see that $dF_2$ and $dF_3$ are linearly independent from each other unless $p_\vartheta = 0$ and $\vartheta = \pi/2$, which corresponds to case (a). Finally, computing
\begin{eqnarray*}
dF_0 &=& \left[ \left( 1 + \frac{2M_H}{r} \right) E + \frac{2M_H}{r} p_r \right] dF_1
 - \left[  \left( 1 - \frac{2M_H}{r} \right) p_r - \frac{2M_H}{r} E \right] dp_r
 - \frac{1}{2r^2} dF_3 \\
 &-& \left[ \frac{M_H}{r}(E + p_r)^2 - \frac{\ell^2}{r^2} \right] \frac{dr}{r}
\end{eqnarray*}
we see that $dF_0$ is linearly independent from $dF_1$, $dF_2$ and $dF_3$ unless
$$
\left( 1 - \frac{2M_H}{r} \right) p_r - \frac{2M_H}{r} E = 0,\qquad
\frac{M_H}{r}(E + p_r)^2 - \frac{\ell^2}{r^2} = 0.
$$
A short calculation reveals that these conditions are equivalent to (b).
\qed

{\bf Remarks}:
\begin{enumerate}
\item In the following, we restrict ourselves to the subset $\Gamma_{accr}$ of the relativistic one-particle phase space $\Gamma$ which consists of the union of those invariant submanifolds $\Gamma_{m,E,\ell_z,\ell}$ corresponding to the cases (I) and (II) discussed above. This is the relevant submanifold describing the accretion of a collisionless kinetic gas into a Schwarzschild black hole.
\item Note that the subset $\Gamma_{accr}$ \emph{does not include the bounded trajectories}, which are contained in those invariant submanifolds $\Gamma_{m,E,\ell_z,\ell}$ for which $\sqrt{8/9} < E/m < 1$ and $\ell_c(E) < \ell < \ell_{ub}(E)$, see App.~\ref{App:EffPot}. In this case, $\Gamma_{m,E,\ell_z,\ell}$ have  topology $\Real\times S^1\times S^1\times S^1$. This case will be considered in separate work~\cite{pRoStZ17}.
\item The subset $\Gamma_{accr}$ does not include those particle trajectories that emanate from the white hole either, since they are not relevant for the accretion of a collisionless gas.
\end{enumerate}

In the next subsection, we shall construct new symplectic coordinates $(Q^\mu,P_\mu)$ on $\Gamma_{accr}$ which are adapted to the invariant submanifolds $\Gamma_{m,E,\ell_z,\ell}$ and which trivialize the Liouville vector field $L$.

\subsection{New symplectic coordinates adapted to the invariant submanifolds}

\begin{figure}[ht]
\centerline{\resizebox{7.5cm}{!}{\includegraphics{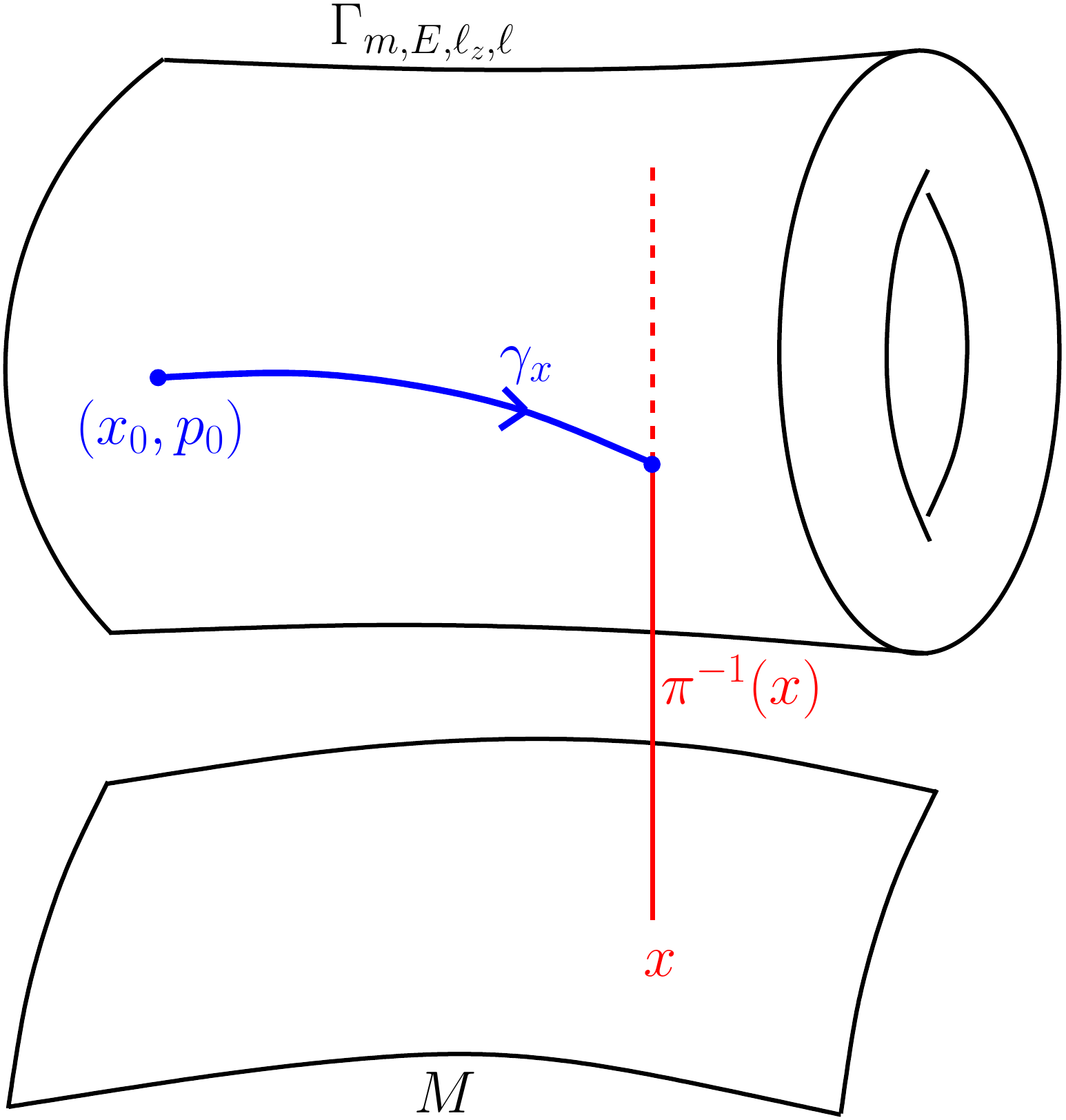}}}
\caption{Illustration showing the invariant submanifold $\Gamma_{m,E,\ell_z\ell}$ and the curve $\gamma_x$ connecting the reference point $(x_0,p_0)$ to a point in the intersection between $\Gamma_{m,E,\ell_z,\ell}$ and the fibre $\pi^{-1}(x)$ over $x\in M$.}
\label{Fig:invariant}
\end{figure}

Following~\cite{Arnold-Book} we construct new symplectic local coordinates $(Q^\mu,P_\mu)$ on $\Gamma_{accr}$ from the generating function
\begin{equation}
S(x; m,E,\ell_z,\ell) := \int\limits_{\gamma_x} \Theta,
\label{Eq:GeneratingFunction}
\end{equation}
where $\Theta$ is the Poincar\'e one-form defined in equation~(\ref{Eq:PoincareForm}) and the line integral is performed over a curve $\gamma_x$ which is confined to the invariant submanifold $\Gamma_{m,E,\ell_z,\ell}$, see figure~\ref{Fig:invariant}. This curve $\gamma_x$ connects a fixed reference point $(x_0,p_0)\in \Gamma_{m,E,\ell_z,\ell}$ to a point in the intersection between $\Gamma_{m,E,\ell_z,\ell}$ and the fibre $\pi^{-1}(x)$ over $x\in M$. We choose the reference point $(x_0,p_0)$ as the one parametrized by $(t,p_t) = (0,-E)$, $(\varphi,p_\varphi) = (0,\ell_z)$, $(\vartheta,p_\vartheta) = (\pi/2,\sqrt{\ell^2 - \ell_z^2})$ and $(r,p_r) = (r_0,p_{r-}(r_0))$, where the function $p_{r-}(r)$ is defined in equation~(\ref{Eq:prpm}) and $r_0 = 2M_H$ in case (I) and $r_0$ corresponds to the turning point in case (II).

Since the restriction of $\Theta$ to $\Gamma_{m,E,\ell_z,\ell}$ is closed, the value of $S(x; m,E,\ell_z,\ell)$ is independent of deformations of the curve $\gamma_x$ within $\Gamma_{m,E,\ell_z,\ell}$ which leave its endpoints fixed. On the other hand, the value of $S(x; m,E,\ell_z,\ell)$ does depend on how many times and in which direction the curve $\gamma_x$ winds around the two cycles $S^1$ of $\Gamma_{m,E,\ell_z,\ell}$. To analyze this dependency in more detail and in order to obtain a more explicit expression for the generating function we recall that in adapted local coordinates the Poincar\'e one-form reads $\Theta = p_\mu dx^\mu$. Consequently,
$$
S(x; m,E,\ell_z,\ell) = -E t + \ell_z\varphi 
 + \int\limits_{r_0}^r p_r dr 
 + \int\limits_{\pi/2}^\vartheta p_\vartheta d\vartheta.
$$
Here, it is important to notice that the last two integrals should be interpreted as line integrals. The first integral is a line integral along the projection $\gamma_{(r,p_r)}$ of the curve $\gamma_x$ onto the $(r,p_r)$-plane, while the second integral is a line integral (oriented clockwise) along the projection $\gamma_{(\vartheta,p_\vartheta)}$ of $\gamma_x$ onto the $(\vartheta,p_\vartheta)$-plane, see figure~\ref{Fig:plot_thpth}.

It is clear that $S$ changes by the additive constant $2\pi\ell_z$ under a full revolution $\varphi\to \varphi + 2\pi$ about the $z$-axis. Likewise, under a full clockwise revolution along the curve $\gamma_{(\vartheta,p_\vartheta)}$, $S$ increases by\footnote{The same integral appears in the discussion of the Kepler problem, see for instance section~10.8 in Ref.~\cite{Goldstein-Book}.}
$$
\Delta S = 4\int\limits_{\pi/2}^{\vartheta_{max}(\ell_z,\ell)} 
 \sqrt{\ell^2 - \frac{\ell_z^2}{\sin^2\vartheta}} d\vartheta = 2\pi(\ell - |\ell_z|),
$$
where $\vartheta_{max}(\ell_z,\ell) > \pi/2$ is the angle corresponding to the turning point, such that $\sin\vartheta_{max}(\ell_z,\ell) = \ell_z/\ell$. It follows from these observations that the quantities $\ell_z$ and $\ell - |\ell_z|$ are action variables~\cite{Arnold-Book}.

The function $S$ defined in equation~(\ref{Eq:GeneratingFunction}) generates new symplectic coordinates $(Q^\mu,P_\mu)$ as follows: the coordinates $P_\mu$ are defined by the conserved quantities:
\begin{eqnarray} \label{P_0}
P_0 &:=& \sqrt{2F_0} = \sqrt{-2H},\\ \label{P_1}
P_1 &:=& F_1 = -p_t,\\ \label{P_2}
P_2 &:=& F_2 = p_\varphi,\\ \label{P_3}
P_3 &:=& \sqrt{F_3} = \sqrt{p_\vartheta^2 + \frac{p_\varphi^2}{\sin^2\vartheta}}, 
\end{eqnarray}
such that each invariant set $\Gamma_{m,E,\ell_z,\ell}$ is parametrized simply by $P_0 = m$, $P_1 = E$, $P_2 = \ell_z$ and $P_3 = \ell$. The corresponding $Q$ variables are defined by $Q^\mu = \partial S/\partial P_\mu$, that is
\begin{eqnarray} \label{Q0}
Q^0 &:=& \frac{\partial S}{\partial m}
 = -m\int\limits_{\gamma_{(r,p_r)}} \frac{dr}{N p_r - \frac{2M_H}{r} E},
\\
Q^1 &:=& \frac{\partial S}{ \partial E}
  = -t + \int\limits_{\gamma_{(r,p_r)}} 
  \frac{\frac{2M_H}{r} p_r + \left( 1 + \frac{2M_H}{r} \right)E}{N p_r - \frac{2M_H}{r} E} dr,
\\
Q^2 &:=& \frac{\partial S}{ \partial \ell_z}
 = \varphi - \ell_z\int\limits_{\gamma_{(\vartheta,p_\vartheta)}}
  \frac{d\vartheta}{p_\vartheta \sin^2 \vartheta},
\\  \label{Q3}
Q^3 &:=& \frac{\partial S}{ \partial\ell}
 = -\ell\int\limits_{\gamma_{(r,p_r)}} \frac{dr}{r^2 \left( N p_r - \frac{2M_H}{r} E \right)} 
 + \ell\int\limits_{\gamma_{(\vartheta,p_\vartheta)}}\frac{d\vartheta}{p_\vartheta}.
\end{eqnarray}
In deriving these equations, we have used the differentials of the relations~(\ref{Eq:thetaPlaneRestriction},\ref{Eq:rPlaneRestriction}). As before, the integrals should be interpreted as line integrals in the $(r,p_r)$-plane and $(\vartheta,p_\vartheta)$-plane, respectively, where the curves connect the reference point to the given points $(r,p_r)$ and $(\vartheta,p_\vartheta)$ along the projections of the invariant submanifold $\Gamma_{m,E,\ell_z,\ell}$ onto these planes. Given $(x,p)\in \Gamma_{accr}$, the invariant submanifold is determined by the parameters $m := P_0(x,p)$, $E := P_1(x,p)$, $\ell_z := P_2(x,p)$ and $\ell := P_3(x,p)$. Note that by virtue of equations~(\ref{Eq:thetaPlaneRestriction},\ref{Eq:rPlaneRestrictionBis}) we have
$$
N p_r - \frac{2M_H}{r} E = \pmÊ\sqrt{E^2 - V_{m,\ell}(r)},\qquad
p_\vartheta = \pm\sqrt{\ell^2 - \frac{\ell_z^2}{\sin^2\vartheta}},
$$
where the correct choice for the sign depends on which part of the curve one is integrating over.

By construction, the function $S$ satisfies
$$
\frac{\partial S}{\partial x^\mu}(x; m,E,\ell_z,\ell) = p_\mu,
$$
with $p = p_\mu dx^\mu$ such that $P_0(x,p) = m$, $P_1(x,p) = E$, $P_2(x,p) = \ell_z$ and $P_3(x,p) = \ell$. Since
$$
\Omega_s = dp_\mu\wedge dx^\mu = \frac{\partial^2 S}{\partial P_\nu\partial x^\mu} dP_\nu\wedge dx^\mu = \frac{\partial Q^\nu}{\partial x^\mu} dP_\nu\wedge dx^\mu = dP_\nu\wedge dQ^\nu,
$$
the coordinates $(Q^\mu,P_\mu)$ are new symplectic coordinates on $\Gamma_{accr}$. Note that the new momenta $P_\mu$ parametrize the invariant four-dimensional submanifolds $\Gamma_{m,E,\ell_z,\ell}$ while the $Q^\mu$'s define local coordinates on $\Gamma_{m,E,\ell_z,\ell}$, with $Q^2$ and $Q^3$ changing by integer multiples of $2\pi$ under revolutions about the azimutal angle $\varphi$ or about the curve $\gamma_{(\vartheta,p_\vartheta)}$. Therefore, $Q^2$ and $Q^3$ can be regarded as angle variables parametrizing the two cycles $S^1$ in $\Gamma_{m,E,\ell_z,\ell}\simeq \Real^2\times S^1\times S^1$.

\subsection{Most general collisionless distribution function describing accretion}

Since in the new symplectic local coordinates $(Q^\mu,P_\mu)$ the one-particle Hamiltonian is $H = -P_0^2/2$, the associated Hamiltonian vector field acquires the simple form
$$
X_H = \frac{\partial H}{\partial P_\mu}\frac{\partial}{\partial Q^\mu}
 - \frac{\partial H}{\partial Q^\mu}\frac{\partial}{\partial P_\mu} 
  = -P_0\frac{\partial}{\partial Q^0}.
$$
Therefore, the Liouville equation~(\ref{Eq:Liouville}) in these coordinates reads,
$$
\frac{\partial f}{\partial Q^0} = 0,
$$
and the most general solution of the Liouville equation can be expressed in closed form,
\begin{equation}
f(x,p) = {\cal F}(Q^1,Q^2,Q^3,P_0,P_1,P_2,P_3),
\label{Eq:SchwCollisionlessf}
\end{equation}
where ${\cal F}$ is an arbitrary function of its argument which is $2\pi$-periodic in $Q^2$ and $Q^3$.

Note that so far, no symmetry assumptions have been made on the distribution function; our construction only relies on the symmetries of the underlying spacetime manifold $(M,g)$ and the fact that the geodesic motion on this manifold is described by an integrable Hamiltonian system. If required, the invariance of $f$ with respect to the symmetry group of $(M,g)$ (or subgroups thereof) can be imposed using the complete lift introduced in section~\ref{SubSec:Symmetries}. The isometry group of the Schwarzschild manifold $(M,g)$ is generated by the Killing vector fields $k := \frac{\partial}{\partial t}$ and the vector fields $\xi_a$, $a=1,2,3$, defined in equations~(\ref{Eq:xi12},\ref{Eq:xi3}). The complete lifts of these generators are given by
$$
\hat{k} = \frac{\partial}{\partial t}
$$
and the vector fields $\hat{\xi}_a$, $a=1,2,3$, defined in equations~(\ref{Eq:hatxi12},\ref{Eq:hatxi3}). In terms of the new symplectic coordinates $(Q^\mu,P_\mu)$ one finds
$$
\hat{k} = -\frac{\partial}{\partial Q^1},\qquad
\hat{\xi}_1 + i\hat{\xi}_2 = e^{i\varphi}(\hat{\eta}_1 + i\hat{\eta}_2),\qquad
\hat{\xi}_3 = \frac{\partial}{\partial Q^2},
$$
where
\begin{eqnarray*}
\hat{\eta}_1 &=& p_\vartheta\left[ \frac{\partial}{\partial P_2} 
  + \ell\int\limits_{\gamma_{(\vartheta,p_\vartheta)}}\frac{d\vartheta}{p_\vartheta^3\sin^2\vartheta}
  \left( \ell_z\frac{\partial}{\partial Q^3} - \ell\frac{\partial}{\partial Q^2} \right) \right]
 - \cot\vartheta\frac{\partial}{\partial Q^2},\\
\hat{\eta}_2 &=& \ell_z\cot\vartheta \left[ \frac{\partial}{\partial P_2} 
  + \ell\int\limits_{\gamma_{(\vartheta,p_\vartheta)}}\frac{d\vartheta}{p_\vartheta^3\sin^2\vartheta}
  \left( \ell_z\frac{\partial}{\partial Q^3} - \ell\frac{\partial}{\partial Q^2} \right) \right]
 - \frac{\ell_z}{p_\vartheta\sin^2\vartheta}\frac{\partial}{\partial Q^2}
 + \frac{\ell}{p_\vartheta}\frac{\partial}{\partial Q^3}.
\end{eqnarray*}
Note also that the Hamiltonian vector field $X_\ell$ associated with the total angular momentum $\ell = P_3$ is
$$
X_\ell = \frac{\partial}{\partial Q^3}.
$$
From these observations it follows immediately that the distribution function is spherically symmetric (that is, invariant with respect to the flows generated by the lifted generators $\hat{\xi}_a$, $a = 1,2,3$) if and only if ${\cal F}$ is independent of $Q^2$, $Q^3$, and $P_2$.

We summarize the main results of this section in the following theorem.

\begin{theorem}
\label{Thm:DistributionSchwarzschild}
On the invariant submanifold $\Gamma_{accr}\subset \Gamma$ of phase space, the most general collisionless distribution function is given by equation~(\ref{Eq:SchwCollisionlessf}), where the function ${\cal F}$ is $2\pi$-periodic in the angle variables $Q^2$ and $Q^3$, and the action-angle-like variables $(Q^\mu,P_\mu)$ are defined in equations~(\ref{P_0}--\ref{Q3}). 

Further, the distribution function is stationary if and only if ${\cal F}$ is independent of $Q^{1}$, axisymmetric if and only if ${\cal F}$ is independent of $Q^2$, and spherically symmetric if and only if ${\cal F}$ is independent of $Q^{2}$, $Q^3$ and $P _{2}$. 
\end{theorem}

{\bf Remarks}:
\begin{enumerate}
\item It follows from the theorem that a stationary and spherically symmetric collisionless distribution function on $\Gamma_{accr}$ only depends on the conserved quantities $m$, $E$ and $\ell$. This result is closely related to Jeans' theorem in the Newtonian case, which states that a stationary, spherically symmetric solution of the Poisson-Liouville system is described by a distribution function depending only on $E$ and $\ell$ (see~\cite{jBwFeH86} for a precise formulation). However, it has been shown that the generalization of Jeans' theorem to the general relativistic case is false in general~\cite{jS99,hAgR06}. The reason that in our setting a stationary and spherically symmetric distribution function must nevertheless be a function of only $(m,E,\ell)$ is due to the fact that we work on a fixed Schwarzschild background (neglecting the self-gravity of the gas) and to our restriction to the subset $\Gamma_{accr}$ of phase space. 
\item Unlike the spherically symmetric case, it is in general not true that a stationary, collisionless distribution function depends only on the conserved quantities arising from the spacetime symmetries. For instance, a stationary axisymmetric distribution function may depend on $Q^3$ apart from the quantities $m$, $E$, $\ell$ and $\ell_z$. 
\end{enumerate}

\section{Particle current density and stress energy-momentum tensor}
\label{Sec:Observables}

In this section, we analyze the relevant observables of a relativistic, collisionless kinetic gas on the Schwarzschild spacetime $(M,g)$. These observables are the particle current density and the stress energy-momentum tensor, and they are defined invariantly as follows:
\begin{equation}
J_x(X) := \int\limits_{\pi^{-1}(x)} p(X) f(x,p)\dvol_x(p),\qquad
T_x(X,Y) := \int\limits_{\pi^{-1}(x)} p(X) p(Y) f(x,p) \dvol_x(p),\qquad
X,Y \in T_x M,
\label{JxX}
\end{equation}
where here $\pi^{-1}(x) = \{ (x,p) : p\in T_x^* M \}\simeq T_x^* M$ is the fibre over $x$ and
$$
\dvol_x(p) := \sqrt{-\det(g^{\mu\nu}(x))} d^4 p
$$
is the invariant volume element on $\pi^{-1}(x)$ induced by the inverse metric $g_x^{-1}$. Using local adapted coordinates $(x^\mu,p_\mu)$ one can rewrite these quantities as
\begin{equation}
J_\mu(x) = \int\limits_{\pi^{-1}(x)} p_\mu f(x,p) \dvol_x(p),\qquad
T_{\mu\nu}(x) = \int\limits_{\pi^{-1}(x)} p_\mu p_\nu f(x,p) \dvol_x(p).
\end{equation}
It can be shown that by virtue of the Liouville equation~(\ref{Eq:Liouville}) these quantities are divergence-free. In the next subsection, we express these quantities more explicitly in terms of dimensionless quantities. The resulting expressions will be used in the next section in order to discuss physical applications regarding the accretion problem.

\subsection{Explicit expressions for the observables}

In order to simplify the computation of the fibre integrals, it is convenient to re-express the new symplectic variables $(Q^\mu,P_\mu)$ in terms of dimensionless quantities. For this purpose, we write
$$
t = M_H\tau,\quad
r = M_H\xi,\quad
p_r = m \pi_\xi,\quad
p_\vartheta = M_H m\pi_\vartheta,
$$
and
$$
E = m\varepsilon,\quad
\ell = M_H m\lambda,\quad
\ell_z = M_H m\lambda_z,
$$
with the new quantities $\tau$, $\xi$, $\pi_\xi$, $\pi_\vartheta$, $\varepsilon$, $\lambda$, $\lambda_z$ being dimensionless. The relevant $Q$ variables are
\begin{eqnarray}
Q^1 &=& M_{H} m\left[ G_{\varepsilon,\lambda}(\xi,\pi_\xi) - \tau \right],
\label{Eq:Q1}\\
Q^2 &=& \varphi - \lambda_z\int\limits_{\gamma_{(\vartheta,\pi_\vartheta)}}
  \frac{d\vartheta}{\pi_\vartheta \sin^2 \vartheta} ,
\label{Eq:Q2}\\
Q^3 &=& -\lambda \int\limits_{\gamma_{(\xi,\pi_\xi)}} 
\frac{d\xi}{\xi^2 \left[ \left( 1 - \frac{2}{\xi} \right) \pi_\xi - \frac{2}{\xi} \varepsilon \right]} 
 + \lambda \int\limits_{\gamma_{(\vartheta,\pi_\vartheta)}}\frac{d\vartheta}{\pi_\vartheta}.
\label{Eq:Q3}
\end{eqnarray}
Here, we have introduced the function
\begin{equation}
G_{\varepsilon,\lambda}(\xi,\pi_\xi) := \int \limits_{\gamma_{(\xi,\pi_\xi)}} 
  \frac{\frac{2}{\xi} \pi_\xi + \left( 1 + \frac{2}{\xi} \right)\varepsilon}
  {\left( 1 - \frac{2}{\xi} \right) \pi_\xi - \frac{2}{\xi} \varepsilon} d\xi,
\label{Eq:G_xpx}
\end{equation}
where the curve $\gamma_{(\xi,\pi_\xi)}$ connects a reference point to the given point $(\xi,\pi_\xi)$ along the set determined by
\begin{equation}
\left( 1 - \frac{2}{\xi} \right) \pi_\xi^2 - \frac{4}{\xi}\varepsilon \pi_\xi 
 - \left( 1 + \frac{2}{\xi} \right)\varepsilon^2 + 1 + \frac{\lambda^2}{\xi^2} = 0,
\label{Eq:xiPlaneRestriction}
\end{equation}
or, equivalently,
\begin{equation}
\pi_\xi = \pi_{\xi\pm}(\xi) 
 := \frac{\frac{2}{\xi}\varepsilon \pm \sqrt{\varepsilon^2 - U_\lambda(\xi)}}{1 - \frac{2}{\xi}}
 = \frac{1 + \frac{\lambda^2}{\xi^2} - \left( 1 + \frac{2}{\xi} \right)\varepsilon^2}
 {\frac{2}{\xi}\varepsilon \mp \sqrt{\varepsilon^2 - U_\lambda(\xi)}},
\label{Eq:pxipm}
\end{equation}
where the dimensionless effective potential is given by
\begin{equation}
U_\lambda(\xi) := \left( 1 - \frac{2}{\xi} \right)\left( 1 + \frac{\lambda^2}{\xi^2} \right).
\label{Eq:Ulambda}
\end{equation}
As explained in the previous section, in case (I) we choose the reference point to lie on the horizon, $(2,\pi_{\xi-}(2))$, while in case (II) we choose the reference point to be the turning point $(\xi_0,\pi_{\xi+}(\xi_0) = \pi_{\xi-}(\xi_0))$, where $U_\lambda(\xi_0) = \varepsilon^2$. The curve $\gamma_{(\vartheta,\pi_\vartheta)}$ appearing in the definitions of $Q^2$ and $Q^3$ connects the reference point $(\vartheta = \pi/2,\pi_\vartheta = \sqrt{\lambda^2-\lambda_z^2})$ to the given point $(\vartheta,\pi_\vartheta)$ clockwise along the set given by
\begin{equation}
\pi_\vartheta^2 + \frac{\lambda_z^2}{\sin^2\vartheta} = \lambda^2.
\end{equation}
As already mentioned previously, $Q^2$ and $Q^3$ increase by $2\pi$ along a full revolution along this curve, so these quantities are angle variables. For the applications in the next section the following result on the behavior of the function $G_{\varepsilon,\lambda}$ which determines the range of the variable $Q^1$ is important.

\begin{figure}[ht]
\centerline{\resizebox{7.0cm}{!}{\includegraphics{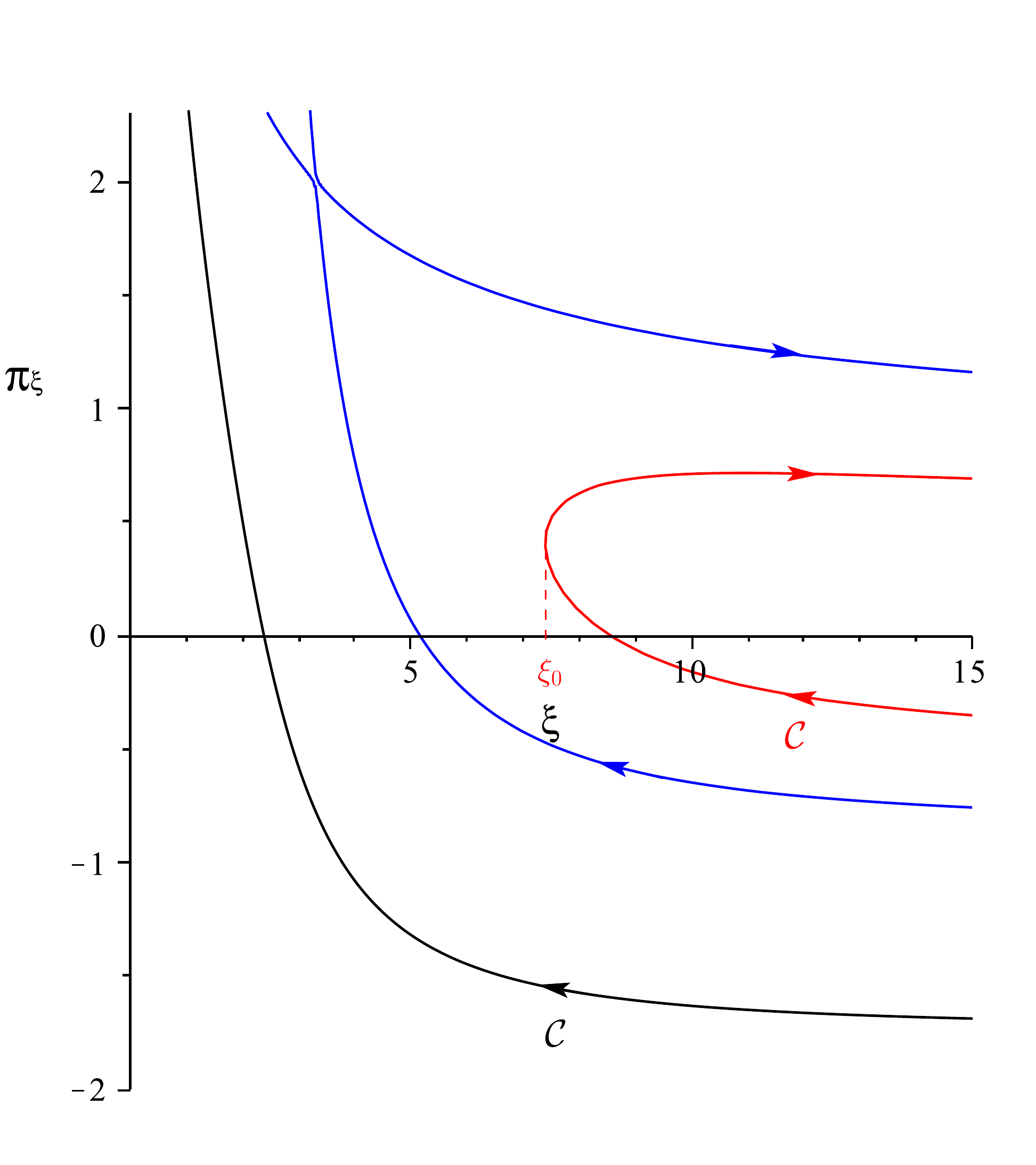}}}
\caption{The curve ${\cal C}$ for the three different cases. Black curve: case (I). Red curve: case (II). Blue curve: the separatrix corresponding to the limit between the two cases.}
\label{Fig:plot_rprBis}
\end{figure}

\begin{lemma}
\label{Lem:Gepslam}
Let $\varepsilon > 1$ and $\lambda\geq 0$, $\lambda\neq \lambda_c(\varepsilon)$ be fixed, and denote by ${\cal C}$ the curve in the $(\xi,\pi_\xi)$-plane determined by equation~(\ref{Eq:xiPlaneRestriction}). Then, the function $G_{\varepsilon,\lambda}: {\cal C}\to \Real$ defined in equation~(\ref{Eq:G_xpx}) is smooth and takes the following values:
\begin{enumerate}
\item[(a)] If $0\leq \lambda < \lambda_c(\varepsilon)$, corresponding to case (I) of absorbed particles, $G_{\varepsilon,\lambda}$ increases monotonously from $-\infty$ to $0$ as one moves along the curve ${\cal C}$ from $\xi = \infty$ to $\xi = 2$.
\item[(b)] If $\lambda > \lambda_c(\varepsilon)$, corresponding to case (II) of the scattered particles, $G_{\varepsilon,\lambda}$ increases monotonously from $-\infty$ to $+\infty$ as one moves clockwise along the curve ${\cal C}$, see figure~\ref{Fig:plot_rprBis}.
\end{enumerate}
\end{lemma} 

\proof We rewrite equation~(\ref{Eq:G_xpx}) as
$$
G_{\varepsilon,\lambda}(\xi,\pi_\xi) = \int\limits_{\gamma_{(\xi,\pi_\xi)}} \iota_{\cal C}^*\omega,
$$
with the one-form $\omega$ on the $(\xi,\pi_\xi)$-plane given by
$$
\omega_{(\xi,\pi_\xi)} = \frac{\frac{2}{\xi} \pi_\xi + \left( 1 + \frac{2}{\xi} \right)\varepsilon}
  {\left( 1 - \frac{2}{\xi} \right) \pi_\xi - \frac{2}{\xi} \varepsilon} d\xi,
$$
and where $\iota_{\cal C}: {\cal C}\subset \Real^2\to \Real^2$ denotes the inclusion map. The one-form $\omega$ is well-defined and smooth except at points where the denominator $v := (1 - 2/\xi) \pi_\xi - 2\varepsilon/\xi$ vanishes.

In case (I), $\pi_\xi$ is given by the expression in equation~(\ref{Eq:pxipm}) corresponding to the minus sign ($\pi_{\xi-}$) along ${\cal C}$, and thus,
$$
G_{\varepsilon,\lambda}(\xi,\pi_\xi) = \int\limits_2^\xi f_-(x) dx,\qquad
f_-(\xi) = -\frac{1}{\sqrt{\varepsilon ^2 - U_{\lambda}(\xi)}}
\frac{\varepsilon^2 \left( 1 + \frac{2}{\xi} \right)
 + \frac{4}{\xi^2}\left(1 + \frac{\lambda^2}{\xi^2} \right)}{\varepsilon + \frac{2}{\xi}\sqrt{\varepsilon ^2 - U_{\lambda}(\xi)}}.
$$
Here, the function $f_-: [2,\infty)\to \Real$ is well-defined, smooth and negative. Further, since $f_-(\xi)\to -\varepsilon/\sqrt{\varepsilon^2-1}$ as $\xi\to +\infty$, statement (a) follows.

In case (II), the denominator vanishes at the turning point $\xi = \xi_0$ of ${\cal C}$. Away from the turning point,
$$
\iota_{\cal C}^*\omega = \left\{ \begin{array}{cc}
f_-(\xi) d\xi, & v < 0,\\
f_+(\xi) d\xi, & v > 0,
\end{array} \right.\qquad
f_\pm(\xi) = \frac{\frac{2}{\xi}\sqrt{\varepsilon ^2-U_{\lambda}(\xi) } \pm \varepsilon}{\left( 1- \frac{2}{\xi}\right)\sqrt{\varepsilon^2 - U_{\lambda}(\xi) }},
$$
which is regular for $\xi\neq \xi_0$. To analyze the behavior in the vicinity of the turning point, we use the new coordinates $(\xi,v)$, in terms of which the curve ${\cal C}$ is parametrized by $v^2 + U_\lambda(\xi) = \varepsilon^2$, so that $2v dv = -U_\lambda'(\xi) d\xi$ along ${\cal C}$. Hence, 
$$
\iota_{\cal C}^*\omega 
 = \frac{1}{1 - \frac{2}{\xi}} \left( \frac{2v}{\xi} + \varepsilon \right) \frac{d\xi}{v}
 = \frac{-2}{1 - \frac{2}{\xi}} \left( \frac{2v}{\xi} + \varepsilon \right)\frac{dv}{U_\lambda'(\xi)},
$$
which is manifestly regular near the turning point, since $U_\lambda'(\xi_0) \neq 0$. Because $f_-(\xi) < 0 < f_+(\xi)$ for all $\xi > \xi_0$ and $f_\pm(\xi)\to \pm\varepsilon/\sqrt{\varepsilon ^2-1}$ as $\xi\to +\infty$, statement (b) follows.
\qed

{\bf Remarks}:
\begin{enumerate}
\item Since $Q^1 = M_H m[G_{\varepsilon,\lambda}(\xi,\pi_\xi) - \tau]$ is constant along the particle trajectories, the function $G_{\varepsilon,\lambda}$ describes the evolution of the dimensionless time coordinate $\tau$ along the trajectories. Observe that in the limit case $\lambda = \lambda_c(\varepsilon)$ (corresponding to the blue curve in figure~\ref{Fig:plot_rprBis}) the saddle point $(\xi_0,\pi_{\xi-}(\xi_0))$ is approached in infinite time, since in this case $U_\lambda'(\xi_0) = 0$.

\item As a consequence of Lemma~\ref{Lem:Gepslam} the variable $Q^1$ is always negative for $\tau\geq 0$ in case (I). In contrast to this, in case (II), the two terms $G_{\varepsilon,\lambda}(\xi,\pi_\xi)$ and $-\tau$ compete against each such that $Q^1$ can be positive or negative. This competition will be important later, when analyzing the asymptotic behavior of the flow as $\tau\to \infty$.
\end{enumerate}

After these remarks concerning the properties of $Q^1$ we return to the computation of the observables. In order to compute them, we find it convenient to introduce an angle $\chi$ such that
\begin{eqnarray*}
\pi_\vartheta &=& \lambda\cos\chi,\\
\frac{\lambda_z}{\sin\vartheta} &=& \lambda\sin\chi.
\end{eqnarray*}
In terms of the dimensionless quantities, the volume element is
\begin{equation}
\dvol_x(p) = \frac{m^3 dm d\varepsilon (\lambda d\lambda) d\chi}
{\xi^2\left| \left(1 - \frac{2}{\xi} \right) \pi_\xi - \frac{2}{\xi}\varepsilon \right|}
 = \frac{m^3 dm d\varepsilon (\lambda d\lambda) d\chi}
{\xi^2\sqrt{\varepsilon^2 - U_\lambda(\xi)}}. 
\label{Eq:FibreMeasure}
\end{equation}
We split the fibre integral into two parts, corresponding to cases (I) and (II) introduced in the previous section. In both cases, the fibre integrals defining the observables range over $m\in (0,\infty)$ and $\chi\in (0,2\pi)$. The ranges for $\varepsilon$ and $\lambda$ are as follows:
\begin{enumerate}
\item[(I)] In this case $\varepsilon\in (1,\infty)$ and $\lambda\in (0,\lambda_c(\varepsilon))$, where $\lambda_c(\varepsilon)$ is the critical angular momentum below which the particles fall into the black hole. Here, one has to choose the solution $\pi_{\xi-}(\xi)$ in equation~(\ref{Eq:pxipm}) which, from the second expression in that equation, is seen to be regular on the horizon $\xi = 2$. 

\item[(II)] In this case the ranges for $\varepsilon$ and $\lambda$ are more complicated since they are restricted by several conditions, some of which depend on the position $\xi$ of the fibre. First, we need $\lambda > \lambda_c(\varepsilon)$ in order for the particle to be reflected at the centrifugal barrier. Next, since we are only considering particles that are incoming from infinity and reflected (as opposed to those that are coming from the white hole region and scattered at the potential before falling into the black hole), we need to ensure that $\xi$ is larger than the location of the maximum of the effective potential, $\xi_{max}(\lambda)$. Finally, $\xi$ needs to lie in the allowed region where $U_\lambda(\xi) \leq \varepsilon^2$.

The precise bounds on $\varepsilon$ and $\lambda$ which fulfill these conditions are given in the next lemma. Note that in this case both solutions $\pi_{\xi+}(\xi)$ and $\pi_{\xi-}(\xi)$ in equation~(\ref{Eq:pxipm}) have to be taken into account and summed over when computing the fibre integral.
\end{enumerate}

\begin{lemma}
Let $\xi > 2$ and $\varepsilon > 1$, and define
\begin{equation}
\varepsilon_{min}(\xi) := \left\{ \begin{array}{rl}
\infty, & \xi\leq 3,\\
\sqrt{\left(1 - \frac{2}{\xi}\right)\left( 1 + \frac{1}{\xi-3} \right)}, & 3 < \xi < 4,\\
1, & \xi \geq 4.
 \end{array} \right.\qquad
\lambda_{max}(\varepsilon,\xi) := \xi\sqrt{ \frac{\varepsilon^2}{1 - \frac{2}{\xi}} - 1 }.
\label{Eq:EminLambdaMaxDef}
\end{equation}
Then, the ranges corresponding to case (II) are:
$$
\varepsilon > \varepsilon_{min}(\xi),\qquad
\lambda_c(\varepsilon) < \lambda < \lambda_{max}(\varepsilon,\xi).
$$
\end{lemma}

\proof First, we note that $U_{\lambda_{max}}(\xi) = \varepsilon^2$, hence the condition $\lambda < \lambda_{max}$ is equivalent to $U_\lambda(\xi) < \varepsilon^2$. Therefore, the inequalities $\lambda_c(\varepsilon) < \lambda < \lambda_{max}(\varepsilon,\xi)$ are necessary for case (II).

In order to justify the first condition $\varepsilon > \varepsilon_{min}(\xi)$ we first note that since $\xi_{max}(\lambda)$ decreases monotonically from $6$ to $3$ as $\lambda^2$ increases from $12$ to $\infty$ (see App.~\ref{App:EffPot}), the condition $\xi > \xi_{max}(\lambda)$ is never satisfied for $\xi\leq 3$. Hence, in this case the integration region is empty. When $3 < \xi\leq 6$ the condition $\xi > \xi_{max}$ is equivalent to $\lambda > \xi/\sqrt{\xi-3}$, which in turn implies
$$
\varepsilon^2\geq U_\lambda(\xi) > \left( 1 - \frac{2}{\xi} \right)\left(1 + \frac{1}{\xi-3} \right).
$$
For $3 < \xi\leq 4$ the right-hand side is equal to $\varepsilon_{min}(\xi)^2$, for $4 < \xi \leq 6$ the right-hand side is smaller than $1$. But since $\varepsilon > 1$ it follows that $\varepsilon > \varepsilon_{min}(\xi)$.

Conversely, if $\varepsilon > \varepsilon_{min}(\xi)$ and $\lambda > \lambda_c(\varepsilon)$, then the monotonicity of $\lambda_c$ and a short calculation reveal that
$$
\lambda^2 > \lambda_c(\varepsilon)^2 > \lambda_c(\varepsilon_{min}(\xi))^2
 = \left\{ \begin{array}{rl}
 \frac{\xi^2}{\xi-3}, & 3 < \xi \leq 4,\\
 16, & 4 < \xi < 6,
 \end{array} \right.
$$
which implies $\lambda > \xi/\sqrt{\xi-3}$ in both cases, and hence $\xi > \xi_{max}(\lambda)$.

Finally, when $\xi \geq 6$, the condition $\xi > \xi_{max}(\xi)$ is automatically satisfied and the only bounds are $\varepsilon > 1$ and $\lambda_c(\varepsilon) < \lambda < \lambda_{max}(\varepsilon,\xi)$.
\qed

After these remarks we are ready to compute the observables. A fibre integral of the form
$$
I(x) := \int\limits_{\pi^{-1}(x)} f(x,p)\dvol_x(p)
$$
is written as
$$
I(x) = I^{(abs)}(x) + I^{(scat)}(x),
$$
with the contribution from the absorbed particles
\begin{equation}
I^{(abs)}(x) = \frac{1}{\xi^2} \int\limits_1^\infty \int\limits_0^{2\pi}
\int\limits_0^{\lambda_c(\varepsilon)} 
\left. {\cal F}_3(G_{\varepsilon,\lambda}(\xi,\pi_\xi) - \tau,Q^2,Q^3,\varepsilon,\lambda\sin\chi\sin\vartheta,\lambda) \right|_{\pi_\xi = \pi_{\xi-}(\xi)}
\frac{\lambda d\lambda d\chi d\varepsilon}{\sqrt{\varepsilon^2 - U_\lambda(\xi)}}
\label{Eq:JgeneralinQ's}
\end{equation}
and the one from the scattered particles
\begin{eqnarray}
I^{(scat)}(x) &=& \frac{1}{\xi^2} \int\limits_{\varepsilon_{min}(\xi)}^\infty \int\limits_0^{2\pi}
\int\limits_{\lambda_c(\varepsilon)}^{\lambda_{max}(\varepsilon,\xi)}
\left. {\cal F}_3(G_{\varepsilon,\lambda}(\xi,\pi_\xi) - \tau,Q^2,Q^3,\varepsilon,\lambda\sin\chi\sin\vartheta,\lambda) \right|_{\pi_\xi = \pi_{\xi+}(\xi)}
\frac{\lambda d\lambda d\chi d\varepsilon}{\sqrt{\varepsilon^2 - U_\lambda(\xi)}}
\nonumber\\
 &+&
 \frac{1}{\xi^2} \int\limits_{\varepsilon_{min}(\xi)}^\infty \int\limits_0^{2\pi}
\int\limits_{\lambda_c(\varepsilon)}^{\lambda_{max}(\varepsilon,\xi)}
\left. {\cal F}_3(G_{\varepsilon,\lambda}(\xi,\pi_\xi) - \tau,Q^2,Q^3,\varepsilon,\lambda\sin\chi\sin\vartheta,\lambda) \right|_{\pi_\xi = \pi_{\xi-}(\xi)}
\frac{\lambda d\lambda d\chi d\varepsilon}{\sqrt{\varepsilon^2 - U_\lambda(\xi)}},
\label{Eq:JgeneralscatQ's}
\end{eqnarray}
where in the expressions for $Q^2$ and $Q^3$ one should substitute $\lambda_ z = \lambda\sin\vartheta\sin\chi$ and $\pi_\vartheta = \lambda\cos\chi$. Here we have introduced the mass momenta of the distribution function, defined by
\begin{equation}
{\cal F}_n\left(G,Q^2,Q^3,\varepsilon,\lambda_z,\lambda \right)
 := \int\limits_0^\infty 
 m^n {\cal F}(M_H m G,Q^2,Q^3,m,m\varepsilon,M_H m\lambda_z,M_H m \lambda) dm,
\qquad n = 0,1,2,3,\ldots
\label{Eq:fMassMoment}
\end{equation}
General expressions for the current density and stress energy-momentum tensor can be computed similarly, taking into account that $p_\tau = -m\varepsilon$, $p_r = m \pi_\xi$, $p_\vartheta = M_H m \lambda\cos\chi$, $p_\varphi = M_H m\lambda\sin\chi\sin\vartheta$. Since the momentum $p$ is proportional to $m$, the current density depends on ${\cal F}_4$ and the stress energy-momentum tensor on ${\cal F}_5$. The fact that the observables only depend on the mass momenta ${\cal F}_n$ of the distribution function is due to the weak equivalence principle.

\subsection{Observables in the spherically symmetric case}

In the spherically symmetric case, the ${\cal F}_n$ are independent of $Q^2$, $Q^3$ and $\lambda_z$, and we obtain the following expressions for the observables associated with the absorbed particles:
\begin{eqnarray}
J_a^{(abs)}(\tau,\xi) &=& \frac{2\pi}{\xi^2} \int\limits_1^\infty \int\limits_0^{\lambda_c(\varepsilon)} 
u_{a-} {\cal F}_4(G_{\varepsilon,\lambda}(\xi,\pi_{\xi-}(\xi)) -\tau,\varepsilon,\lambda) 
\frac{\lambda d\lambda d\varepsilon}{\sqrt{\varepsilon^2 - U_\lambda(\xi)}},
\qquad a = \tau,r,
\label{Eq:Jain}\\
T_{ab}^{(abs)}(\tau,\xi) &=& \frac{2\pi}{\xi^2} \int\limits_1^\infty \int\limits_0^{\lambda_c(\varepsilon)} 
u_{a-} u_{b-} {\cal F}_5(G_{\varepsilon,\lambda}(\xi,\pi_{\xi-}(\xi)) -\tau,\varepsilon,\lambda) 
\frac{\lambda d\lambda d\varepsilon}{\sqrt{\varepsilon^2 - U_\lambda(\xi)}},
\qquad a,b = \tau,r,
\label{Eq:Tabin}\\
T_{\vartheta\vartheta}^{(abs)}(\tau,\xi) &=& M_H^2\frac{\pi}{\xi^2} \int\limits_1^\infty
 \int\limits_0^{\lambda_c(\varepsilon)} 
\lambda^2 {\cal F}_5(G_{\varepsilon,\lambda}(\xi,\pi_{\xi-}(\xi)) -\tau,\varepsilon,\lambda) 
\frac{\lambda d\lambda d\varepsilon}{\sqrt{\varepsilon^2 - U_\lambda(\xi)}},
\label{Eq:Tthetathetain}
\end{eqnarray}
with $T_{\varphi\varphi}^{(abs)} = \sin^2\vartheta\, T_{\vartheta\vartheta}^{(abs)}$ and the remaining components of $J_\mu^{(abs)}$ and $T_{\mu\nu}^{(abs)}$ being zero. Here we have introduced the shorthand notation $(u_{\tau\pm},u_{r\pm}) = (-\varepsilon,\pi_{\xi\pm}(\xi))$ and for simplicity we have removed the entries $Q^2$, $Q^3$ and $\lambda_z$ from ${\cal F}_n$. The nonvanishing components of the scattering part are
\begin{eqnarray}
J_a^{(scat)}(\tau,\xi) &=& \frac{2\pi}{\xi^2} \int\limits_{\varepsilon_{min}(\xi)}^\infty 
\int\limits_{\lambda_c(\varepsilon)}^{\lambda_{max}(\varepsilon,\xi)} 
\sum\limits_\pm\left[ u_{a\pm}
{\cal F}_4(G_{\varepsilon,\lambda}(\xi,\pi_{\xi\pm}(\xi)) - \tau,\varepsilon,\lambda) \right]\frac{\lambda d\lambda d\varepsilon}{\sqrt{\varepsilon^2 - U_\lambda(\xi)}},
\label{Eq:Jascat}\\
T_{ab}^{(scat)}(\tau,\xi) &=& \frac{2\pi}{\xi^2} \int\limits_{\varepsilon_{min}(\xi)}^\infty
\int\limits_{\lambda_c(\varepsilon)}^{\lambda_{max}(\varepsilon,\xi)} 
\sum\limits_\pm\left[ u_{a\pm} u_{b\pm} 
{\cal F}_5(G_{\varepsilon,\lambda}(\xi,\pi_{\xi\pm}(\xi)) - \tau,\varepsilon,\lambda) \right]
\frac{\lambda d\lambda d\varepsilon}{\sqrt{\varepsilon^2 - U_\lambda(\xi)}},
\label{Eq:Tabscat}\\
T_{\vartheta\vartheta}^{(scat)}(\tau,\xi) &=& M_H^2\frac{\pi}{\xi^2} \int\limits_{\varepsilon_{min}(\xi)}^\infty
\int\limits_{\lambda_c(\varepsilon)}^{\lambda_{max}(\varepsilon,\xi)} 
\lambda^2 \sum\limits_\pm
\left[ {\cal F}_5(G_{\varepsilon,\lambda}(\xi,\pi_{\xi\pm}(\xi)) - \tau,\varepsilon,\lambda) 
\right]\frac{\lambda d\lambda d\varepsilon}{\sqrt{\varepsilon^2 - U_\lambda(\xi)}},
\label{Eq:Tthetathetascat}
\end{eqnarray}
and $T_{\varphi\varphi}^{(scat)} = \sin^2\vartheta\, T_{\vartheta\vartheta}^{(scat)}$, with $\sum_\pm$ indicating the sum over the contributions from the two signs in $\pi_{\xi\pm}$.

\subsection{Example of a simple, steady-state, spherically symmetric isotropic gas}
\label{SubSec:ExSSSI}

As an explicit example, we consider a steady-state, spherically symmetric, collisionless gas consisting of identical particles of rest mass $m > 0$. Furthermore, we assume that in the asymptotic region the gas is isotropic. These assumptions imply that the distribution function must have the form
\begin{equation}
f(x,p) = \delta\left( P_0(x,p) - m \right) \left. f_\infty(\varepsilon) \right|_{\varepsilon = P_1(x,p)/m},
\label{Eq:SSSSIG}
\end{equation}
with $f_\infty: (1,\infty)\to \Real$ a smooth, non-negative function which is bounded and decays sufficiently fast at infinity (see below). The associated mass momenta are
$$
{\cal F}_n(G,Q^2,Q^3,\varepsilon,\lambda_z,\lambda)
 = m^n f_\infty(\varepsilon),\qquad n = 0,1,2,3,\ldots
$$
Since the ${\cal F}_n$ only depend on the energy variable $\varepsilon$, the integrals over the total angular momentum $\lambda$ can be computed explicitly in equations~(\ref{Eq:Jain}--\ref{Eq:Tthetathetascat}). Using the integral identities in App.~\ref{App:integrals}, we obtain
\begin{eqnarray}
\left( \begin{array}{l} J^\tau \\ J^r \end{array} \right)^{(abs)}(\xi)
 &=& \frac{2\pi m^4}{\xi^2}\int\limits_1^\infty
\frac{\lambda_c(\varepsilon)^2}{\sqrt{\varepsilon^2 - U_0(\xi)} + \sqrt{\varepsilon^2 - U_c(\xi)}} 
\left( \begin{array}{l} v^\tau \\ v^r \end{array} \right) f_\infty(\varepsilon) d\varepsilon,
\label{Eq:JainSS}\\
\left( \begin{array}{ll} T^\tau{}_\tau & T^\tau{}_r  \\ T^r{}_\tau & T^r{}_r \end{array} \right)^{(abs)}
(\xi)
 &=& \frac{2\pi m^5}{\xi^2}\int\limits_1^\infty
\frac{\lambda_c(\varepsilon)^2}
{\sqrt{\varepsilon^2 - U_0(\xi)} + \sqrt{\varepsilon^2 - U_c(\xi)}} 
\left( \begin{array}{ll}
 v^\tau v_\tau & v^\tau v_r  + \frac{2}{\xi} W\\
 v^r v_\tau & v^r v_r + \left( 1- \frac{2}{\xi} \right) W
\end{array} \right) f_\infty(\varepsilon) d\varepsilon,
\label{Eq:TabinSS}\\
(T^\vartheta{}_\vartheta)^{(abs)}(\xi) &=& \frac{\pi m^5}{3\xi^4}\int\limits_1^\infty
\lambda_c(\varepsilon)^4\frac{2\sqrt{\varepsilon^2 - U_0(\xi)} + \sqrt{\varepsilon^2 - U_c(\xi)}}
{\left( \sqrt{\varepsilon^2 - U_0(\xi)} + \sqrt{\varepsilon^2 - U_c(\xi)} \right)^2} f_\infty(\varepsilon)
d\varepsilon,
\label{Eq:TthetathetainSS}
\end{eqnarray}
where we have introduced the shorthand notation $U_c(\xi) := U_{\lambda_c(\varepsilon)}(\xi)$, defined the quantity
$$
W := -\frac{1}{4}\left( \frac{1}{\varepsilon + \sqrt{\varepsilon^2 - U_0(\xi)}} 
 -\frac{1 + \frac{\lambda_c(\varepsilon)^2}{\xi^2}}{\varepsilon + \sqrt{\varepsilon^2 - U_c(\xi)}} 
\right)^2 + \frac{1}{3\xi^4}\frac{\lambda_c(\varepsilon)^4}{\left( \sqrt{\varepsilon^2 - U_0(\xi)} + \sqrt{\varepsilon^2 - U_c(\xi)} \right)^2},
$$
and the two-vector
$$
\left( \begin{array}{l} v^\tau \\ v^r \end{array} \right)
 = \left( \begin{array}{l}
 \varepsilon + \frac{1}{\xi}\frac{1}{\varepsilon + \sqrt{\varepsilon^2 - U_0(\xi)}}
 + \frac{1}{\xi}\frac{1 + \frac{\lambda_c(\varepsilon)^2}{\xi^2}} 
 {\varepsilon + \sqrt{\varepsilon^2 - U_c(\xi)}} \\
-\frac{1}{2}\left( \sqrt{\varepsilon^2 - U_0(\xi)} + \sqrt{\varepsilon^2 - U_c(\xi)} \right)
\end{array} \right),
$$
with corresponding co-vector
$$
\left( \begin{array}{l} v_\tau \\ v_r \end{array} \right)
 = \left( \begin{array}{l}
-\varepsilon \\
-\varepsilon + \frac{1}{2}\frac{1}{\varepsilon + \sqrt{\varepsilon^2 - U_0(\xi)}}
 + \frac{1}{2}\frac{1 + \frac{\lambda_c(\varepsilon)^2}{\xi^2}}
 {\varepsilon + \sqrt{\varepsilon^2 - U_c(\xi)}}
\end{array} \right).
$$

For the terms corresponding to the scattered particles, we find
\begin{eqnarray}
\left( \begin{array}{l} J^\tau \\ J^r \end{array} \right)^{(scat)}(\xi)
 &=& \frac{4\pi m^4}{\left( 1 - \frac{2}{\xi} \right)^2}\int\limits_{\varepsilon_{min}(\xi)}^\infty
\left( \begin{array}{r} 1 \\ 0 \end{array} \right)
 \varepsilon\sqrt{\varepsilon^2 - U_c(\xi)} f_\infty(\varepsilon) d\varepsilon,
\label{Eq:JascatSS}\\
\left( \begin{array}{ll} T^\tau{}_\tau & T^\tau{}_r \\ T^r{}_\tau & T^r{}_r \end{array} \right)^{(scat)}(\xi)
 &=& \frac{4\pi m^5}{\left( 1 - \frac{2}{\xi} \right)^2}\int\limits_{\varepsilon_{min}(\xi)}^\infty
\left( \begin{array}{ll}
-\varepsilon^2 & \frac{2}{3\xi}\frac{4\varepsilon^2 - U_c(\xi)}{1 - \frac{2}{\xi}} \\
 0 & \frac{1}{3}(\varepsilon^2 - U_c(\xi) )
\end{array} \right) \sqrt{\varepsilon^2 - U_c(\xi)} f_\infty(\varepsilon)  d\varepsilon,
\label{Eq:TabscatSS}\\
(T^\vartheta{}_\vartheta)^{(scat)}(\xi) &=& \frac{2\pi m^5}{\left( 1 - \frac{2}{\xi} \right)^2}
\int\limits_{\varepsilon_{min}(\xi)}^\infty 
\left[ \left( 1 - \frac{2}{\xi} \right)\frac{\lambda_c(\varepsilon)^2}{\xi^2}
 + \frac{2}{3}( \varepsilon^2 - U_c(\xi) ) \right]
\sqrt{\varepsilon^2 - U_c(\xi)} f_\infty(\varepsilon) d\varepsilon.
\label{Eq:TthetathetascatSS}
\end{eqnarray}
Physical applications of these results will be discussed in the next section; for the moment, we just like to emphasize that the expressions $J_\mu^{(abs)}$ and $T_{\mu\nu}^{(abs)}$ are manifestly regular on the horizon $\xi = 2$, while the contributions $J_\mu^{(scat)}$ and $T_{\mu\nu}^{(scat)}$ for the scattered particles vanish inside the photon sphere, that is, for $\xi < 3$ (see the definition of $\varepsilon_{min}(\xi)$ in equation~(\ref{Eq:EminLambdaMaxDef})).

The integrals in equations~(\ref{Eq:JainSS}--\ref{Eq:TthetathetascatSS}) are well-defined as long as the function $f_\infty$ decays sufficiently fast as $\varepsilon\to \infty$. For example, this is the case for exponential decay, i.e. if there are constants $\alpha,\beta > 0$ such that
$$
0\leq f_\infty(\varepsilon) \leq \alpha e^{-\beta\varepsilon},\qquad \varepsilon\geq 1.
$$

\subsection{Comparison with the perfect fluid case}
\label{SubSec:FluidComparison}

We conclude this section with a comparison between the properties of the observables of a collisionless gas with those of an isotropic perfect fluid. In the latter case, the particle current density and the stress energy-momentum tensor are given by
\begin{equation}
J^\mu = n u^\mu,\qquad
T^\mu{}_\nu = \rho u^\mu u_\nu + p \left( u^\mu u_\nu  + \delta^\mu{}_\nu \right),
\label{Eq:JTFluid}
\end{equation}
with $n$, $\rho$, $p$ and $u^\mu$ the particle density, energy density, pressure and four-velocity of the fluid, respectively. The conditions that characterize the observables~(\ref{Eq:JTFluid}) can be formulated as follows.
\begin{enumerate}
\item[(i)] $J^\mu = n u^\mu$ is a timelike eigenvector of $T^\mu{}_\nu$, that is $T^\mu{}_\nu u^\nu = -\rho u^\mu$. An important consequence of this property is that the stress energy-momentum tensor can be diagonalized:
\begin{equation}
T^\mu{}_\nu = \rho e_0^\mu e_{0\nu}
 + p_1 e_1^\mu e_{1\nu} + p_2 e_2^\mu e_{2\nu} + p_3 e_3^\mu e_{3\nu},
\label{Eq:TmunuDecomp}
\end{equation}
with $e_0 = u$ and $e_1$,  $e_2$, $e_3$ an orthonormal frame perpendicular to $u$. In this case, $p_1$, $p_2$ and $p_3$ are called the {\em principle pressures}. In the spherically symmetric case, $e_1$ can be chosen in the radial direction, and then $p_1 = p_{rad}$ describes the radial pressure and $p_2 = p_3 = p_{tan}$ the tangential one.
\item[(ii)] The principal pressures are equal to each other.
\end{enumerate}

In order to compare the properties of the kinetic gas to those of the perfect fluid, let us analyze these conditions for the spherically symmetric accretion case described in the previous subsection. First, regarding the scattering part given in equations~(\ref{Eq:JascatSS},\ref{Eq:TabscatSS},\ref{Eq:TthetathetascatSS}) we observe that the current density is proportional to the Killing vector field $k = \frac{\partial}{\partial t}$. Furthermore, it is seen from equation~(\ref{Eq:TabscatSS}) that $k$ is also an eigenvector of $T^\mu{}_\nu^{(scat)}$, such that condition (i) is satisfied. To verify the second condition, we introduce the orthonormal frame
\begin{eqnarray}
e_0 &=&\frac{M_H}{\sqrt{1-\frac{2}{\xi}}}\frac{\partial}{\partial \tau},
\label{Eq:e0Static}\\
e_1 &=&\frac{M_H}{\sqrt{1-\frac{2}{\xi}}}\left[
 \frac{2}{\xi} \frac{\partial}{\partial \tau} + \left(1-\frac{2}{\xi} \right) \frac{\partial}{\partial \xi}
 \right],
\label{Eq:e1Static}\\
e_2 &=&\frac{M_H}{\xi}\frac{\partial}{\partial \vartheta},
\label{Eq:e2Static}\\
e_3 &=&\frac{M_H}{\xi\sin\vartheta}\frac{\partial}{\partial \varphi},
\label{Eq:e3Static}
\end{eqnarray}
such that $e_0$ is parallel to $J^{scat}$. It is simple to verify that this frame diagonalizes $T^\mu{}_\nu^{(scat)}$, and the associated principle pressures are
\begin{eqnarray}
p_{rad}^{(scat)} &=& \frac{4\pi m^5}{\left( 1 - \frac{2}{\xi} \right)^2}
\int\limits_{\varepsilon_{min}(\xi)}^\infty 
\frac{1}{3}(\varepsilon^2 - U_c(\xi) )^{3/2} f_\infty(\varepsilon)  d\varepsilon,\\
p_{tan}^{(scat)} &=& p_{rad}^{(scat)} + \frac{2\pi m^5}{1 - \frac{2}{\xi}}
\int\limits_{\varepsilon_{min}(\xi)}^\infty 
\frac{\lambda_c(\varepsilon)^2}{\xi^2}(\varepsilon^2 - U_c(\xi) )^{1/2} f_\infty(\varepsilon) 
d\varepsilon.
\end{eqnarray}
We see that in general, $p_{tan}^{(scat)}$ is larger than $p_{rad}^{(scat)}$, so that the principal pressures are different from each other. In the asymptotic limit $\xi\to \infty$ this difference converges to zero and the gas is fluid-like. For later purposes, we also note that the triad frame $\{ e_1,e_2,e_3 \}$ is Fermi-transported along the Killing observers, such that it defines a non-rotating frame for these observers.

Regarding the contributions from the absorbed particles described by equations~(\ref{Eq:JainSS},\ref{Eq:TabinSS},\ref{Eq:TthetathetainSS}), in general not even the first condition (i) is satisfied, as we will see in an explicit example in the next section. Further, in general $J^{(abs)}$ does not point in the same direction as $J^{(scat)}$. These comments should make clear that the observables belonging to a kinetic gas have a much richer structure than those belonging to a perfect fluid.

\section{Applications to the accretion problem}
\label{Sec:Accretion}

In this section, we apply our results to the accretion of a relativistic, collisionless kinetic gas into a Schwarzschild black hole. We discuss two important results. The first one, discussed in section~\ref{SubSec:SSS}, provides the most general steady-state, spherically symmetric accretion solution for a collisionless gas which, at infinity, is given by an equilibrium configuration with some given temperature. From this we compute the relevant physical quantities, like the accretion and compression rates as well as the particle and energy densities and the principle pressures in the asymptotic region and on the horizon.

The second result, discussed in section~\ref{SubSec:FlowStability}, considers dynamical nonlinear perturbations of the spherical, steady-state configurations which arises from rather general (possibly nonspherical) initial data for the distribution function. In particular, we derive an asymptotic stability result implying the stability of the spherically symmetric, steady-state configurations.

\subsection{Spherically symmetric, steady-state configurations with given temperature at infinity}
\label{SubSec:SSS}

Here, we consider the physical scenario in which gas particles are accreted from a reservoir of identical massive and spinless particles at spatial infinity. Like in section~\ref{SubSec:ExSSSI}, we assume that the gas is steady-state, spherically symmetric and collisionless. Additionally, we suppose that the reservoir is isotropic and that some physical process in the past drove it to thermodynamic equilibrium. Therefore, we can characterize the state of the gas at infinity by an equilibrium distribution function with given temperature $T > 0$. However, we emphasize that it does not make sense to associate a local temperature to the gas, since we neglect collisions between the gas particles so that at finite radius one does not expect to gas to be in local thermodynamic equilibrium. Hence, $T$ should be regarded as an asymptotic parameter only.

As a consequence of our assumptions, the distribution function has the form
\begin{equation}
f(x,p) = \alpha\delta\left( P_0(x,p) - m \right) \left. e^{-\beta E} \right|_{E = P_1(x,p)},
\label{Eq:fEquilibrium}
\end{equation}
with $\alpha > 0$ a normalization constant, $m > 0$ the mass of the gas particles and $\beta = (k_B T)^{-1} > 0$ the inverse temperature.\footnote{As we discuss in~\cite{pRoS17} the distribution function defined by equation~(\ref{Eq:fEquilibrium}) does not describe a local equilibrium state at finite radius, though it does so at infinity.} The particle current density and stress energy-momentum tensor are given by the expressions in equations~(\ref{Eq:JainSS}--\ref{Eq:TthetathetascatSS}) with the function $f_\infty(\varepsilon)$ replaced with $\alpha e^{-m\beta\varepsilon}$. In the following, we discuss the physical content of these results.

First, we note that using the inverse metric~(\ref{Eq:SchwarzschildInv}) we can easily compute the particle and energy fluxes through a sphere of constant areal radius $r$,\footnote{The coordinate-invariant definition of these quantities is $j_n = 4\pi r^2 dr({\bf J})$ and $j_\varepsilon = -4\pi r^2 dr({\bf T}(k))$, where $r$ is the areal radius, ${\bf J} = J^\mu\partial_\mu$ the current density vector and ${\bf T}(k) := T^\mu{}_\nu k^\nu\partial_\mu$ the contraction of the stress energy-momentum tensor with the Killing vector field $k$.}
\begin{eqnarray} \label{j_n}
j_n &:=& 4\pi r^2 J^r = -4\pi^2\alpha M_H^2 m^4
\int\limits_1^\infty \lambda_c(\varepsilon)^2 e^{-z\varepsilon} d\varepsilon,
\\ \label{j_epsilon}
j_\varepsilon &:=& -4\pi r^2 T^r{}_\tau = -4\pi^2\alpha M_H^2 m^5
\int\limits_1^\infty \varepsilon\lambda_c(\varepsilon)^2 e^{-z\varepsilon} d\varepsilon,
\end{eqnarray}
where we recall that $\lambda_c(\varepsilon)$ is the critical angular momentum calculated in App.~\ref{App:EffPot}. Here and in the following, the dimensionless parameter $z$ is defined by
$$
z := m\beta = \frac{m c^2}{k_B T},
$$
where for convenience we have reintroduced the speed of light $c$. For typical astrophysical applications this quantity is very large, since the thermal energy is much smaller than the rest energy of the particles. For gas accreted from the interstellar medium, for instance, $z$ is of the order $10^9$~\cite{Shapiro-Book}. As expected from the conservation laws $\nabla_\mu J^\mu = 0$ and $\nabla_\mu(-T^\mu{}_\nu k^\nu) = 0$ the flux quantities $j_n$ and $j_\varepsilon$ are constant. Note that only the absorbed particles contribute to these fluxes; the contributions from the in- and outgoing particles that are scattered at the potential cancel out, such that $(J^r)^{(scat)} = (T^r{}_\tau)^{(scat)} = 0$.

Next, we analyze the observables in the asymptotic region $\xi\to \infty$. In this case, it is only the scattered particles that yield a nonvanishing contribution, and due to our assumptions on the distribution function, the gas in this region behaves as an isotropic, special relativistic perfect fluid~\cite{fJ11a,wI63}. Indeed, taking the limit $\xi\to \infty$ in equations~(\ref{Eq:JascatSS}--\ref{Eq:TthetathetascatSS}) and noticing that $U_c(\xi)\to 1$ one finds
$$
\lim\limits_{r\to\infty} J^\mu = n_\infty u_\infty^\mu,\qquad
\lim\limits_{r\to\infty} T^{\mu\nu} 
 = \left(\varepsilon_\infty + p_\infty \right) u_\infty^\mu u_\infty^\nu + p_\infty\eta^{\mu\nu},
$$
with the four-velocity at infinity given by $u_\infty = k = \frac{\partial}{\partial\tau}$ and $\eta^{\mu\nu}$ denoting the components of the inverse Minkowski metric. The particle density $n_\infty$, energy density $\varepsilon_\infty$ and pressure $p_\infty$ at infinity can be expressed in terms of the modified Bessel functions of the second kind:
$$
K_l(z) = \int\limits_0^\infty e^{-z\cosh\psi}\cosh(l\psi) d\psi
 = \frac{z^l}{1\cdot 3\cdots (2l-1)}\int\limits_0^\infty
e^{-z\cosh\psi} \sinh^{2l}\psi\, d\psi,\quad z > 0,\quad l = 1,2,3,\ldots
$$
as
\begin{eqnarray} \label{n_infty}
n_\infty(z) &=& 4\pi\alpha m^4\frac{K_2(z)}{z},\\ 
\varepsilon_\infty(z)
 &=& 4\pi\alpha m^5 \left[ \frac{K_1(z)}{z} + 3\frac{ K_2(z) }{z^2} \right],\qquad
p_\infty(z) = 4\pi\alpha m^5 \frac{K_2(z)}{z^2},
\end{eqnarray}
and the ideal gas equation $p_\infty = k_B T n_\infty$ is satisfied.

At the horizon, $(J^\mu)^{(scat)}$ and $(T^\mu{}_\nu)^{(scat)}$ vanish, and thus we only have the contributions from the absorbed particles. The particle and energy densities $n_H$ and $\rho_H$ and the radial and tangential pressures $p_{rad}$ and $p_{tan}$ are determined by $J^\mu = n_H u_H^\mu$ ($g_{\mu\nu} u_H^\mu u_H^\nu = -1$) and the decomposition~(\ref{Eq:TmunuDecomp}) which, for a kinetic gas, is guaranteed to exist~\cite{Synge2-Book,oStZ13}. Evaluating the expressions in equations~(\ref{Eq:JainSS}--\ref{Eq:TthetathetainSS}) at $\xi = 2$ and noting that $U_\lambda(2) = 0$ one finds
\begin{eqnarray}
\left. \left( \begin{array}{l}  J^\tau \\ J^r \end{array} \right) \right|_{\xi = 2} 
 &=& \frac{1}{4}\pi\alpha m^4\int\limits_1^\infty \frac{\lambda_c(\varepsilon)^2}{\varepsilon}
 \left( \begin{array}{l} v^\tau \\ v^r \end{array} \right) e^{-z\varepsilon} d\varepsilon,
\label{J-Horizon}\\
\left. \left( \begin{array}{ll}  
 T^\tau{}_\tau & T^\tau{}_r \\  T^r{}_\tau & T^r{}_r
 \end{array} \right) \right|_{\xi = 2} 
 &=& \frac{1}{4}\pi\alpha m^5\int\limits_1^\infty \frac{\lambda_c(\varepsilon)^2}{\varepsilon}
 \left( \begin{array}{ll}  
  v^\tau v_\tau & v^\tau v_r + W \\
  v^r v_\tau & v^r v_r 
 \end{array}\right) e^{-z\varepsilon} d\varepsilon,\qquad
W = \frac{1}{3\varepsilon^2}\left( \frac{\lambda_c(\varepsilon)}{4} \right)^4,
\label{T-Horizon}\\
\left. T^\vartheta{}_\vartheta \right|_{\xi = 2} 
 &=& \frac{1}{64}\pi\alpha m^5\int\limits_1^\infty \frac{\lambda_c(\varepsilon)^4}{\varepsilon} 
 e^{-z\varepsilon} d\varepsilon,
\label{Ttheta-Horizon}
\end{eqnarray}
with
$$
\left( \begin{array}{l} v^\tau \\ v^r \end{array} \right)
 = \varepsilon\left( \begin{array}{l} 
 1 + \frac{1}{2\varepsilon^2}\left( 1 + \frac{\lambda_c(\varepsilon)^2}{8} \right) \\ -1
\end{array} \right),\qquad
\left( \begin{array}{l} v_\tau \\ v_r \end{array} \right)
 = -\varepsilon\left( \begin{array}{l} 
 1 \\ 1 - \frac{1}{2\varepsilon^2}\left( 1 + \frac{\lambda_c(\varepsilon)^2}{8} \right)
\end{array} \right).
$$
These are still rather complicated expressions. To obtain simpler expressions which are easier to interprete, we take the limit $z\to \infty$ (remember that for typical astrophysical applications $z$ is very large) and obtain (see App.~\ref{App:Limits} for details)
\begin{eqnarray}
\frac{\rho_H}{n_H} &=& \frac{m c^2}{2\sqrt{3}} \left( 3+\sqrt{\frac{31}{3}} \right) 
\approx 1.79 m c^2,
\label{Eq:nH}\\
\frac{p_{rad}}{n_H} &=& \frac{mc^2}{2\sqrt{3}} \left(-3+\sqrt{\frac{31}{3}} \right)
\approx 0.0619 m c^2,
\label{Eq:pradH}\\
\frac{p_{tan}}{n_H} &=& \frac{m c^2}{\sqrt{3}}\approx 0.577 m c^2.
\label{Eq:ptanH}
\end{eqnarray}
We note that the tangential pressure is almost an order of magnitude larger than the radial one, showing that the collisionless kinetic gas behaves very differently than an isotropic perfect fluid near the horizon. This difference is probably due to the fact that most gas particles have nonvanishing angular momenta and do not collide. Moreover, the four-velocity and timelike eigenvector $e_0$ of $(T^\mu{}_\nu)$ are
\begin{equation}
u_H = \frac{1}{\sqrt{3}}\left[ \frac{5}{2}\frac{\partial}{\partial \tau} 
 - \frac{\partial}{\partial r} \right],\qquad
e_0 = \left( \frac{31}{3} \right)^{-1/4}\left[
 \left( 1 + \frac{1}{2}\sqrt{\frac{31}{3}} \right)\frac{\partial}{\partial \tau} 
 - \frac{\partial}{\partial r} \right].
\end{equation}
Although the value of $1 + \sqrt{31/3}/2 \simeq 2.61$ is quite close to $5/2$, this calculation shows that $u$ and $e_0$ are not parallel, and thus condition (i) in section~\ref{SubSec:FluidComparison} for the observables to be fluid-like is violated.

Finally, we compute the mass accretion rate $\dot{M} := m j_n$ and compression ratio $n_H/n_\infty$ of the gas. For large $z$ we find (see App.~\ref{App:Limits})
\begin{equation}
\lim\limits_{z\to \infty}\frac{1}{\sqrt{2\pi z}}\frac{\dot{M}}{n_\infty}
 = - 4m c r_H^2,\qquad
\lim\limits_{z\to \infty}\frac{1}{\sqrt{2\pi z}}\frac{n_H}{n_\infty} 
 = \frac{\sqrt{3}}{\pi},
\end{equation}
where $r_H = 2G M_H/c^2$ is the Schwarzschild radius of the black hole. Therefore, both the accretion rate and the compression ratio scale like $z^{1/2}$ for large values of $z$, and hence they are smaller by a factor of $z$ compared to the corresponding quantities in the Michel model~\cite{fM72,ZelNovik-Book,Shapiro-Book,eCoS15a}, describing the spherical steady-state accretion of a polytropic perfect fluid.

A more detailed analysis regarding the behavior and physical properties of the observables of the spherical steady-state model as a function of radius and the asymptotic temperature is given in Ref.~\cite{pRoS17}.

\subsection{Stability of the flow}
\label{SubSec:FlowStability}

The main goal of this subsection is to study the stability of the spherical, steady-state accretion flows discussed in the previous subsection. To this end, we consider time-dependent, possibly nonspherical and nonlinear perturbations of the distribution function arising from fairly  general initial data on the initial time slice $\tau = 0$. According to the results in the previous sections, such initial data can be parametrized by two functions ${\cal F}_i^{(abs)}(G,Q^2,Q^3,\varepsilon,\lambda_z,\lambda)$ and ${\cal F}_i^{(scat)}(G,Q^2,Q^3,\varepsilon,\lambda_z,\lambda)$ of the six variables $G$, $Q^2$, $Q^3$, $\varepsilon$, $\lambda_z$ and $\lambda$, see equations~(\ref{Eq:Q1}--\ref{Eq:Q3}) for the definitions of $G,Q^2,Q^3$ as functions of $(x,p)$.\footnote{For simplicity, we assume that all the gas particles have the same positive rest mass $m > 0$. For solutions describing particles with different masses, one simply replaces the distribution function with the appropriate mass momenta, see equation~(\ref{Eq:fMassMoment}).} Here, we recall that the range of these variables is:
$$
R^{(abs)} := \{ (G,Q^2,Q^3,\varepsilon,\lambda_z,\lambda)\in \Real^6
 : -\infty < G \leq 0, 1 < \varepsilon < \infty, 
|\lambda_z| < \lambda, 0\leq \lambda < \lambda_c(\varepsilon) \}
$$
and
$$
R^{(scat)} := \{ (G,Q^2,Q^3,\varepsilon,\lambda_z,\lambda)\in \Real^6
 : -\infty < G \leq \infty, 1 < \varepsilon < \infty, 
|\lambda_z| < \lambda,  \lambda > \lambda_c(\varepsilon)\},
$$
respectively. Therefore, ${\cal F}_i^{(abs)}: R^{(abs)}\to \Real$ and ${\cal F}_i^{(scat)}: R^{(scat)}\to \Real$ are non-negative functions which, in addition, are $2\pi$-periodic in $Q^2$ and $Q^3$.

The solution generated by this data is:
\begin{equation}
f(x,p) = \left\{ \begin{array}{rl}
 {\cal F}_i^{(abs)}(G_{\varepsilon, \lambda}(\xi,\pi_{\xi}) - \tau,Q^2,Q^3, 
 \varepsilon,\lambda_z,\lambda), & \hbox{if $\lambda < \lambda_c(\varepsilon)$}\\
 {\cal F}_i^{(scat)}(G_{\varepsilon, \lambda}(\xi,\pi_{\xi}) - \tau,Q^2,Q^3, 
 \varepsilon,\lambda_z,\lambda), & \hbox{if $\lambda > \lambda_c(\varepsilon)$}
\end{array} \right.,
\label{Eq:fSolution}
\end{equation}
where the function $G_{\varepsilon, \lambda}(\xi,\pi_{\xi})$ is given in equation~(\ref{Eq:G_xpx}). Note that the solution is stationary if and only if the functions ${\cal F}_i^{(abs)}$ and ${\cal F}_i^{(scat)}$ are independent of $G$. The observables are given by fibre integrals over this distribution function, as explained and worked out in the previous section. Here, we analyze the asymptotic behavior of these observables along certain future-directed timelike curves.

We assume that the initial functions ${\cal F}_i^{(abs)}$ and ${\cal F}_i^{(scat)}$ satisfy the following conditions:
\begin{enumerate}
\item[(i)] ${\cal F}_i^{(abs)}: R^{(abs)}\to \Real$ and ${\cal F}_i^{(scat)}: R^{(scat)}\to \Real$ are non-negative, measurable functions which are $2\pi$-periodic in $Q^2$ and $Q^3$. Furthermore, we require these functions to be uniformly bounded from above by an equilibrium distribution function, that is, there exist positive constants $\alpha$ and $\beta$ such that
$$
0\leq {\cal F}_i^{(abs,scat)}(G,Q^2,Q^3,\varepsilon,\lambda_z,\lambda)
 \leq \alpha e^{-\beta\varepsilon}
$$
for all $(G,Q^2,Q^3,\varepsilon,\lambda_z,\lambda)\in R^{(abs,scat)}$.
\item[(ii)] There exists a function $f_{-\infty}: (1,\infty)\to \Real$ such that
$$
\lim\limits_{G\to -\infty} {\cal F}_i^{(abs)}(G,Q^2,Q^3,\varepsilon,\lambda_z,\lambda) 
 = f_{-\infty}(\varepsilon)
$$
for all $0\leq Q^2,Q^3\leq 2\pi$, $\varepsilon > 1$, and $|\lambda_z| < \lambda < \lambda_c(\varepsilon)$.
\item[(iii)] There exists a function $f_{+\infty}: (1,\infty)\to \Real$ such that in the limits $G\to \pm\infty$ the function ${\cal F}_i^{(scat)}$ converges uniformly in $(Q^2,Q^3,\lambda_z,\lambda)$ to the functions $f_{\pm\infty}$:
\begin{eqnarray}
&& \lim\limits_{G\to -\infty} \sup_{\substack{Q^2,Q^3\in [0,2\pi]\\
|\lambda_z| < \lambda, \lambda > \lambda_c(\varepsilon)}}
|{\cal F}_i^{(scat)}(G,Q^2,Q^3,\varepsilon,\lambda_z,\lambda) - f_{-\infty}(\varepsilon) | = 0,\\
&& \lim\limits_{G\to +\infty} \sup_{\substack{Q^2,Q^3\in [0,2\pi]\\
|\lambda_z| < \lambda, \lambda > \lambda_c(\varepsilon)}}
|{\cal F}_i^{(scat)}(G,Q^2,Q^3,\varepsilon,\lambda_z,\lambda) - f_{+\infty}(\varepsilon) | = 0,\
\end{eqnarray}
for all $\varepsilon > 1$.
\end{enumerate}

Condition (i) guarantees the existence of the fibre integrals defining the particle current density $J_\mu$ and the stress energy-momentum tensor $T_{\mu\nu}$, as follows from the analysis in the last section. Conditions (ii) and (iii) imply that in the asymptotic region $\xi\to \infty$, where $G\to \pm \infty$, the initial distribution function converges to distribution functions $f_{\pm\infty}(\varepsilon)$ which only depend on the energy $\varepsilon$. Here, the plus (minus) signs refer to particles which, at $\tau = 0$, have positive (negative) radial velocity. For an equilibrium distribution function which does not discriminate between infalling and outgoing particles, one chooses $f_{-\infty} = f_{+\infty}$, but for generality we shall not necessarily impose this condition.

In the following, we analyze the behavior of the observables $J_\mu$ and $T_{\mu\nu}$ associated with the solution~(\ref{Eq:fSolution}) of the Liouville equation. These observables are given by integrals of the form given in equations~(\ref{Eq:JgeneralinQ's},\ref{Eq:JgeneralscatQ's}). The main results of this section are described in the following two theorems.

\begin{theorem}
\label{Thm:Stability1}
Consider the solution $f: \Gamma_{accr}\to \Real$ of the Liouville equation belonging to initial data satisfying the conditions (i),(ii), and (iii) above. Then, along the world line of a future-directed static observer $\gamma(t)$,
\begin{equation}
\lim\limits_{t\to\infty} J_\alpha(\gamma(t)) = J_\alpha[f_{-\infty}],\qquad
\lim\limits_{t\to\infty} T_{\alpha\beta}(\gamma(t)) = T_{\alpha\beta}[f_{-\infty}],
\end{equation}
where $J_\alpha$ and $T_{\alpha\beta}$ are the components of the current density and stress energy-momentum tensor with respect to the non-rotating frame $\{ e_0, e_1, e_2, e_3 \}$ defined in equations~(\ref{Eq:e0Static}--\ref{Eq:e3Static}), and $J_\alpha[f_{-\infty}] = J_\alpha^{(abs)} + J_\alpha^{(scat)}$ and $T_{\alpha\beta}[f_{-\infty}] = T_{\alpha\beta}^{(abs)} + T_{\alpha\beta}^{(scat)}$ are the corresponding quantities for a steady-state, spherical configurations as in equations~(\ref{Eq:JainSS}--\ref{Eq:TthetathetascatSS}) with $f_\infty$ replaced with $f_{-\infty}$.
\end{theorem}

\begin{theorem}
\label{Thm:Stability2}
Consider the solution $f: \Gamma_{accr}\to \Real$ of the Liouville equation belonging to initial data satisfying the conditions (i),(ii), and (iii) above. Let $\gamma(t)$ be the world line of a future-directed timelike observer with constant asymptotic radial velocity with respect to static observers,
$$
v_{obs,\infty} := \lim\limits_{t\to \infty} \frac{dr}{dt}(t)
 = \lim\limits_{\tau\to \infty} \frac{d\xi}{d\tau}(\tau)
$$
satisfying $0 < v_{obs,\infty} < 1$ meaning that at large times, the observer moves away from the black hole with constant radial velocity.

Then, with respect to the static frame $\{ e_0, e_1, e_2, e_3 \}$ defined in equations~(\ref{Eq:e0Static}--\ref{Eq:e3Static}), we have
\begin{eqnarray}
&& \lim\limits_{t\to\infty}\left( \begin{array}{l} J^0 \\ J^1 \end{array} \right)(\gamma(t))
 = 4\pi m^4\int\limits_1^\infty \left( \begin{array}{r} 1  \\  0 \end{array} \right) \varepsilon \sqrt{\varepsilon^2-1}f_{-\infty}(\varepsilon) d\varepsilon
 + \pi m^4 \int\limits_{1}^{\varepsilon_{obs,\infty}}  \left( \begin{array}{r} 
 2\varepsilon\left[ \sqrt{\varepsilon^2-1} - a(\varepsilon) \right] \\  
\varepsilon^2 -1 - a(\varepsilon)^2  \end{array} \right) \Delta f_{\infty}(\varepsilon) d\varepsilon,
\label{Eq:JascatLimit}
\end{eqnarray}
\begin{eqnarray}
\lim\limits_{t\to\infty}\left( \begin{array}{ll} T^0{}_0 & T^0{}_1 \\ T^1{}_0 & T^1{}_1 \end{array} \right)(\gamma(t))
 &=& 4\pi m^5\int\limits_1^\infty
\left( \begin{array}{ll}
-\varepsilon^2 & 0 \\ 0 & \frac{1}{3}(\varepsilon^2 - 1)
\end{array} \right) \sqrt{\varepsilon^2 - 1} f_{-\infty}(\varepsilon) d\varepsilon
\nonumber\\
 &+& \pi m^5\int\limits_1^ {\varepsilon_{obs,\infty}} 
\left( \begin{array}{ll}
2\varepsilon^2 \left[ a(\varepsilon) - \sqrt{\varepsilon^2-1} \right] & 
\varepsilon\left[ \varepsilon^2 -1 - a(\varepsilon)^2 \right]  \\
-\varepsilon\left[ \varepsilon^2 -1 - a(\varepsilon)^2 \right]  &
-\frac{2}{3}\left[ a(\varepsilon)^3 - (\varepsilon^2-1)^{3/2} \right]
\end{array} \right) \Delta f_{\infty}(\varepsilon) d\varepsilon,
\label{Eq:TabscatLimit}
\end{eqnarray}
\begin{eqnarray}
\lim\limits_{t\to\infty}T^2{}_2(\gamma(t)) = \lim\limits_{t\to\infty} T^3{}_3(\gamma(t))
 &=& \frac{4\pi m^5}{3}\int\limits_1^\infty (\varepsilon^2 -1)^{3/2} f_{-\infty}(\varepsilon) d\varepsilon
\nonumber\\
 &+& \frac{\pi m^5}{3}\int\limits_1^{\varepsilon_{obs,\infty}}
[a(\varepsilon) - \sqrt{\varepsilon^2-1}]^2 [a(\varepsilon) + 2\sqrt{\varepsilon^2-1}]
\Delta f_{\infty}(\varepsilon) d\varepsilon,
\label{Eq:TthetathetascatLimit}
\end{eqnarray}
where $\varepsilon_{obs,\infty} := (1 - v_{obs,\infty}^2)^{-1/2}$, $a(\varepsilon) := (\varepsilon^2 - 1)/(\varepsilon v_{obs,\infty})$, and where we have defined the difference
$$
\Delta f_{\infty} := f_{+\infty} - f_{-\infty}.
$$
The limits of the remaining components vanish.
\end{theorem}

{\bf Remarks}:
\begin{enumerate}
\item Therefore, in the limit $t\to \infty$, the tetrad components of the observables $J_\alpha$ and $T_{\alpha\beta}$ converge pointwise to the sum of two expressions, the first one being independent of the observer's radial velocity and thus identical to the ones obtained from spherically symmetric stationary accretion with boundary condition $f_{-\infty}(\varepsilon)$. The other expression depends on the observer's asymptotic radial velocity and the difference $\Delta f_{\infty}$ between the in- and outgoing distribution functions at infinity. For the particular case where $f_{+\infty} = f_{-\infty}$ this second expression vanishes, and the pointwise limit of the observables $J_\alpha$ and $T_{\alpha\beta}$ is independent of the observer.

\item From the moving observer's point of view, it makes sense to ask how the results given in Theorem~\ref{Thm:Stability2} change if instead of the static frame $\{ e_0, e_1, e_2, e_3 \}$ defined in equations~(\ref{Eq:e0Static}--\ref{Eq:e3Static}), a non-rotating (Fermi-propagated) frame $\{ \hat{e}_0, \hat{e}_1, \hat{e}_2, \hat{e}_3 \}$ along the observer is considered. The relation between the two frames is given by a Lorentz transformation of the form
$$
\hat{e}_\alpha(t)  = \Lambda(t)^\delta{}_\alpha e_\delta(t),\qquad
\Lambda(t)^\delta{}_\alpha := [\Lambda_{rot}(t)]^\beta{}_\alpha 
[\Lambda_{boost}(t)]^\delta{}_\beta,
$$
where $\Lambda_{rot}(t)$ is an appropriate rotation and $\Lambda_{boost}(t)$ a boost such that $u = [\Lambda_{boost}(t)]^\delta{}_0 e_\delta(t)$ is the observer's four-velocity. Under such a transformation, the tetrad components of the observables change according to
$$
J^\alpha = \Lambda^\alpha{}_\beta\hat{J}^\beta,\qquad
T^\alpha{}_\beta = \Lambda^\alpha{}_\gamma \Lambda_\beta{}^\delta 
\hat{T}^\gamma{}_\delta,
$$
where $\Lambda_\beta{}^\delta$ denotes the components of the inverse transposed of the matrix $\Lambda$. Hence, the limits of $\hat{J}^\alpha$ and $\hat{T}^\alpha{}_\beta$ as $t\to \infty$ exist if and only if $\Lambda(t)^\alpha{}_\beta$ has a well-defined limit for $t\to \infty$. This will be the case, for example, if the observer has an asymptotic constant velocity in the radial direction, in which case
$$
\lim\limits_{t\to \infty} (\Lambda_{boost}(t)^\alpha{}_\beta)
 = \left( \begin{array}{llll}
\varepsilon_{obs,\infty} & \varepsilon_{obs,\infty} v_{obs,\infty} & 0 & 0 \\
\varepsilon_{obs,\infty} v_{obs,\infty} & \varepsilon_{obs,\infty} & 0 & 0 \\
 0 & 0 & 1 & 0 \\
 0 & 0 & 0 & 1
\end{array} \right).
$$

\item The proofs of our stability theorems require the understanding of two key issues. The first one is the behavior of the distribution function $f(x,p)$ for large values of $\tau$. This involves analyzing the function $G_{\varepsilon, \lambda}(\xi,\pi_{\xi}) - \tau$ along the world line of the observer, where $G_{\varepsilon, \lambda}(\xi,\pi_{\xi})$ is given by the line integral in equation~(\ref{Eq:G_xpx}). For static observers this issue is trivial since $G_{\varepsilon, \lambda}(\xi,\pi_{\xi})$ is constant in $\xi$ (for fixed values of the energy $\varepsilon$ and angular momentum $\lambda$). However, for moving observers the two terms $G_{\varepsilon, \lambda}(\xi,\pi_{\xi})$ and $\tau$ compete against each other, and thus this issue requires a detailed analysis.

The second issue is about the validity of passing the limit $\tau \to \infty$ under the integrals over the momentum $p$. In order to handle this issue, we use Lebesgue's dominated convergence theorem\footnote{See, for instance Ref.~\cite{Royden}.}. This issue is also much more delicate for moving observers than for static ones, since in the former case the fibre and the volume form on it are not fixed, so that it is more difficult to find a uniform bound.
\end{enumerate}

\proofof{Theorem~\ref{Thm:Stability1}}
In the static case, the function $G_{\varepsilon, \lambda}(\xi,\pi_{\xi})$ is constant along the world line of the observer (for each fixed values of the energy $\varepsilon$ and the angular momentum $\lambda$). Therefore, it follows from our assumptions that
$$
\lim\limits_{\tau\to \infty} {\cal F}_i^{(abs,scat)}(G_{\varepsilon, \lambda}(\xi,\pi_{\xi}) - \tau, Q^2,Q^3,\varepsilon,\lambda_z,\lambda)
 = f_{-\infty}(\varepsilon)
$$
for any values of $\xi > 2$, $\varepsilon > 1$, $Q^2,Q^3\in [0,2\pi]$, and $|\lambda_z| < \lambda$ and $\pi_\xi$ given by one of the two expressions $\pi_{\xi\pm}(\xi)$ defined in equation~(\ref{Eq:pxipm}).

The observables $J_\alpha$ and $T_{\alpha\beta}$ are given by expressions of the same form as defined in equations~(\ref{Eq:JgeneralinQ's},\ref{Eq:JgeneralscatQ's}) with the additional weights $p_\alpha$ and $p_\alpha p_\beta$, respectively. With respect to the non-rotating frame $\{ e_0, e_1, e_2, e_3 \}$ defined in equations~(\ref{Eq:e0Static}--\ref{Eq:e3Static}), the components of the momentum are:
\begin{eqnarray}
p_0 &=& \frac{-m \varepsilon}{\sqrt{1-\frac{2}{\xi}}},
\label{P0}\\
p_1 &=& \frac{m}{\sqrt{1-\frac{2}{\xi}}}
\left[ -\frac{2\varepsilon}{\xi} + \left(1-\frac{2}{\xi} \right) \pi_{\xi} \right],
\label{P1}\\
p_2 &=& \frac{m\pi_{\vartheta}}{\xi} = \frac{m\lambda}{\xi}\cos\chi,
\label{P2}\\
p_3 &=& \frac{m \lambda_{z}}{\xi \sin \vartheta} = \frac{m\lambda}{\xi}\sin\chi.
\label{P3}
\end{eqnarray}
From these expressions and equation~(\ref{Eq:pxipm}) it follows the existence of positive constants $c_1$ and $c_2$ (depending on $\xi > 2$) such that
$$
|p_\alpha| \leq m\left[ c_1\varepsilon + c_2\lambda \right]
$$
for all $\alpha = 0,1,2,3$. Therefore, the integrands in $J_\alpha$ and $T_{\alpha\beta}$ are bounded by the functions
$$
H_s(\varepsilon,\lambda) 
 := \alpha m^{3+s} (c_1\varepsilon + c_2\lambda)^s e^{-\beta\varepsilon}
\frac{\lambda}{\xi^2\sqrt{\varepsilon^2 - U_\lambda(\xi)}}
$$
with $s=1$ and $s=2$ for $J_\alpha$ and $T_{\alpha\beta}$, respectively. These functions are Lebesgue-integrable on the desired ranges, for
$$
\int\limits_1^\infty\int\limits_0^{2\pi}\int\limits_0^{\lambda_c(\varepsilon)} 
H_s(\varepsilon,\lambda) d\lambda d\chi d\varepsilon
 \leq 2\pi\alpha m^{3+s} \int\limits_1^\infty
 \left( \int\limits_0^{\lambda_c(\varepsilon)}
  \frac{\lambda d\lambda}{\xi^2\sqrt{\varepsilon^2 - U_\lambda(\xi)}} \right) 
  (c_1\varepsilon + c_2\lambda_c(\varepsilon))^s e^{-\beta\varepsilon} d\varepsilon.
$$
According to App.~\ref{App:integrals} the integral over $\lambda$ gives
$$
\frac{1}{1 - \frac{2}{\xi}} \left[ \sqrt{\varepsilon^2 - U_0(\xi)} - \sqrt{\varepsilon^2 - U_c(\xi)} \right] \leq c_3\varepsilon.
$$
Since $\lambda_c(\varepsilon) \leq c_4\varepsilon$ (see App.~\ref{App:EffPot}), it follows that $H_s$ is integrable over the range corresponding to case (I). A similar argument shows that $H_s$ is also integrable over the range associated with the scattered particles (case (II)).

Now we can use Lebesgue's dominated convergence theorem in order to pass the limit below the integral, and we conclude that $J_\alpha$ and $T_{\alpha\beta}$ converge to the corresponding stationary expressions computed in section~\ref{SubSec:ExSSSI} with $f_\infty$ replaced with $f_{-\infty}$.
\qed

\proofof{Theorem~\ref{Thm:Stability2}}
As mentioned before, the proof for moving observers is more complicated than in the static case due to two issues. The first issue is that $G_{\varepsilon, \lambda}(\xi,\pi_{\xi})$ is not constant anymore, since the assumption implies that $\xi\to \infty$ along the observer's world line. Hence, depending on the values of the constants $\varepsilon$ and $\lambda$ the expression
$$
G_{\varepsilon, \lambda}(\xi,\pi_{\xi+}) - \tau
$$
might converge to $-\infty$ or $\infty$ along the world line of the observer (cf. Lemma~\ref{Lem:Gepslam}). The second issue is the fact that the fibre and the volume element change in time. As a consequence, the measure (see equation~(\ref{Eq:FibreMeasure}))
$$
\frac{\lambda d\lambda d\varepsilon d\chi}{\xi^2\sqrt{\varepsilon^2 - U_\lambda(\xi)}}
$$
appearing in the definition of the observables is not constant anymore, and the application of Lebesgue's theorem would require finding a uniform ($\xi$-independent) Lebesgue-integrable upper bound for the integrand. To bypass this problem, we change the integration variable $\lambda$ to the new variable
$$
v := \sqrt{\varepsilon^2 - U_{\lambda}(\xi)},
$$
which represents the radial velocity of the particles. In terms of this new variable,
$$
\frac{\lambda d\lambda d\varepsilon d\chi}{\xi^2\sqrt{\varepsilon^2 - U_\lambda(\xi)}}
 = -\frac{dv d\varepsilon d\chi}{1 - \frac{2}{\xi}},
$$
so the measure becomes trivial, up to the factor $1 - 2/\xi$ which converges to $1$ along the observer's world line. The expressions for the current density have the form (cf.~(\ref{Eq:JgeneralinQ's},\ref{Eq:JgeneralscatQ's}))\footnote{Note that the lower limit $\varepsilon_{min}(\xi)$ in the $\varepsilon$-integral in $J_\delta^{(scat)}$ can be replaced by $1$ since $\xi\to \infty$.}
\begin{eqnarray*}
J_\delta^{(abs)}(x) &=& \frac{m^3}{1-\frac{2}{\xi}}
 \int\limits_0^{2\pi} \int\limits_1^\infty \int\limits_0^{\infty} 
p_{\delta-} {\cal F}_{-}^{(abs)} 
{\bf 1}_{\left(v_{c}(\varepsilon,\xi), \sqrt{\varepsilon^2 -1 + \frac{2}{\xi}} \right)}(v) dv d\varepsilon d\chi,
\label{Eq:Jdeltaabs}\\
J_\delta^{(scat)}(x) &=& \frac{m^3}{1 - \frac{2}{\xi}}\int\limits_0^{2\pi}\int\limits_1^\infty \int\limits_0^\infty 
\left[ p_{\delta-} {\cal F}_{-}^{(scat)} + p_{\delta+} {\cal F}_{+} ^{(scat)}\right] 
{\bf 1}_{(0,v_c(\varepsilon,\xi))}(v) dv d\varepsilon d\chi,
\label{Eq:Jdeltascat}
\end{eqnarray*}
where we have abbreviated ${\cal F}_{-}^{(abs)} := {\cal F}_i^{(abs)}(G_{\varepsilon,\lambda}(\xi,\pi_{\xi-}(\xi)) -\tau,Q^2,Q^3,\varepsilon,\lambda\sin\chi\sin\vartheta,\lambda)$ and ${\cal F}_\pm^{(scat)} := {\cal F}_i^{(scat)}(G_{\varepsilon,\lambda}(\xi,\pi_{\xi\pm}(\xi)) -\tau,Q^{2},Q^{3},\varepsilon,\lambda, \lambda\sin\chi \sin\vartheta)$, where we have defined $v_c(\varepsilon,\xi) := \sqrt{\varepsilon^2 - U_{\lambda_c(\varepsilon)}(\xi)}$, and where $\lambda$ should be substituted by the expression
\begin{equation}
\lambda(v,\xi) :=  \xi\sqrt{\frac{\varepsilon^2 - v^2}{1-\frac{2}{\xi}} - 1}.
\label{Eq:lambdavxi}
\end{equation}
Finally, in order to work on a fixed integration domain, we have introduced the indicator function ${\bf 1}_{(a,b)}$ with respect to the interval $(a,b)$, which is defined by:
$$
{\bf 1}_{(a,b)}(v) = \left\{ \begin{array}{lcc}
             1 & \hbox{if} & v\in (a,b),\\
             0 & \hbox{if} & v\notin (a,b).
\end{array} \right.
$$

According to the hypothesis, both integrands are bounded by a constant times
$$
|p_\delta| e^{-\beta\varepsilon} {\bf 1}_{(0,\varepsilon)}(v).
$$
Here, the components of the momentum, $p_\delta$, with respect to the static frame defined in equations~(\ref{Eq:e0Static}--\ref{Eq:e3Static}) can be bounded using the explicit expressions given in equations~(\ref{P0}--\ref{P3}), where we notice that (see equation~(\ref{Eq:lambdavxi}))
$$
\left| \frac{\lambda}{\xi} \right| \leq c_5 \varepsilon
$$
and
$$
|\pi_{\xi\pm}| = \left| \frac{\frac{2}{\xi}\varepsilon \pm v}{1 - \frac{2}{\xi}} \right| \leq c_6\varepsilon + c_7 v
$$
for some positive constants $c_5$, $c_6$ and $c_7$ which are independent of $\varepsilon$, $v$ and $\xi$. Consequently, both integrands are bounded by a constant times
$$
\varepsilon e^{-\beta\varepsilon} {\bf 1}_{(0,\varepsilon)}(v),
$$
which is integrable. Therefore, we can apply Lebesgue's dominated convergence theorem and pass the limit $\tau\to \infty$ below the integral. In the following, we analyze these limits.

Let us start with the contribution from the absorbed particles, $J_\delta^{(abs)}$. In this case, both endpoints of the interval $(v_{c}(\varepsilon,\xi), \sqrt{\varepsilon^2 -1 + \frac{2}{\xi}})$ converge to $\sqrt{\varepsilon^2-1}$ when $\xi\to \infty$. Consequently, in the limit $\xi\to \infty$, the integrand converges pointwise to zero almost everywhere and hence
$$
\lim\limits_{t\to \infty} J_\delta^{(abs)}(\gamma(t)) = 0.
$$

\begin{figure}[ht]
\centerline{\resizebox{8.5cm}{!}{\includegraphics{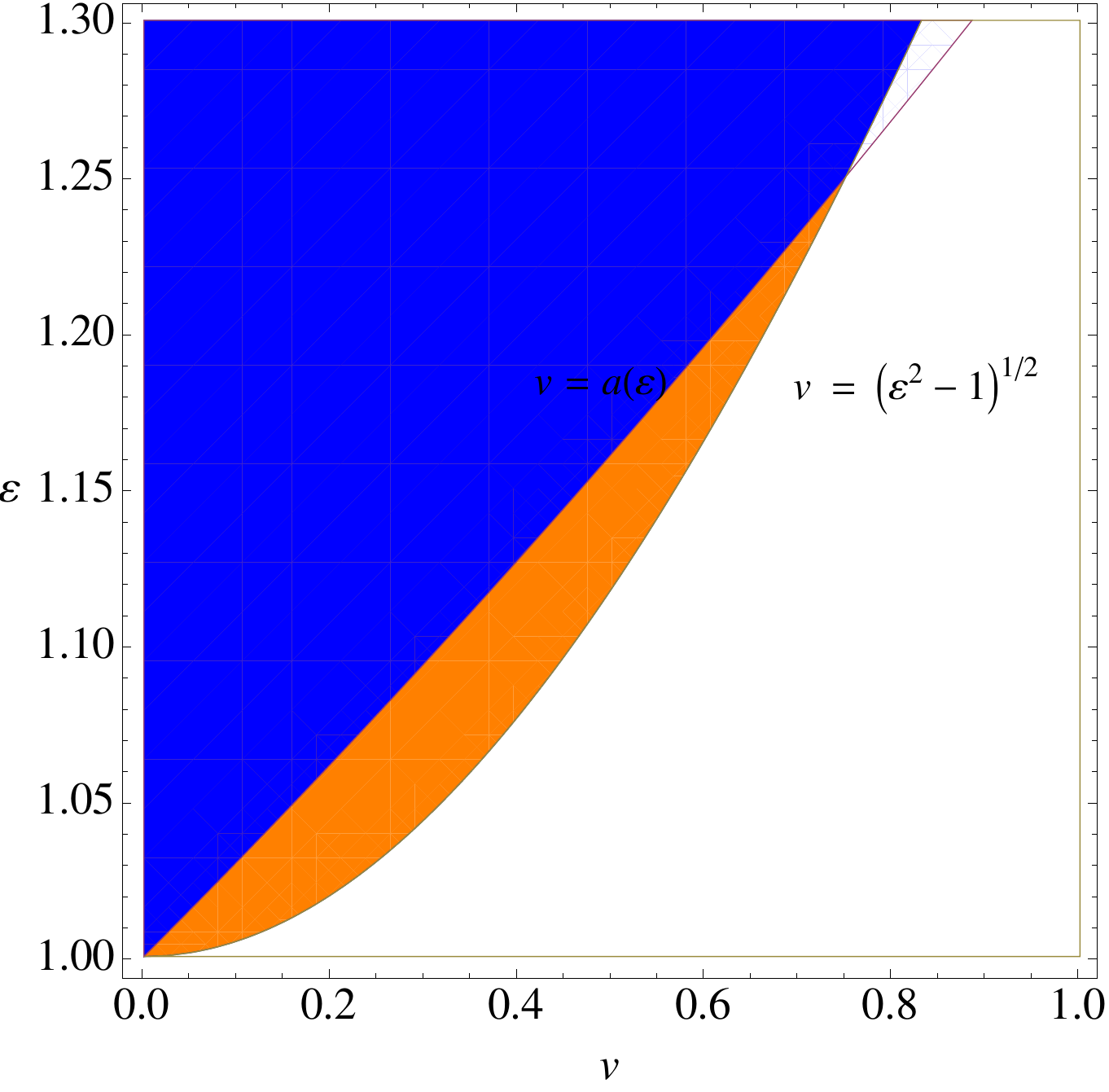}}}
\caption{The three regions in the $(\varepsilon,v)$ diagram which are delimited by the two curves $v = \sqrt{\varepsilon^2-1}$ and $v = a(\varepsilon) = (\varepsilon^2-1)/(\varepsilon v_{obs,\infty})$. In the white region, the integrand defining the observables converges to zero. In the orange region the expression $G_{\varepsilon,\lambda(v,\xi(\tau))}(\xi(\tau),\pi_{\xi+}(\tau)) -\tau$ tends to $+\infty$, and in the blue region it tends to $-\infty$.}
\label{Fig:regions}
\end{figure}

Next, we analyze the limit of the integrand of $J_\delta^{(scat)}$. For this, we first notice that $v_c(\varepsilon,\xi)\to \sqrt{\varepsilon^2-1}$ as $\xi\to \infty$, and hence
$$
\lim_{\xi \to \infty}{\bf 1}_{\left(0,v_{c}(\varepsilon,\xi) \right)}\left(v \right)= \left\{ \begin{array}{lcr}
             1 & \hbox{if} & 0 < v < \sqrt{\varepsilon^2 -1},\\
             0 & \hbox{if} & v > \sqrt{\varepsilon^2 -1}
\end{array} \right.,
$$ 
Therefore, the limit is zero in the region $v > \sqrt{\varepsilon^2-1}$ (see figure~\ref{Fig:regions}). Consequently, in what follows, we may focus our attention on the complementary region $v^2 < \varepsilon^2 - 1$.
The next step consists in determining the limit values of ${\cal F}_\pm^{(scat)}$. For this, we need to understand the limit
$$
\lim\limits_{\tau\to\infty} \left[ G_{\varepsilon, \lambda(v,\xi(\tau))}(\xi(\tau), \pi_{\xi\pm}(\tau)) 
 - \tau \right],\qquad \pi_{\xi\pm}(\tau) := \pi_{\xi\pm}(\xi(\tau)).
$$
The result is given in the following

\begin{lemma}[cf. figure~\ref{Fig:regions}]
\label{Lem:mtauG}
Let $\varepsilon > 1$ and $0 < v < \sqrt{\varepsilon^2 -1}$. Then,
\begin{eqnarray*}
\lim\limits_{\tau\to\infty} \left[ G_{\varepsilon,\lambda(v,\xi(\tau))}(\xi(\tau),\pi_{\xi-}(\tau)) -\tau \right] 
 &=& -\infty,\\
\lim\limits_{\tau\to\infty} \left[ G_{\varepsilon,\lambda(v,\xi(\tau))}(\xi(\tau),\pi_{\xi+}(\tau)) -\tau \right]
 &=& \left\{ \begin{array}{ll}
 -\infty & \hbox{if } v < a(\varepsilon),\\
 +\infty & \hbox{if } v > a(\varepsilon),
\end{array} \right.
\end{eqnarray*}
where we recall that $a(\varepsilon) = (\varepsilon^2-1)/(\varepsilon v_{obs,\infty})$.
\end{lemma}

\proof We defer the proof of this technical lemma to App.~\ref{App:mtauG}.
\qed

With the help of Lemma~\ref{Lem:mtauG} it is now straightforward to complete the computation of the asymptotic limit of $J_\delta^{(scat)}$ along the observer's path. Using assumption (iii) of the theorem we obtain
$$
\lim\limits_{t\to \infty}ÊJ_\delta^{(scat)}(\gamma(t)) = m^3
\int\limits_0^{2\pi}\int\limits_1^\infty \int\limits_0^{\sqrt{\varepsilon^2-1}}
\left\{  \lim\limits_{\tau\to \infty} p_{\delta-} f_{-\infty}(\varepsilon)
  + \lim\limits_{\tau\to \infty} p_{\delta+} \left[ f_{-\infty}(\varepsilon) 
  + \Delta f_\infty(\varepsilon) {\bf 1}_{(a(\varepsilon),\infty)}(v) \right] \right\} 
dv d\varepsilon d\chi,
$$
where we recall that $\Delta f_{\infty} := f_{+\infty} - f_{-\infty}$. The limits of the components of the momentum can be obtained using equations~(\ref{P0}--\ref{Eq:lambdavxi}):
$$
\lim\limits_{\tau\to \infty} p_{0\pm} = -m\varepsilon,\quad
\lim\limits_{\tau\to \infty} p_{1\pm} = \pm m v,\quad
\lim\limits_{\tau\to \infty} p_{2\pm} = m\cos\chi\sqrt{\varepsilon^2 - v^2 - 1},\quad
\lim\limits_{\tau\to \infty} p_{3\pm} = m\sin\chi\sqrt{\varepsilon^2 - v^2 - 1}.
$$
From this, it is simple to obtain the result given in equation~(\ref{Eq:JascatLimit}). The proof for the convergence of the stress energy-momentum tensor proceeds in a similar way.
\qed

\section{Conclusions}
\label{Sec:Conclusions}

In this work, we have considered a relativistic, collisionless kinetic gas propagating on the curved spacetime describing a Schwarzschild black hole. First, we provided a brief review for the formal structure of the theory needed in this article, including the symplectic structure, the Hamiltonian formulation of the Liouville vector field, and the lift of Killing vector fields on the cotangent bundle associated with the spacetime manifold. Next, using standard tools from the theory of Hamiltonian mechanics, we derived the most general distribution function describing the accretion of a collisionless gas into a Schwarzschild black hole (see Theorem~\ref{Thm:DistributionSchwarzschild}). Unlike the previous derivation in~\cite{oStZ14b} where the same problem was solved in the Kerr case, here we directly exploited the integrability of the Hamiltonian system and defined the generating function leading to the new symplectic coordinates in which the Liouville vector field is trivialized through an integral over the Poincar\'e one-form. Furthermore, we provided explicit expressions for the relevant physical observables on the spacetime manifold, including the particle current density and the stress energy-momentum tensor.

After these general remarks on the structure of the general solution of the Liouville equation on a Schwarzschild background, we turned our attention to applications for the accretion problem. We started with the simple case in which the gas flow is assumed to be spherically symmetric and steady-state, and showed that these assumptions yield a one-particle distribution function which depends only on the mass $m$, the energy $E$ and the total angular momentum $\ell$ of the particle. The value of $\ell$ plays an important role in distinguishing those particle that fall into the black hole from those that are reflected at the centrifugal barrier. As we exhibited, the former particles contribute to the accretion rate but not to the particle density $n_\infty$ at infinity, while the reflected particles yield a positive $n_\infty$ but do not contribute to the accretion rate. Next, we provided an explicit example in which the distribution function describes a gas in thermodynamic equilibrium at some temperature $T$ at infinity, and computed the observables as a function of the inverse temperature $\beta = 1/(k_B T)$. In the low temperature limit $\beta\to \infty$ we verified that our results coincide with those in~\cite{Shapiro-Book,ZelNovik-Book} for the accretion rate that were obtained based on Newtonian calculations. Additionally, we computed the energy density and radial and tangential pressures $p_{rad}$ and $p_{tan}$ at the horizon for $\beta\to \infty$ and showed that $p_{tan}$ is almost an order of magnitude larger than $p_{rad}$. This provides a partial explanation for the fact that the accretion is much less intense than in the Bondi-Michel case of a perfect, polytropic fluid case where $p_{tan} = p_{rad}$.

In the final part of this work, we studied the nonlinear stability of the steady-state, spherically symmetric accretion flows. To this end, we specified initial conditions for the distribution function on a constant time slice satisfying the following properties: first, the initial distribution function ${\cal F}_i$ is non-negative and bounded from above by an equilibrium distribution function. Second, in the asymptotic region ${\cal F}_i$ converges (in a suitable sense) to a function $f_\infty(E)$ depending only on the energy of the particles. Under these assumptions, we proved that outside of the horizon the observables, including the current density and the stress energy-momentum tensor, converge pointwise to the corresponding observables associated with the stationary, spherical flow described by $f_\infty(E)$ along the world lines of static observers (see Theorem~\ref{Thm:Stability1}). Similar results hold for the convergence along the world lines of timelike, nonstatic observers with asymptotic constant and positive radial velocity (see Theorem~\ref{Thm:Stability2}). This proves an asymptotic stability result for the steady, spherical flows of collisionless matter. We emphasize that the initial data is not required to be spherically symmetric for our result to hold. Of course, the physical explanation for this result is that almost all the gas particles either disperse or fall into the black hole, so that after infinite time only those particles emanating from the reservoir in the asymptotic region with distribution $f_\infty(E)$ are ``seen" by the observer.

There are several important questions that have been left unanswered in this work and we would like to conclude this article by mentioning some of these points. First, although our stability result shows that (under the hypotheses made in this article) the kinetic gas relaxes in time to a stationary spherical configuration, a physically relevant problem is to determine the time scale associated with this relaxation process. Here, the distribution of the gas particles moving on or near the trajectories corresponding to the critical value of the angular momentum $\ell_c(E)$ (i.e. the value of $\ell$ below which the particle falls into the black hole and above which the particle is scattered at the potential barrier) should play a critical role. Indeed, a ``fine-tuned" distribution of particles whose angular momenta is precisely $\ell = \ell_c(E)$ will not decay in time since the particles take an infinite time to reach the unstable equilibrium point, corresponding to the unstable circular orbit. Of course, this example does not contradict our stability result since the corresponding distribution function $f$ has the form of a Dirac-delta-distribution in momentum space; hence it does not satisfy our assumptions. However, an interesting question is to consider a smooth distribution function (satisfying our assumptions) which is strongly peaked about the critical value $\ell = \ell_c(E)$. The instability of the equilibrium point implies that the peak will disperse in time, and it is precisely this dispersion process that should be responsible for the decay and dictate its rate.

Next, we emphasize that we have restricted ourselves to the invariant submanifold $\Gamma_{accr}$ of the relativistic phase space corresponding to unbounded trajectories. Thus, the results derived so far in this article apply to distribution functions which have their support in the region of phase space corresponding to such unbounded trajectories. Due to the linearity of the Liouville equation, the complementary case of bounded trajectories can be analyzed separately; hence, as long as the self-gravity and collisions between the gas particles can be neglected, our main results will not be influenced by the contributions from the bounded trajectories. In particular, these contributions affect neither the accretion rate nor the expressions for the particle and energy densities and pressures at the horizon or at infinity, and hence our results for the accretion and compression rates persist. Likewise, our stability results in Theorems~\ref{Thm:Stability1} and \ref{Thm:Stability2} will not be altered if a contribution due to gas particles moving on bounded trajectories is added, as long as the observables $J_\alpha$ and $T_{\alpha\beta}$ in these theorems are restricted to the contributions from the unbounded trajectories. However, an interesting question is whether or not the full observables including the contributions from gas particles on bounded orbits still relax to a stationary configuration along the world lines of static observers. This question is much more difficult to answer since in this case the gas particles do not disperse.

Finally, it should be interesting to explore the situation where collisions are taken into account and see how they affect the accretion rate. The analysis of this case is much more demanding than the one considered in the present article, since it requires solving the relativistic Boltzmann equation including the collision term on the curved Schwarzschild background geometry. Furthermore, in the collisional case, it is not possible to analyze the different cases (absorbed vs. scattered vs. bounded trajectories) separately since collisions might mix trajectories of different types. Nevertheless, it is our hope that the new symplectic coordinates introduced in the this article should simplify the analysis since they trivialize the transport part of the Boltzmann equation. We hope to address some of these issues in future work.


\acknowledgments

It is our pleasure to thank H\r{a}kan Andr{\'{e}}asson, Francisco Astorga, Dar\'io N\'u\~nez, Gerhard Rein, Manuel Tiglio, and Thomas Zannias for fruitful and stimulating discussions. We also thank Aftab Ahmad, James Edwards, and Thomas Zannias for reading a previous version of the manuscript and making suggestions. This research was supported in part by CONACyT Grants No. 577742 and No. 271904, and by a CIC Grant to Universidad Michoacana.

\appendix
\section{Properties of the effective potential}
\label{App:EffPot}

In this appendix, we briefly discuss the qualitative properties of the effective potential $V_{m,\ell}$ defined in equation~(\ref{Eq:Vmell}) describing the geodesics in the Schwarzschild spacetime. Since this is discussed in standard textbooks on general relativity (see for example~\cite{Straumann-Book}) here we focus on the properties needed for the analysis in this article.

\begin{figure}[ht]
\centerline{\resizebox{8.5cm}{!}{\includegraphics{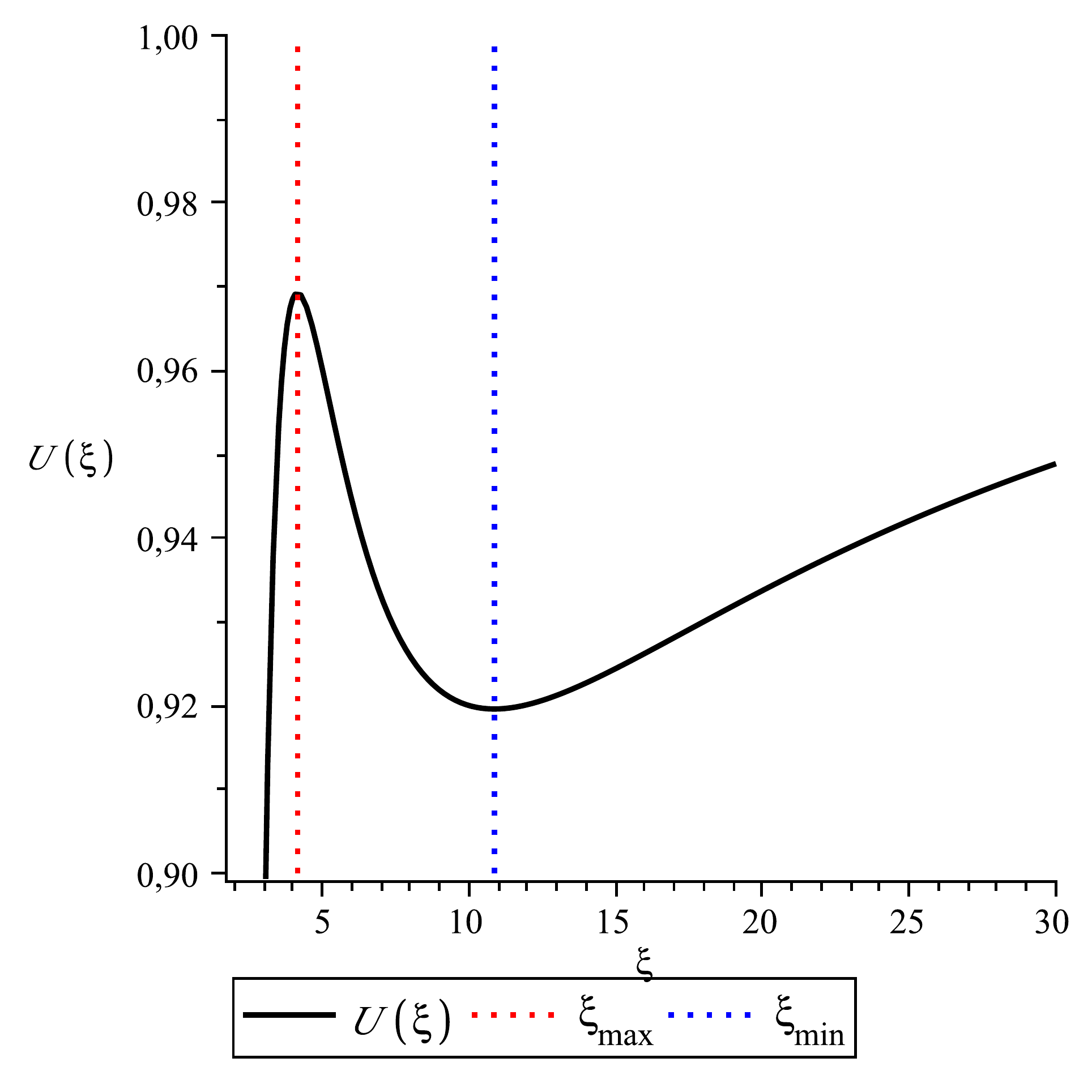}}}
\caption{The effective potential $U(\xi) = U_\lambda(\xi)$ as a function of $\xi$ for $\lambda = \sqrt{15}$.}
\label{Fig:potential}
\end{figure}

In terms of dimensionless variables, we have $V_{m,\ell}(r) = m^2 U_\lambda(\xi)$ with
\begin{equation}
U_\lambda(\xi) := \left( 1 - \frac{2}{\xi} \right)\left( 1 + \frac{\lambda^2}{\xi^2} \right)
 = 1 - \frac{2}{\xi} + \frac{\lambda^2}{\xi^2} - \frac{2\lambda^2}{\xi^3},\qquad
\xi = \frac{r}{M_H}.
\label{Eq:UlambdaBis}
\end{equation}
For $\lambda^2 \leq 12$ the function $U_\lambda : [2,\infty)\to \Real$ is a strictly monotonically increasing function from $0$ to $1$. When $\lambda^2 > 12$ this function has a local maximum (describing the potential barrier), which is given by
\begin{equation}
\xi_{max} = \frac{\lambda^2}{2}\left[ 1 - \sqrt{1 - \frac{12}{\lambda^2}} \right],\qquad
U_\lambda(\xi_{max}) = \frac{8}{9} + \frac{\lambda^2 - 12}{9\xi_{max}},
\label{Eq:ximax}
\end{equation}
and a local minimum (describing a potential well) given by
\begin{equation}
\xi_{min} = \frac{\lambda^2}{2}\left[ 1 + \sqrt{1 - \frac{12}{\lambda^2}} \right],\qquad
U_\lambda(\xi_{min}) = \frac{8}{9} + \frac{\lambda^2 - 12}{9\xi_{min}},
\label{Eq:ximin}
\end{equation}
see figure~\ref{Fig:potential}. As $\lambda^2$ grows from $12$ to $\infty$, $\xi_{max}$ decreases monotonically from $6$ to $3$ (the radius of the photon sphere) while $\xi_{min}$ increases monotonically from $6$ (the innermost stable circular orbit) to $\infty$. Accordingly, $U_\lambda(\xi_{max})$ grows monotonically from $8/9$ to $\infty$ while $U_\lambda(\xi_{min})$ grows monotonically from $8/9$ to $1$.

Given $\varepsilon > \sqrt{8/9}$, the critical value $\lambda_c(\varepsilon) > \sqrt{12}$ for the total angular momentum for which the maximum of the potential barrier is exactly equal to $\varepsilon^2$ is given by
$$
\lambda_c(\varepsilon)^2 = \frac{12}{1 - 4\alpha - 8\alpha^2 + 8\alpha\sqrt{\alpha^2 + \alpha}},
\qquad \alpha := \frac{9}{8}\varepsilon^2 - 1 > 0.
$$

As $\varepsilon$ increases from $\sqrt{8/9}$ to $\infty$, $\lambda_c(\varepsilon)$ increases monotonically from $\sqrt{12}$ to $\infty$. For large $\alpha$ we have $\lambda_c(\varepsilon)^2 = 24\alpha[ 1 + {\cal O}(\alpha^{-1}) ]$. For the particular value $\varepsilon = 1$, corresponding to the case of particles with vanishing radial velocity at infinity, we have $\lambda_c(\varepsilon) = 4$ and the maximum is located at $\xi_{max} = 4$.

Likewise, given $\varepsilon \in (\sqrt{8/9},1)$, there is an upper bound for the angular momentum $\lambda_{ub}(\varepsilon)$ corresponding to the case where the minimum of the potential well is equal to $\varepsilon^2$, which is given by
$$
\lambda_{ub}(\varepsilon)^2 = \frac{12}{1 - 4\alpha - 8\alpha^2 - 8\alpha\sqrt{\alpha^2 + \alpha}}.
$$
As $\varepsilon$ increases from $\sqrt{8/9}$ to $1$, $\lambda_{ub}(\varepsilon)$ increases monotonically from $\sqrt{12}$ to $\infty$.

With these observations we can classify the orbits in the following way. For $\sqrt{8/9} < \varepsilon < 1$ the particles are scattered at the effective potential and fall into the black hole when $\lambda < \lambda_c(\varepsilon)$, while the orbits are bounded when $\lambda_c(\varepsilon) < \lambda < \lambda_{ub}(\varepsilon)$. For $\varepsilon > 1$ the particles are falling into the black hole when $\lambda < \lambda_c(\varepsilon)$ while they are scattered at the potential when $\lambda > \lambda_c(\varepsilon)$.

\section{Some useful integral identities}
\label{App:integrals}

In this appendix, we summarize some of the integral identities used in the derivation of equations~(\ref{Eq:JainSS}--\ref{Eq:TthetathetascatSS}). Introducing the shorthand notation $s_{1,2} := \sqrt{\varepsilon^2 - U_{\lambda_{1,2}}(\xi)}$ we have, for any $0 \leq \lambda_1 < \lambda_2 \leq \lambda_{max}$,
\begin{eqnarray*}
\int\limits_{\lambda_1}^{\lambda_2} 
\frac{\lambda d\lambda}{\sqrt{\varepsilon^2 - U_\lambda(\xi)}}
 &=& -\frac{\xi^2}{1 - \frac{2}{\xi}}(s_2 - s_1)
 = \frac{\lambda_2^2 - \lambda_1^2}{s_1 + s_2},
\\
\int\limits_{\lambda_1}^{\lambda_2} \pi_{\xi\pm}(\xi)
\frac{\lambda d\lambda}{\sqrt{\varepsilon^2 - U_\lambda(\xi)}}
 &=& \frac{\lambda_2^2 - \lambda_1^2}{1 - \frac{2}{\xi}}\left[
  \frac{2\varepsilon}{\xi}\frac{1}{s_1 + s_2} \pm \frac{1}{2} \right],
\\
\int\limits_{\lambda_1}^{\lambda_2} \pi_{\xi\pm}^2(\xi)
\frac{\lambda d\lambda}{\sqrt{\varepsilon^2 - U_\lambda(\xi)}}
 &=& \frac{\lambda_2^2 - \lambda_1^2}{s_1 + s_2}
\left[ \left( \frac{\frac{2\varepsilon}{\xi} \pm s_1}{1 - \frac{2}{\xi}} \right)
\left( \frac{\frac{2\varepsilon}{\xi} \pm s_2}{1 - \frac{2}{\xi}} \right)
 + \frac{1}{3}\left( \frac{s_2 - s_1}{1 - \frac{2}{\xi}} \right)^2 \right],
\\
\int\limits_{\lambda_1}^{\lambda_2} \lambda^2
\frac{\lambda d\lambda}{\sqrt{\varepsilon^2 - U_\lambda(\xi)}}
 &=& \frac{1}{3}\frac{\lambda_2^2 - \lambda_1^2}
 {(s_1 + s_2)^2}\left[ \lambda_1^2(s_1 + 2s_2) + \lambda_2^2(s_2 + 2s_1) \right].
\end{eqnarray*}
For the particular in which $\pi_\xi = \pi_{\xi-}(\xi)$, the second and third identities can also be rewritten as
\begin{eqnarray*}
\int\limits_{\lambda_1}^{\lambda_2} \pi_{\xi-}(\xi)
\frac{\lambda d\lambda}{\sqrt{\varepsilon^2 - U_\lambda(\xi)}}
 &=& -\frac{\lambda_2^2 - \lambda_1^2}{s_1 + s_2}
\left[ \varepsilon - \frac{1}{2}\frac{1 + \frac{\lambda_1^2}{\xi^2}}{\varepsilon + s_1}
 - \frac{1}{2}\frac{1 + \frac{\lambda_2^2}{\xi^2}}{\varepsilon + s_2} \right],
\\
\int\limits_{\lambda_1}^{\lambda_2} \pi_{\xi-}^2(\xi)
\frac{\lambda d\lambda}{\sqrt{\varepsilon^2 - U_\lambda(\xi)}}
 &=& \frac{\lambda_2^2 - \lambda_1^2}{s_1 + s_2}
\left[ \left( \varepsilon - \frac{1 + \frac{\lambda_1^2}{\varepsilon^2}}{\varepsilon + s_1} \right)
 \left( \varepsilon - \frac{1 + \frac{\lambda_2^2}{\varepsilon^2}}{\varepsilon + s_2} \right)
 + \frac{1}{3\xi^4}\left( \frac{\lambda_2^2 - \lambda_1^2}{s_1 + s_2}
 \right)^2 \right],
\end{eqnarray*}
which shows that the corresponding expressions are regular on the horizon $\xi =2$.

\section{Explicit expressions in the low temperature limit}
\label{App:Limits}

In this appendix, we provide some details about the calculations required for understanding the low temperature limit $z = m\beta\to \infty$ in section~\ref{SubSec:SSS}. We start with the computation of the ratio between the mass accretion rate and the energy density at infinity, that is,
$$
\frac{\dot{M}}{n_{\infty}} = \frac{m j_n}{n_{\infty}}.
$$
Using equation~(\ref{j_n}) and equation~(\ref{n_infty}) for the particle density at infinity, which we rewrite as
$$
n_\infty = 4\pi\alpha m^4
\int\limits_1^\infty e^{-z\varepsilon} \sqrt{\varepsilon ^2 -1 }\varepsilon d\varepsilon,
$$
we obtain
$$
\frac{m j_n}{n_{\infty}} = - \pi m M_H^2  \frac{ 
\int\limits_1^\infty e^{-z(\varepsilon -1)}\lambda_c(\varepsilon)^2 d\varepsilon }{  \int \limits _{1}^{\infty} e^{-z(\varepsilon -1)} \sqrt{\varepsilon ^2 -1 }\varepsilon d\varepsilon}
 = -\pi m M_H^2 \sqrt{z}  \frac{ 
\int\limits_0^\infty e^{-x}\lambda_c(1+\frac{x}{z})^2 dx }{  \int \limits _{0}^{\infty} e^{-x} \sqrt{2x + \frac{x^2}{z}}\left(1+\frac{x}{z} \right) dx},
$$
where we have applied the variable substitution $\varepsilon\mapsto x=z(\varepsilon -1)$ in the last step. Dividing both sides of the equation by $\sqrt{z}$, taking the limit $z\to\infty$ and recalling that $\lambda _{c}\left( 1\right) = 4$ we obtain
$$
\lim\limits_{z\to\infty} \frac{1}{\sqrt{z}}\frac{\dot{M}}{n_{\infty}} 
 = -16\pi m M_H^2\frac{ \int\limits_0^\infty e^{-x} dx }
 { \int \limits _{0}^{\infty} e^{-x} \sqrt{ 2x } dx} = - 16\sqrt{2\pi} m M_H^2.
$$

Next, to determine the particle density at the horizon, we use a similar computation to find from equation~(\ref{J-Horizon}),
$$
\lim\limits_{z\to\infty} \frac{1}{\sqrt{z}}\frac{J^\tau}{n_\infty} = \frac{5}{\sqrt{2\pi}},\qquad
\lim\limits_{z\to\infty} \frac{1}{\sqrt{z}}\frac{J^r}{n_\infty} = -\sqrt{\frac{2}{\pi}}.
$$
Using $n_H = \left. \sqrt{-g_{\mu\nu} J^\mu J^\nu}\right|_{\xi=2}$ we find
\begin{equation}
\lim\limits_{z\to\infty} \frac{1}{\sqrt{z}}\frac{n_H}{n_\infty} = \sqrt{\frac{6}{\pi}}.
\label{Eq:nHninfty}
\end{equation}

Finally, in order to compute the energy density and principle pressures at the horizon, we rewrite equation~(\ref{T-Horizon}) in the form
$$
T^\tau{}_\tau = -A - B,\quad
T^\tau{}_r = -A + H,\quad
T^ r{}_\tau = A,\quad
T^r{}_r  = A - B,\quad
T^\vartheta {}_\vartheta = T^\varphi {}_{\varphi} = D,
$$
with 
\begin{eqnarray}
A &=&\frac{1}{4}\pi\alpha m^5\int\limits_1^\infty \lambda_c(\varepsilon)^2 \varepsilon e^{-z\varepsilon}  d\varepsilon ,\\
B &=&\frac{1}{4}\pi\alpha m^5\int\limits_1^\infty \frac{\lambda_c(\varepsilon)^2}{2\varepsilon} \left(1+\frac{\lambda ^2_{c}(\varepsilon)}{8} \right) e^{-z\varepsilon} d\varepsilon ,\\
H &=&\frac{1}{4}\pi\alpha m^5\int\limits_1^\infty \frac{\lambda_c(\varepsilon)^2}{4 \varepsilon ^3} \left( 1 +\frac{\lambda ^2_{c}(\varepsilon)}{4} + \frac{\lambda ^4_{c}(\varepsilon)}{48}\right) e^{-z\varepsilon} d\varepsilon ,\\
D &=& \frac{1}{64}\pi\alpha m^5\int\limits_1^\infty \frac{\lambda_c(\varepsilon)^4}{\varepsilon}
e^{-z\varepsilon} d\varepsilon.
\end{eqnarray}
In terms of these quantities, the eigenvalues of the matrix $T^\mu{}_\nu$ are
$$
\lambda_\pm = -B \pm \sqrt{A H},\qquad D.
$$
Dividing the expressions for $A$, $B$, $H$ and $D$ by $\sqrt{z} n_\infty$ and taking the limit $z\to \infty$ we obtain
$$
\lim\limits_{z\to \infty}\frac{1}{\sqrt{z}}\frac{A}{n_\infty} 
 = \lim\limits_{z\to \infty}\frac{1}{\sqrt{z}}\frac{D}{n_\infty} = \sqrt{\frac{2}{\pi}}m,\qquad
\lim\limits_{z\to \infty}\frac{1}{\sqrt{z}}\frac{B}{n_\infty} = \frac{3}{\sqrt{2\pi}}m, \qquad
\lim\limits_{z\to \infty}\frac{1}{\sqrt{z}}\frac{H}{n_\infty} = \frac{31}{12}\sqrt{\frac{2}{\pi}}m,
$$
from which
\begin{eqnarray*}
\lim\limits_{z\to \infty}\frac{1}{\sqrt{z}} \frac{\rho_H}{n_{\infty}}
 &=& \lim\limits_{z\to \infty}\frac{1}{\sqrt{z}} \frac{-\lambda_-}{n_{\infty}}
  = \frac{1}{\sqrt{2\pi}} \left( 3 + \sqrt{\frac{31}{3}} \right)m,\\
\lim\limits_{z\to \infty}\frac{1}{\sqrt{z}} \frac{p_{rad}}{n_{\infty}}
 &=& \lim\limits_{z\to \infty}\frac{1}{\sqrt{z}} \frac{\lambda_+}{n_{\infty}}
  = \frac{1}{\sqrt{2\pi}} \left(-3 + \sqrt{\frac{31}{3}} \right)m,\\
\lim\limits_{z\to \infty}\frac{1}{\sqrt{z}} \frac{p_{tan}}{n_{\infty}}
 &=& \lim\limits_{z\to \infty}\frac{1}{\sqrt{z}} \frac{D}{n_{\infty}}
  = \sqrt{\frac{2}{\pi}} m.
\end{eqnarray*}
Together with equation~(\ref{Eq:nHninfty}) this yields the results given in equations~(\ref{Eq:nH},\ref{Eq:pradH},\ref{Eq:ptanH}).

\section{Proof of Lemma~\ref{Lem:mtauG}}
\label{App:mtauG}

To prove Lemma~\ref{Lem:mtauG} we need to determine the limits
$$
\lim\limits_{\tau\to\infty} \left[ G_{\varepsilon,\lambda(v,\xi(\tau))}(\xi(\tau),\pi_{\xi\pm}(\tau))
 - \tau \right]
$$
for fixed $0 < v < \sqrt{\varepsilon^2 - 1}$, where the function $G_{\varepsilon,\lambda}: {\cal C}\to \Real$ is defined in Lemma~\ref{Lem:Gepslam}, and where $\lambda(v,\xi)$ is defined in equation~(\ref{Eq:lambdavxi}). For the lower sign, $\pi_{\xi-}(\tau)$, the limit is $-\infty$ since in this case $G_{\varepsilon,\lambda(v,\xi(\tau))}(\xi(\tau),\pi_{\xi-}(\tau))\leq 0$. This proves the first statement of the lemma.

The other case is more delicate since here $G_{\varepsilon,\lambda(v,\xi(\tau))}(\xi(\tau),\pi_{\xi+}(\tau))$ is positive and thus this term may dominate $-\tau$ as $\tau \to \infty$. To analyze this case we write
\begin{equation}
G_{\varepsilon, \lambda(v,\xi(\tau))}(\xi(\tau), \pi_{\xi+}(\tau)) - \tau
= \tau \left[ \frac{\xi(\tau)}{\tau}
\frac{1}{\xi(\tau)}\int \limits_{\xi_0}^{\xi(\tau)} \frac{\frac{2}{x}\sqrt{\varepsilon^2 - U_{\lambda}(x)} + \varepsilon}{\left(1-\frac{2}{x} \right)\sqrt{\varepsilon^2 - U_{\lambda}(x)}} dx - 1
 \right]_{\lambda = \lambda(v,\xi(\tau))},
\label{Eq:mtauG}
\end{equation}
with $\xi_0$ the radius of the turning point, where $\sqrt{\varepsilon^2 - U_\lambda(\xi_0)}$ vanishes. In order to understand the behavior of the integral in equation~(\ref{Eq:mtauG}) for large $\tau$, it is convenient to change the integration variable $x$ to the new variable $\bar{v} := \sqrt{\varepsilon^2-U_{\lambda}(x)}$ which ranges from $0$ to $v$. More precisely, we consider for fixed $\xi = \xi(\tau) > \xi_0$ and $\lambda = \lambda(v,\xi)$ the auxiliary function
\begin{eqnarray*}
H_\lambda : (\xi_0,\xi_{min}) &\to& (0,\sqrt{\varepsilon^2 - U_\lambda(\xi_{min})})\\
x &\mapsto& \sqrt{\varepsilon^2 - U_\lambda(x)},
\end{eqnarray*}
where $\xi_{min}$ is the location of the minimum of the effective potential $U_\lambda$, see equation~(\ref{Eq:ximin}) in App.~\ref{App:EffPot}. Since by definition of $\xi_0$ and $\xi_{min}$ the function $H_\lambda$ is monotonously increasing, it is invertible. Its derivative is
$$
\frac{dH_\lambda}{dx}(x) = -\frac{1}{2H_\lambda(x)} \frac{dU_\lambda}{dx}(x) > 0,
\qquad \xi_0 < x < \xi_{min}.
$$
In order to use the variable substitution $\bar{v} := H_\lambda(x)$ in the integral on the right-hand side of equation~(\ref{Eq:mtauG}) we need to make sure that $\xi = \xi(\tau)$ lies in the required interval $(\xi_0,\xi_{min})$. In order to see that this condition is satisfied for large enough $\xi$, recall equation~(\ref{Eq:ximin}) which shows that for large values of $\lambda$ we have $\xi_{min} = \xi_{min}(\lambda) \simeq \lambda^2$. Furthermore, it follows immediately from equation~(\ref{Eq:lambdavxi}) that
\begin{equation}
\lim_{\xi\to\infty} \frac{\lambda(v,\xi)}{\xi} = \sqrt{\varepsilon^2 - v^2 - 1}.
\label{Eq:Limit1}
\end{equation}
Hence,
$$
\lim\limits_{\xi\to \infty} \frac{\xi_{min}(\lambda(v,\xi))}{\xi^2}
 = \lim\limits_{\xi\to \infty} \frac{\lambda(v,\xi)^2}{\xi^2}
 \frac{\xi_{min}(\lambda(v,\xi))}{\lambda(v,\xi)^2} = \varepsilon^2 - v^2 - 1 > 0,
$$
which guarantees that for large enough $\xi$ we have  $\xi < \xi_{min}$. Therefore, for such large $\xi$ the variable substitution $\bar{v} := H_\lambda(x)$ is allowed and we can write
\begin{equation}
\frac{1}{\xi}\left. \int \limits_{\xi_0}^{\xi} \frac{\frac{2}{x}\sqrt{\varepsilon^2 - U_{\lambda}(x)} + \varepsilon}{\left(1-\frac{2}{x} \right)\sqrt{\varepsilon^2 - U_{\lambda}(x)}} dx 
\right|_{\lambda = \lambda(v,\xi)}
 = -\frac{2}{\xi}\int \limits_0^v \left.
  \frac{\frac{2}{x} \bar{v} + \varepsilon}{\left( 1 - \frac{2}{x} \right)\frac{dU_\lambda}{dx}(x)} 
\right|_{x = H_\lambda^{-1}(\bar{v}), \lambda = \lambda(v,\xi)} d\bar{v}.
\label{Eq:GDivxi}
\end{equation}

In order to compute the required limit for $\xi\to \infty$ we use:

\begin{lemma}
\label{Lem:Limits}
Let $\varepsilon > 1$ and $0 < v < \sqrt{\varepsilon^2 - 1}$. Then, 
\begin{equation}
\lim_{\lambda\to \infty} \frac{H_\lambda^{-1}(\bar{v})}{\lambda} 
 = \frac{1}{\sqrt{\varepsilon^2 - \bar{v}^2 - 1}}
\label{Eq:Limit2}
\end{equation}
for all $0\leq \bar{v} \leq v$.
\end{lemma}

\proof Set $x := H_\lambda^{-1}(\bar{v})$. By definition, $x > \xi_0 > 3$, and hence
$$
\varepsilon^2 - \bar{v}^2 = U_\lambda(x) 
 = \left( 1 - \frac{2}{x} \right)\left( 1 + \frac{\lambda^2}{x^2} \right) 
 \geq \frac{1}{3}\left( 1 + \frac{\lambda^2}{x^2} \right).
$$
This implies that $\lambda/x$ is bounded as $\lambda\to \infty$, which in turn implies that $x\to \infty$ as $\lambda\to \infty$. Consequently,
$$
\lim\limits_{\lambda\to \infty} \frac{\lambda^2}{x^2} 
 = \lim\limits_{\lambda\to \infty}\frac{\varepsilon^2 - \bar{v}^2}{1 - \frac{2}{x}} - 1
 = \varepsilon^2 - \bar{v}^2 - 1
$$
which proves the lemma.
\qed

With the help of the limits given in equations~(\ref{Eq:Limit1},\ref{Eq:Limit2}) we find
$$
\lim\limits_{\xi\to \infty} \xi\left. \frac{dU_\lambda}{dx}(x)
 \right|_{x = H_\lambda^{-1}(\bar{v}), \lambda = \lambda(v,\xi)}
= \lim\limits_{\xi\to \infty} \frac{\xi}{\lambda(v,\xi)}
\left[ -\frac{2\lambda^3}{x^3} + \frac{2\lambda}{x^2} + \frac{6\lambda^3}{x^4} \right]_{x = H_\lambda^{-1}(\bar{v}), \lambda = \lambda(v,\xi)}
 = -2\frac{(\varepsilon^2 - \bar{v}^2 - 1)^{3/2}}{(\varepsilon^2 - v^2 - 1)^{1/2}}.
$$
Hence, it follows from equation~(\ref{Eq:GDivxi}) that
$$
\lim\limits_{\xi\to \infty} \frac{1}{\xi}\left. \int \limits_{\xi_0}^{\xi} \frac{\frac{2}{x}\sqrt{\varepsilon^2 - U_{\lambda}(x)} + \varepsilon}{\left(1-\frac{2}{x} \right)\sqrt{\varepsilon^2 - U_{\lambda}(x)}} dx 
\right|_{\lambda = \lambda(v,\xi)}
 =  \varepsilon
\int\limits_0^v\frac{(\varepsilon^2 - v^2 - 1)^{1/2}}{(\varepsilon^2 - \bar{v}^2 - 1)^{3/2}} d\bar{v}
 = \varepsilon \frac{v}{\varepsilon^2 - 1}.
$$
Therefore, using equation~(\ref{Eq:mtauG}) we obtain
\begin{eqnarray*}
\lim\limits_{\tau\to \infty} 
\frac{1}{\tau}\left[ G_{\varepsilon, \lambda(v,\xi(\tau))}(\xi(\tau), \pi_{\xi+}(\tau)) - \tau \right]
 &=& \lim\limits_{\tau\to \infty}\left[ \frac{\xi(\tau)}{\tau}
\frac{1}{\xi(\tau)}\int \limits_{\xi_0}^{\xi(\tau)} \frac{\frac{2}{x}\sqrt{\varepsilon^2 - U_{\lambda}(x)} + \varepsilon}{\left(1-\frac{2}{x} \right)\sqrt{\varepsilon^2 - U_{\lambda}(x)}} dx - 1
 \right]_{\lambda = \lambda(v,\xi(\tau))}Ê\\
 &=& v_{obs,\infty}\frac{\varepsilon v}{\varepsilon^2 - 1} - 1 = \frac{v}{a(\varepsilon)} - 1
\end{eqnarray*}
for all $0 < v < \sqrt{\varepsilon^2 - 1}$. This implies the second statement of Lemma~\ref{Lem:mtauG}.
\qed

\bibliographystyle{unsrt}
\bibliography{../References/refs_kinetic}

\end{document}